\documentclass[final]{cvpr}

\usepackage{times}
\usepackage{epsfig}
\usepackage{graphicx}
\usepackage{amsmath}
\usepackage{amssymb}
\usepackage{array} 
\usepackage{booktabs} %
\usepackage{makecell}
\usepackage{diagbox}

\usepackage{enumitem} %
\setlist{nosep} %

\usepackage{multirow}
\usepackage[pagebackref=true,breaklinks=true,colorlinks,bookmarks=false]{hyperref}

\def\cvprPaperID{7679} %
\def\confYear{CVPR 2021}

\long\def\comment#1{}

\begin{document}

\title{Unsupervised Deep Video Denoising}

\author{Dev Yashpal Sheth$^{1*}$, \hspace{0.1cm} Sreyas Mohan$^{2}$\thanks{equal contribution.}, \hspace{0.1cm} Joshua L. Vincent$^{3}$, \hspace{0.1cm} Ramon Manzorro$^{3}$, \hspace{0.1cm}Peter A. Crozier$^{3}$, \\Mitesh M. Khapra$^{1,6}$, \hspace{0.3cm}Eero P. Simoncelli$^{2,4,5}$, \hspace{0.3cm}Carlos Fernandez-Granda$^{2,5}$ \vspace{0.2cm} \\ 
$^{1}$Indian Institute of Technology Madras, India, 
$^{2}$Center for Data Science, New York University, \\
$^{3}$School for Engineering of Matter, Transport, and Energy, ASU, \\
$^{4}$Center for Neural Science, NYU and Flatiron Institute, Simons Foundation, \\
$^{5}$Courant Institute of Mathematical Sciences, NYU,
$^{6}$Robert Bosch Center for Data Science and AI.
}

\maketitle

\begin{abstract}

Deep convolutional neural networks (CNNs) for video denoising are typically trained with supervision, assuming the availability of clean videos. However, in many applications, such as microscopy, noiseless videos are not available. To address this, we propose an Unsupervised Deep Video Denoiser (UDVD\footnote{See \href{https://sreyas-mohan.github.io/udvd/}{https://sreyas-mohan.github.io/udvd/} for code and more results.}), a CNN architecture designed to be trained exclusively with noisy data. The performance of UDVD is comparable to the supervised state-of-the-art, even when trained only on a single short noisy video. We demonstrate the promise of our approach in real-world imaging applications by denoising raw video, fluorescence-microscopy and electron-microscopy data. In contrast to many current approaches to video denoising, UDVD does not require explicit motion compensation. This is advantageous because motion compensation is computationally expensive, and can be unreliable when the input data are noisy. A gradient-based analysis reveals that UDVD automatically adapts to local motion in the input noisy videos. Thus, the network learns to perform implicit motion compensation, even though it is only trained for denoising.

\end{abstract}

\section{Introduction}

Video denoising is a fundamental problem in image processing, as well as an important preprocessing step for computer vision tasks. 
Convolutional neural networks (CNNs) \cite{lecun2015deep} provide current state-of-the-art solutions for this problem \cite{dvdnet, fastdvdnet, xue2019video,rawvideo,f2f, mf2f, vnlnet, videnn}. These networks are typically trained using a database of clean videos, which are corrupted with simulated noise. However, in applications such as microscopy, noiseless ground truth videos are often not available. To address this issue, we propose a  method to train a video denoising CNN without access to supervised data, which we call Unsupervised Deep Video Denoising (UDVD). 
UDVD is inspired by the ``blind-spot'' technique, recently introduced for unsupervised still image denoising~\cite{n2n, n2v, noise2self,blindspotnet}, in which a CNN is trained to estimate each {\em noisy} pixel from the surrounding spatial neighborhood \emph{without including the pixel itself}. Here, we propose a blind-spot architecture that processes the surrounding spatio-temporal neighborhood to denoise videos. 

\begin{figure*}[ht]
    \def\f1ht{0.33\textwidth}%
    \centering 
    \begin{tabular}{ >{\centering\arraybackslash}m{0.31\textwidth}
     >{\centering\arraybackslash}m{0.31\textwidth} 
     >{\centering\arraybackslash}m{0.31\textwidth}}
     \centering
     
     \hspace{2mm}  \footnotesize{(a) Clean frame, PSNR / SSIM} &   \footnotesize{(b) Noisy Input, 19.06 / 0.279} &  \footnotesize{(c) Supervised (FastDVDnet), 31.73 / 0.873}  \\

     \includegraphics[width=\f1ht]{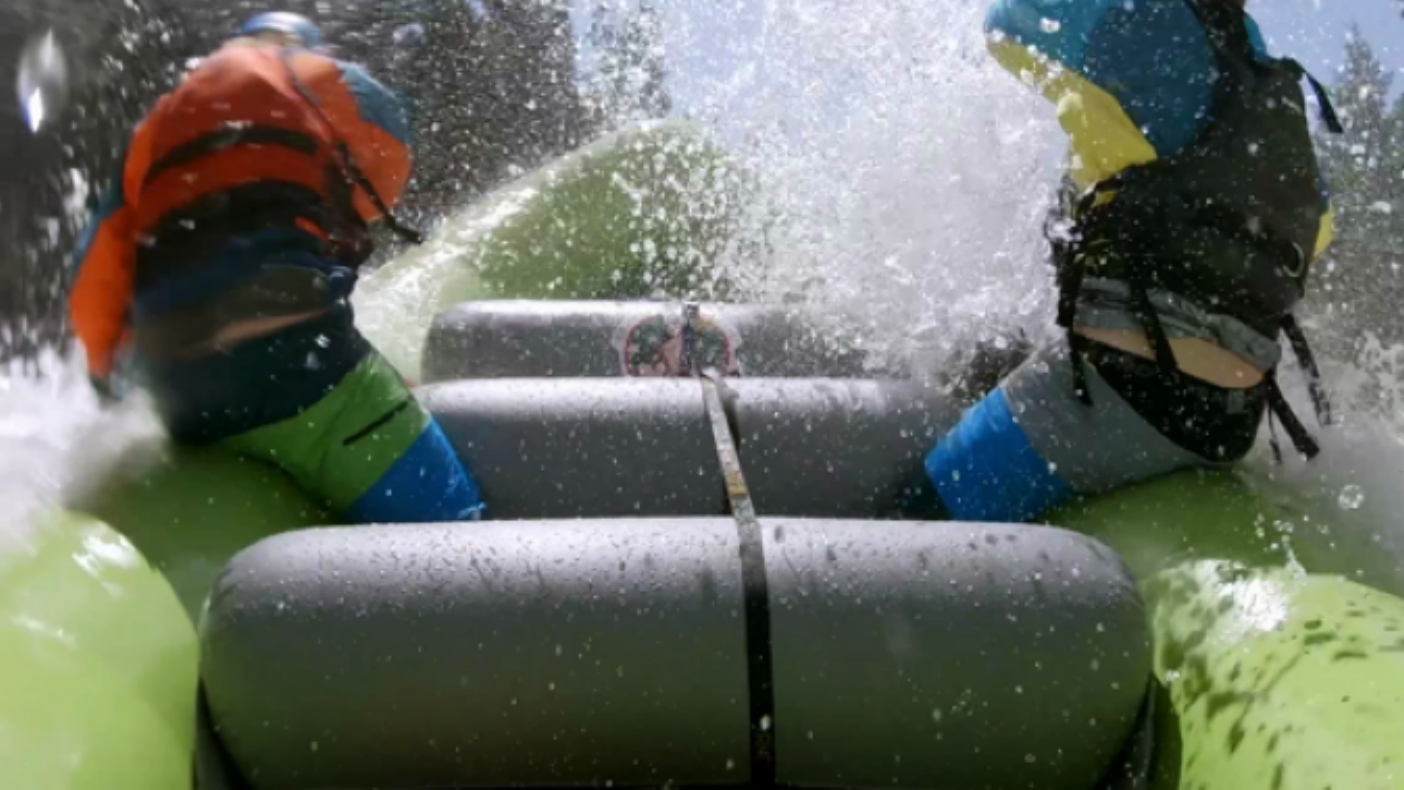} &
     \includegraphics[width=\f1ht]{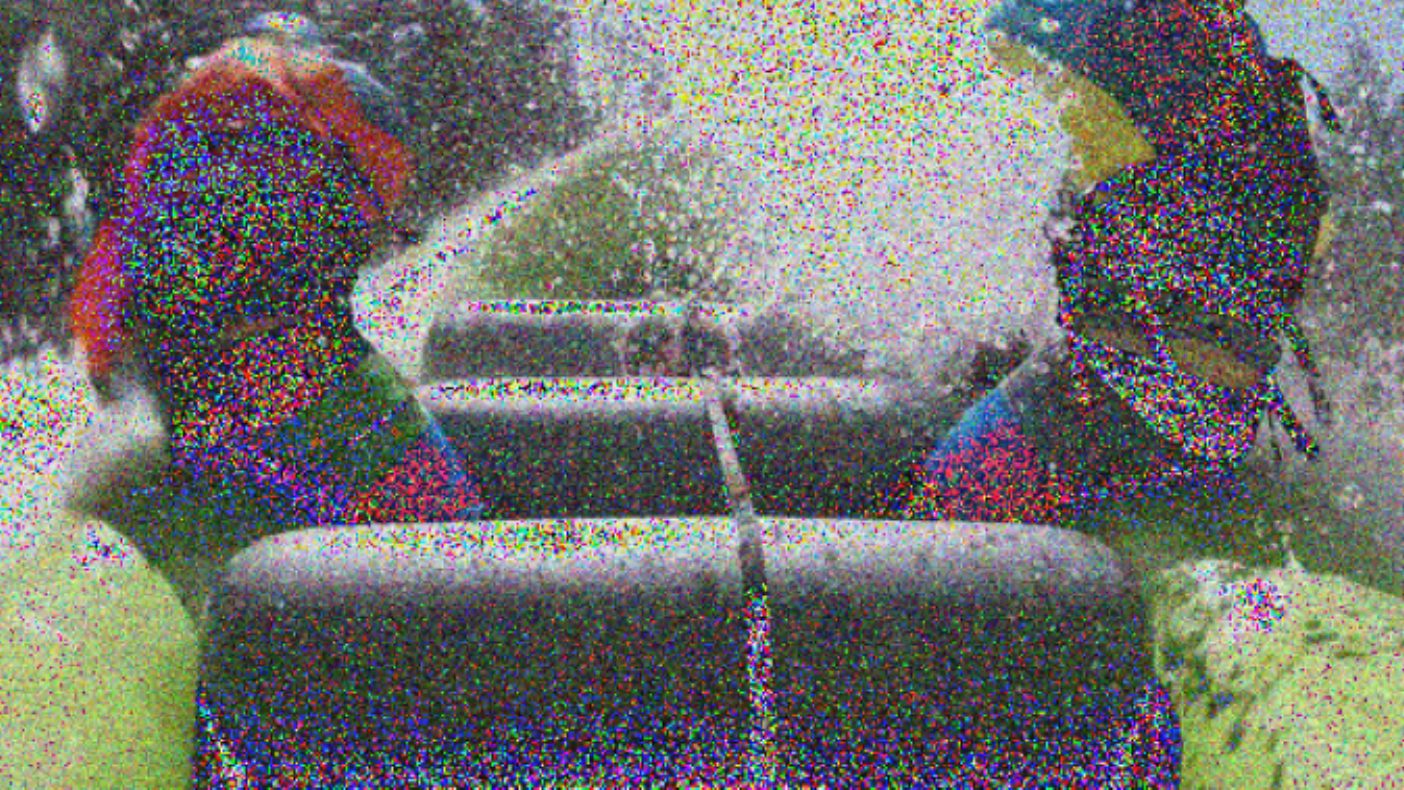} &
     \includegraphics[width=\f1ht]{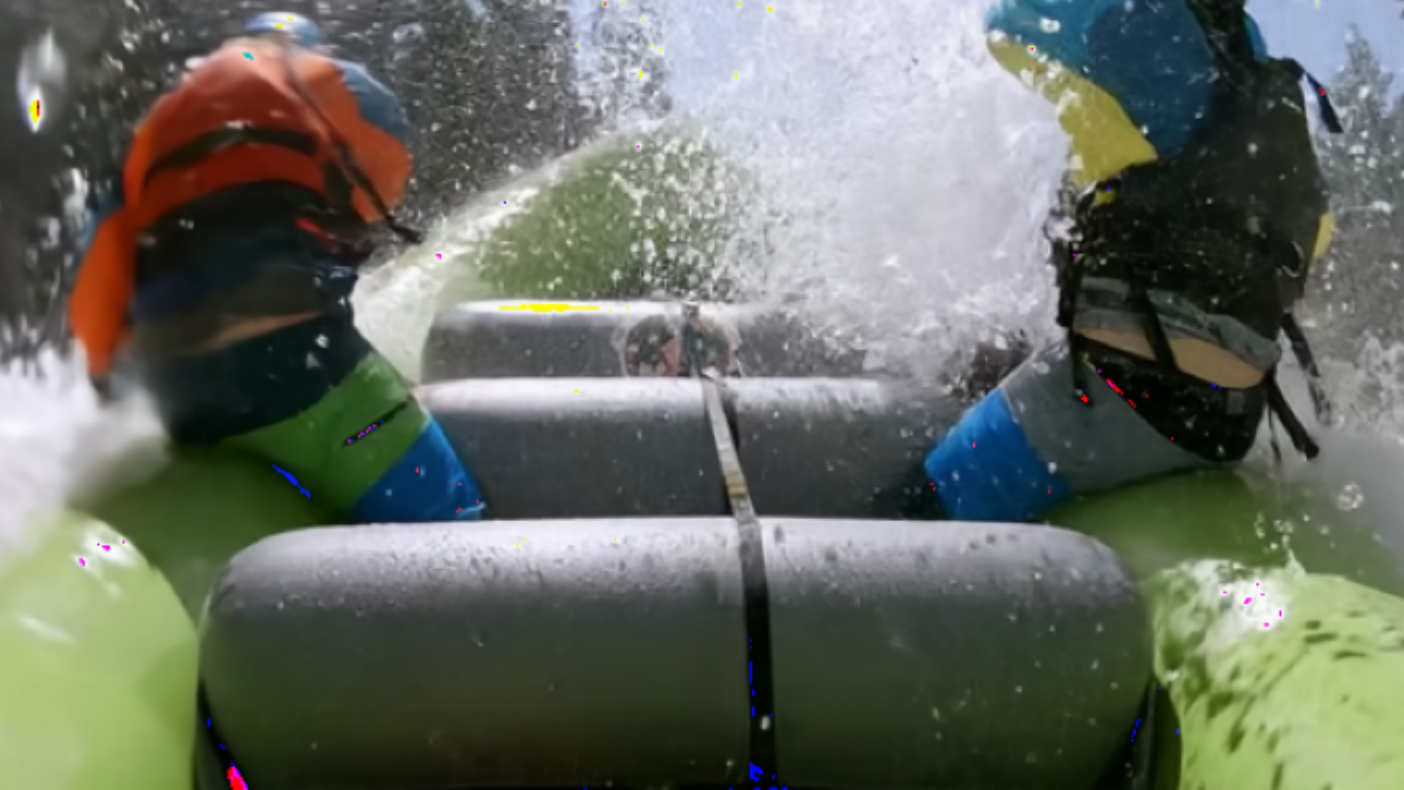} \\
     
     \hspace{2mm}  \footnotesize{(d) Unsupervised (MF2F) %
     , 30.35 / 0.825} &   \footnotesize{(e) UDVD, 31.62 / 0.869} & \footnotesize{(f) UDVD-S, 31.39 / 0.865}  \\
     
     \includegraphics[width=\f1ht]{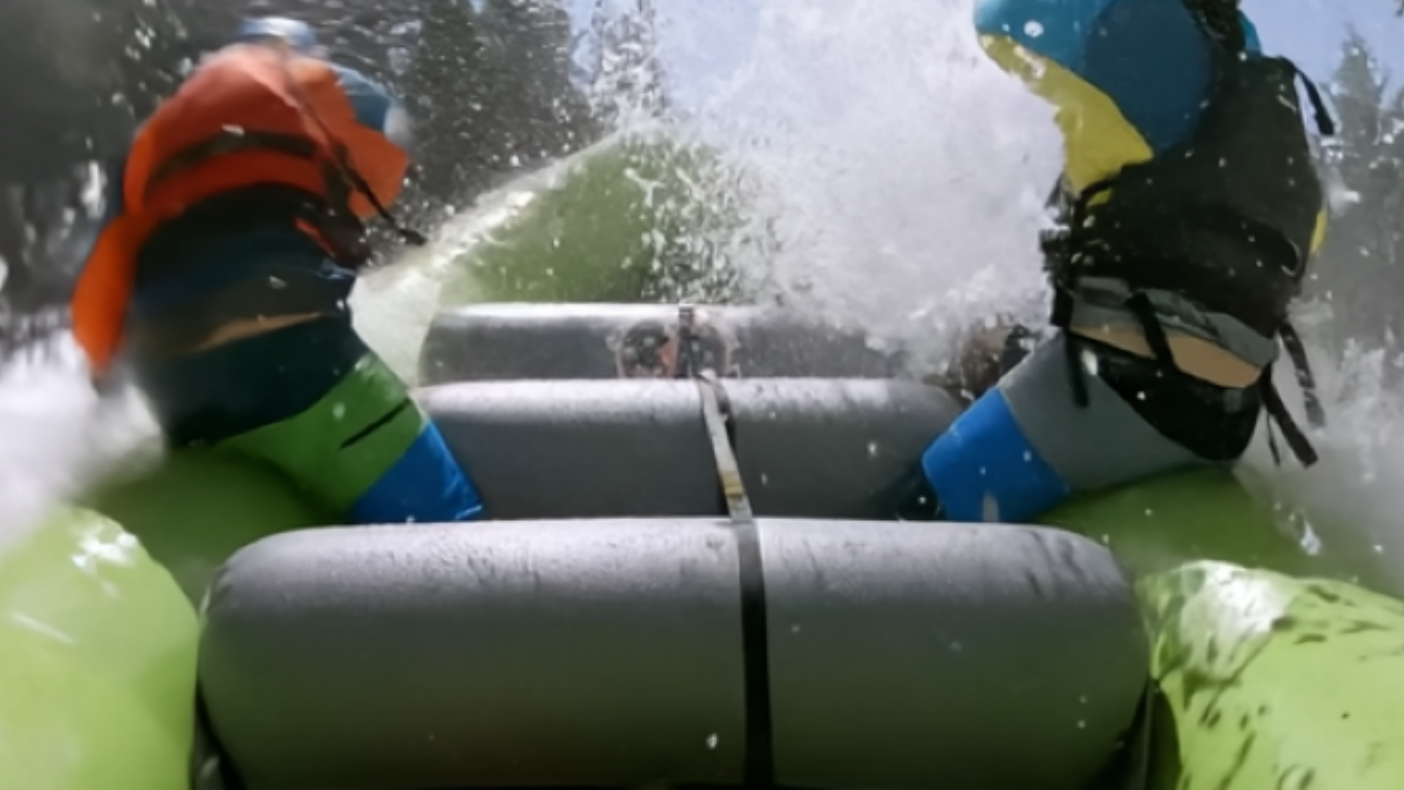} &
     \includegraphics[width=\f1ht]{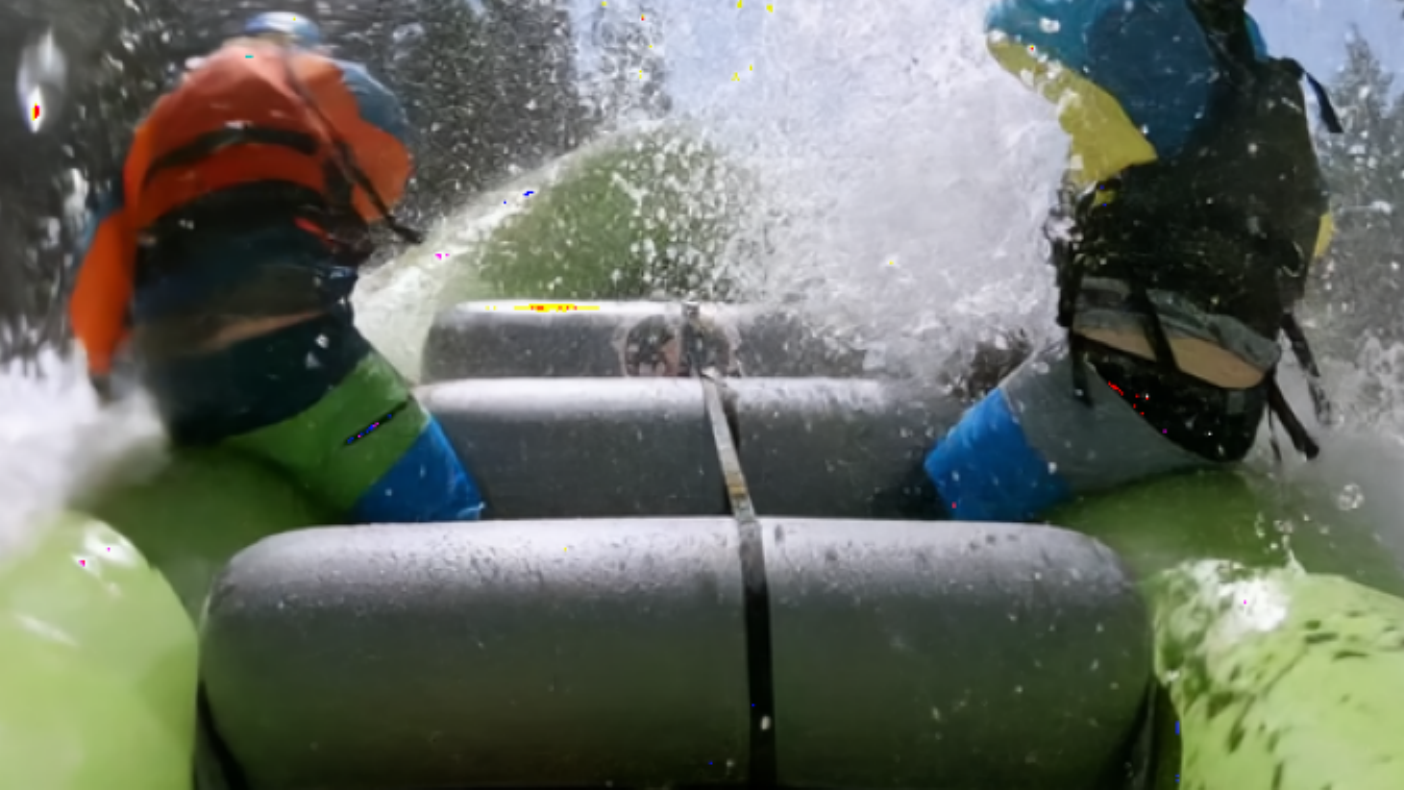} &
     \includegraphics[width=\f1ht]{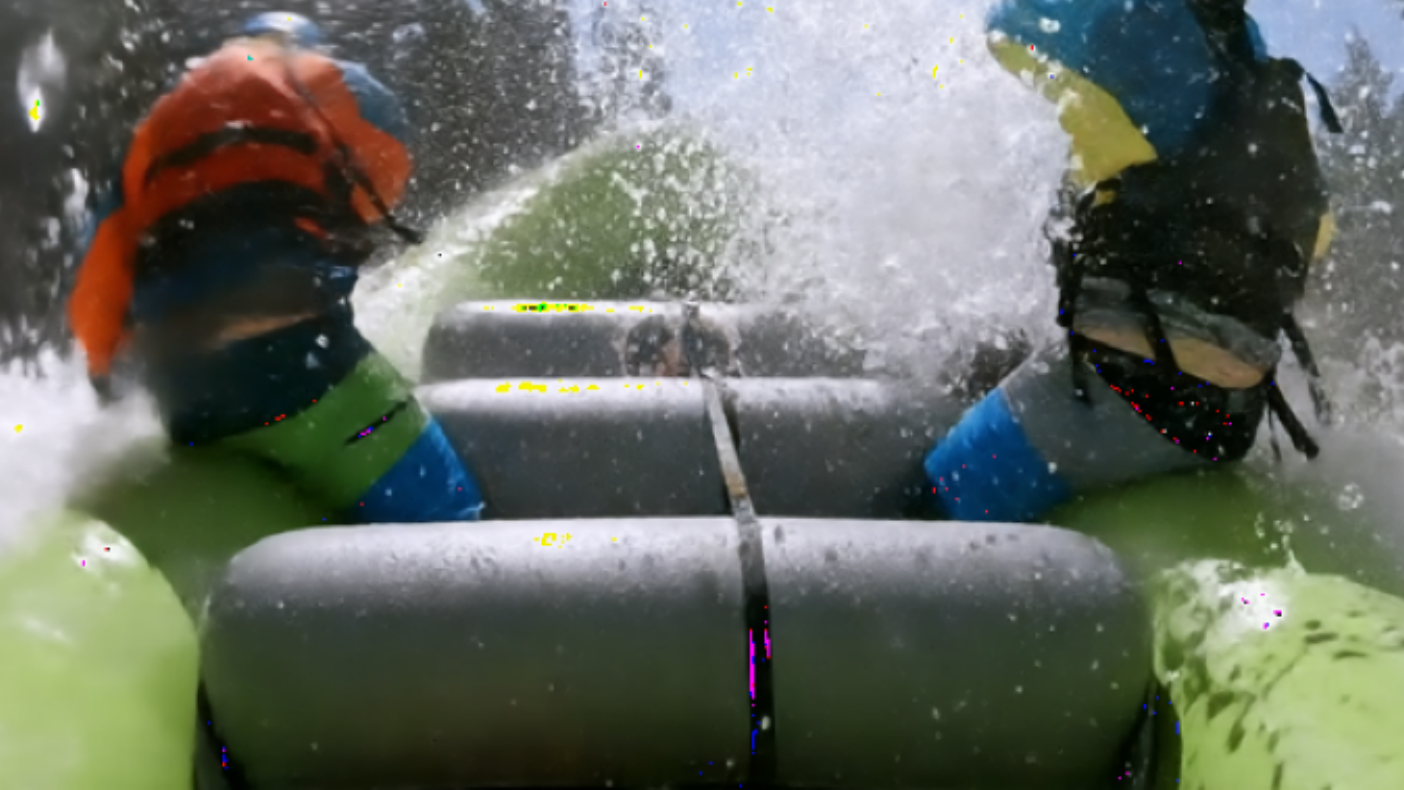} \\
    
     \end{tabular}
     
\caption{\textbf{Unsupervised denoising matches the performance of supervised denoising}. 
Frame from a video in the Set8 dataset denoised using different approaches. (a) Clean frame. (b) Frame corrupted with Gaussian noise of standard deviation $30$ (relative to intensity range [0-255]). (c) FastDVDnet~\cite{fastdvdnet}, a supervised method trained on the DAVIS dataset. (d) MF2F~\cite{mf2f}, an unsupervised method which fine-tunes a pre-trained FastDVDnet on the noisy video (e) Our proposed unsupervised method (UDVD), which uses five frames to denoise each frame, trained on the DAVIS dataset. (f) UDVD trained only on the noisy video itself. Performance is quantified using PSNR / SSIM \cite{ssim}, respectively. The corresponding videos, as well as additional examples, are included in Section C of the supplementary material.
}

\label{fig:comparison}
\end{figure*}

We show that UDVD is competitive with the current supervised state-of-the-art on standard benchmarks, despite not having access to ground-truth clean videos during training (see Figure~\ref{fig:comparison}). Moreover, when combined with aggressive data augmentation and early stopping, it can produce high-quality denoising even when trained exclusively on a single \emph{brief} noisy video sequence (as few as 30 frames), outperforming unsupervised video denoising techniques (e.g. F2F\cite{f2f} and MF2F~\cite{mf2f}) which are pre-trained with supervision. Finally, methods based on pre-training are not suitable for imaging applications where clean data is unavailable. In contrast, we demonstrate that UDVD can effectively denoise three different real-world datasets: raw videos from surveillance cameras, fluorescence-microscopy videos of cells, and electron-microscopy videos of catalytic nanoparticles.

The state-of-the-art performance of UDVD is unexpected. Nearly all existing approaches to video denoising~\cite{liu2010high, arias2018video, buades2016patch, vbm4d}, including those based on deep CNNs~\cite{dvdnet, xue2019video, f2f, F2F_extension, Yu_2020_CVPR_Workshops}, use estimates of optical flow to adaptively compensate for the motion of objects in the video. Conventional wisdom suggest that ignoring such motion should lead to denoising results in which moving content is blurred. Contrary to this intuition, UDVD and some recent state-of-the-art supervised methods for video denoising~\cite{fastdvdnet, vnlnet, videnn} yield excellent empirical performance without explicit estimation of optical flow. \emph{How can is this achieved?} We use a gradient-based analysis to show that both UDVD and supervised CNNs perform spatio-temporal \textit{adaptive} filtering, which is aligned with underlying motion. Thus, these CNNs are  \emph{automatically performing implicit motion compensation}. To quantify this, we demonstrate that it is possible to estimate optical flow accurately from the network gradients, even though the network architectures are not designed to account for optical flow, and the models receive no optical-flow information during training.

\noindent \textbf{Our Contributions}:
\begin{itemize}[leftmargin=*]
    \item A novel blind-spot architecture/objective for unsupervised video denoising, which achieves performance competitive with state-of-the-art supervised methods. 
    \item A training paradigm using aggressive data augmentation (time and space reversal) and early stopping to achieve state-of-the-art performance from training \emph{on a single brief noisy video}. 
    \item A demonstration of our method's effectiveness in denoising real-world electron and fluorescence microscopy data, as well as raw videos. Unlike most existing methods for unsupervised video denoising, our proposed method does not require pre-training, which is key in real-world imaging applications.
    \item An analysis of the denoising mechanism learned by UDVD, demonstrating that it performs implicit motion compensation even though it is only trained for denoising. We apply the analysis to supervised networks, showing that the same conclusion holds. 
\end{itemize}

\section{Background and Related Work}
\label{sec:related_work}

\noindent \textbf{Traditional and CNN-based video denoising.} Traditional techniques for single image denoising include nonlinear filtering~\cite{tomasi1998bilateral,milanfar2012tour}, sparse prior methods~\cite{ksvd,donoho1995,simoncelli1996,
chang2000adaptive,portilla2003image,bm3d}, and nonlocal means~\cite{nlb}; many of which have been extended to videos~\cite{liu2010high,arias2018video,vbm4d,buades2016patch}.
In order to exploit the spatio-temporal structure of 
the video, these methods typically employ motion compensation based on estimates of 
optical flow. 

In the last five years, data-driven methods based on deep CNNs \cite{lecun2015deep} have outperformed all other techniques in image~\cite{dncnn, cbdnet, chen2016trainable} and video denoising~\cite{dvdnet,xue2019video,fastdvdnet,rawvideo}. The CNNs are trained to minimize the mean squared error between the network output and ground truth using large databases of natural images/videos. Many deep-learning techniques also perform explicit motion compensation. DVDnet~\cite{dvdnet} applies an image-denoising CNN to each input frame, estimates the optical flow from the denoised frames using DeepFlow~\cite{deepflow} (a CNN pre-trained for this purpose), warps the frames using the flow estimate to align their content, and finally processes the registered frames with a CNN. 
Ref.~\cite{xue2019video} applies a similar pipeline, but jointly trains an optical-flow module with the denoising CNN.

\noindent \textbf{Video denoising without motion compensation.} Three recent methods perform video denoising without explicit motion estimation.  VNLnet~\cite{vnlnet} uses a non-local search algorithm to find self-similar patches in the input video, and then uses a CNN to process the patches. ViDeNN~\cite{videnn} consists of a first stage that denoises each frame using a CNN, and a second stage that exploits temporal structure by using the frames, $(t-1)$, $t$ and $t+1$ to produce the denoised $t$th frame. FastDVDnet~\cite{fastdvdnet} uses UNet~\cite{ronneberger2015u} blocks, trained end to end, to denoise each frame using five contiguous frames. These methods achieve state-of-the-art performance without any explicit motion compensation, similar to our proposed UDVD. In this work we show that such CNNs actually performs \emph{implicit} motion estimation, which can be revealed through a gradient-based analysis.

\noindent \textbf{Unsupervised denoising.} Noise2Noise~(N2N) is an unsupervised image-denoising technique where a CNN is trained on pairs of noisy images corresponding to the same clean image~\cite{n2n}. Frame2Frame~(F2F)~\cite{f2f} exploits this approach to fine-tune a pretrained image-denoising CNN with noisy data. The idea is to register contiguous frames using the optical flow (obtained from TV-L1~\cite{tvl1}), and treat them as noisy realizations of the same clean image. This scheme is extended to have a trainable flow estimation module in \cite{Yu_2020_CVPR_Workshops}, additional optical-flow consistency in \cite{F2F_extension} and to use multiple noisy frames as input in Multi-Frame2Frame~(MF2F) \cite{mf2f}.

\begin{figure*}[ht]
\begin{center}
    \includegraphics[width=0.92\textwidth]{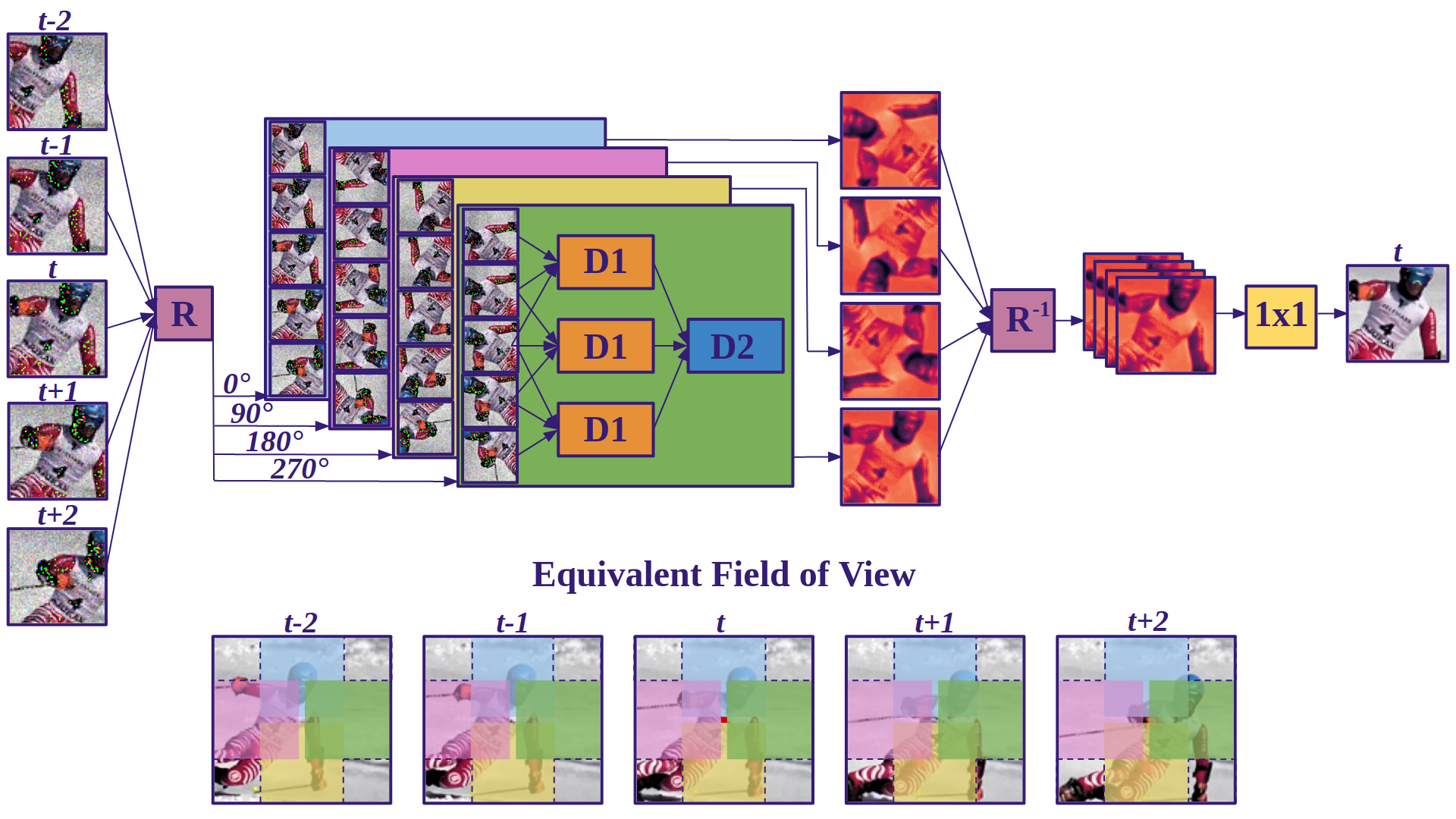}
\end{center}
\caption{\textbf{Unsupervised Deep Video Denoising (UDVD) Network Architecture}. The network takes $5$ consecutive noisy frames as input and produces a denoised central frame as output. We rotate the input frames by multiples of $90^{\circ}$ and process them in four separate branches with shared parameters, each containing asymmetric convolutional filters that are \emph{vertically causal}. As a result, the branches produce outputs that only depend on the pixels above ($0^{\circ}$ rotation, blue region), to the left ($90^{\circ}$, pink region), below ($180^{\circ}$, yellow region) or to the right ($270^{\circ}$, green region) of the output pixel. 
Each branch consists of a cascade of 2 Unet-style blocks (D1 and D2) to combine information over frames. These outputs are then \emph{derotated} and linearly combined (using a  $1\times 1$ convolutions) followed by a ReLU nonlinearity to produce the final output. The resulting ``field of view'' is depicted at the bottom with each color representing the contribution of the corresponding branch.}
\label{fig:network_arch}
\end{figure*}

Using the N2N framework to perform unsupervised video denoising requires warping adjoining frames, which in turn requires explicit motion compensation, and accurate occlusion estimation. In addition, the assumption that contiguous frames can be registered may not hold, particularly if the motion speeds in the video are large relative to the frame rate or local intensity changes are not due to translation. In order to bypass these issues, we develop a blind-spot network that trains denoising CNNs by fitting the noisy data directly. The CNN is trained to estimate each noisy pixel value using the surrounding spatio-temporal neighborhood, but without taking into account the noisy pixel itself in order to avoid the trivial identity solution. This ``blind spot'' can be enforced through architecture design~\cite{blindspotnet}, or by masking~\cite{noise2self,n2v}. For still images, several variations of this approach have been shown to provide effective denoising for natural images and noisy images from fluorescence microscopy~\cite{pn2v, fpn2v, poissongaussian}.

\section{Unsupervised Deep Video Denoising}
\label{sec:methodology}
In this section we describe our proposed architecture (see Figure~\ref{fig:network_arch} for a detailed diagram). 

\noindent \textbf{Multi-frame blind-spot architecture.} Our CNN maps five contiguous noisy frames to a denoised estimate of the middle frame. Building on the ``blind spot'' idea proposed in~\cite{blindspotnet} for single-image denoising, we design the architecture so that each output pixel is estimated from a spatio-temporal neighbourhood that does not include the pixel itself. 
We rotate the input frames by multiples of $90^{\circ}$ and process them through four separate branches containing asymmetric convolutional filters that are \emph{vertically causal}. As a result, the branches produce outputs that only depend on the pixels above ($0^{\circ}$ rotation), to the left ($90^{\circ}$), below ($180^{\circ}$) or to the right ($270^{\circ}$) of the output pixel. These partial outputs are then \emph{derotated} and combined using a three-layered cascade of $1\times 1$ convolutions and nonlinearities to produce the final output. The resulting field of view does not include the pixel being denoised, as depicted at the bottom of Figure~\ref{fig:network_arch}. 

UDVD processes the video in two stages as shown in Figure~\ref{fig:network_arch}, similar to previously proposed networks for supervised video denoising~\cite{dvdnet, videnn, fastdvdnet}. A first stage, consisting of three UNets~\cite{ronneberger2015u} (D1 in the diagram) with shared parameters, maps each group of three contiguous frames (i.e. $(t-2, t-1, t)$, $(t-1, t, t+1)$ and $(t, t+1, t+2)$) to a separate feature map. These features are then mapped to a single output using another UNet (D2). See Suppl. A for a detailed description of the architecture. 

\noindent \textbf{Bias-free architecture.} 
Inspired by~\cite{biasfree}, we remove all additive terms from the convolutional layers in UDVD. This provides automatic generalization to varying noise levels not encountered during training, and facilitates our proposed analysis to interpret the denoising mechanisms learned by the network (see Section~\ref{sec:results} and \ref{sec:analysis}).

\noindent \textbf{Using the missing pixel.} The denoised value generated by the proposed architecture at each pixel is computed without using the noisy observation at that location. This avoids overfitting -- i.e. learning the trivial identity map that minimizes the mean-squared error cost function -- but ignores important information provided by the noisy pixel. In the special case of Gaussian additive noise, we can use this information via a precision-weighted average between the network output and the noisy pixel value. Following~\cite{blindspotnet, pn2v}, the weights in the average are derived by assuming a Gaussian distribution for the error in the blind-spot estimates of the color pixel values. Specifically, we model the distribution of the three color channels of a pixel $x \in \mathcal{R}^3$ given the noisy neighbourhood $\Omega_y$ as $p(x | \Omega_y) = \mathcal{N}(\mu_x, \Sigma_x)$, where $\mu_x \in \mathcal{R}^3$ and $\Sigma_x \in \mathcal{R}^3$ represent the mean vector and covariance matrix. Let $y = x + \eta$, $\eta \sim \mathcal{N}(0, \sigma^{2} I_{3})$ be the observed noisy pixel. We integrate the information in the noisy pixel with the UDVD output by computing the mean of the posterior $p(x | y, \Omega_y)$, given by 
\begin{equation}
\label{eq:post_mean}
E[x|y] = (\Sigma_x^{-1} + \sigma^{-2} I )^{-1} (\Sigma_x^{-1} \mu_x + \sigma^{-2} y ).   
\end{equation}
See Suppl.~A for more details. 
The CNN architecture is trained to estimate the mean and covariance of this distribution at each pixel by maximizing the log likelihood of the noisy data:
\begin{equation}
\label{eq:loss}
\begin{split}
    \mathcal{L}(\mu_x, \Sigma_x) & = \frac{1}{2}[(y - \mu_x)^T(\Sigma_x + \sigma^2 I)^{-1}(y - \mu_x)]\\
    & \quad + \frac{1}{2}\log|\Sigma_x + \sigma^2 I |.
\end{split}
\end{equation}
When the noise process is unknown, we simply minimize the MSE between the denoised output and noisy video, and ignore the center pixel (see Suppl.~A for more details).   

\noindent \textbf{Data augmentation and early stopping.} In supervised denoising with simulated noise, training can rely on the generation of a virtually unlimited set of fresh noise realizations, which prevents overfitting. In the unsupervised setting, this is not possible, which makes it more challenging to train models that can denoise short video sequences. To address this, we (a) leverage data augmentation strategies: spatial flipping and time reversal, and (b) perform early stopping by monitoring the mean squared error between the network output and noisy frames on a held-out set of frames. These strategies make it possible to train UDVD with short video sequences (as few as 30 frames), while achieving denoising performance that is on par with or superior to both unsupervised and supervised networks trained on much larger datasets (see Figure~\ref{fig:comparison}, Table~\ref{tab:single_video} and Suppl. D).

\begin{table*}[ht]
    \centering
    \footnotesize{
    \begin{tabular}{ccccccccccc}
        \toprule

        \multicolumn{1}{l}{\phantom} &
        \multicolumn{1}{l}{\phantom} &
        \multicolumn{2}{c}{Traditional}    &
        \multicolumn{3}{c}{Supervised CNN}   &
        \multicolumn{3}{c}{Unsupervised CNN (UDVD)}    \\ 
        \cmidrule(lr){3-4}
        \cmidrule(lr){5-7}
        \cmidrule(lr){8-10}
        
         \multicolumn{1}{c}{test set} &
        \multicolumn{1}{c}{$\sigma$} &
        \multicolumn{1}{c}{VNLB } &
        \multicolumn{1}{c}{VBM4D}     &
        \multicolumn{1}{c}{VNLnet} &
        \multicolumn{1}{c}{DVDnet} &
        \multicolumn{1}{c}{FastDVDnet} & 
        \multicolumn{1}{c}{$1$ frame} & 
        \multicolumn{1}{c}{$3$ frames} & 
        \multicolumn{1}{c}{$5$ frames} \\

        \midrule
        \multirow{3}{*}{DAVIS} 
        & $30$ & 33.73 & 31.65 & -     & \textbf{34.08} & 34.06 & 32.80 & 33.48 & \textbf{33.92} \\
        & $40$ & 32.32 & 30.05 & 32.32 & \textbf{32.86} & 32.80 & 31.48 & 32.20 & \textbf{32.68} \\
        & $50$ & 31.13 & 28.80 & 31.43 & \textbf{31.85} & 31.83 & 30.47 & 31.20 & \textbf{31.70} \\
        \midrule
        \multirow{3}{*}{Set8} 
        & $30$ & 31.74 & 30.00 & -     & \textbf{31.79} & 31.60 & 30.91 & 31.62 & \textbf{32.01} \\
        & $40$ & 30.39 & 28.48 & \textbf{30.55} & \textbf{30.55} & 30.37 & 29.63 & 30.42 & \textbf{30.82} \\
        & $50$ & 29.24 & 27.33 & 29.47 & \textbf{29.56} & 29.42 & 28.65 & 29.47 & \textbf{29.89} \\
        \bottomrule
    \end{tabular}
    }
    \vspace{0.3cm}
    \caption{\textbf{Denoising results on natural video datasets}. All networks are trained on the DAVIS train set. Performance values are PSNR of each trained network averaged over held-out test data. UDVD, operating on $5$ frames, outperforms the supervised methods on Set8 and is competitive on the DAVIS test set.  Unsupervised denoisers with more temporal frames show a consistent improvement in denoising performance. DVDnet and FastDVDnet are trained using varying noise levels ($\sigma \in [0, 55]$) and VNLnet is trained and evaluated on each specified noise level. All UDVD networks are trained \emph{only} at $\sigma=30$, showing that they generalize well on unseen noise levels. See Sections C and F in the supplementary material for additional results. The PSNR values for all methods except UDVD are taken from \cite{fastdvdnet}.}
    \label{tab:test_data}
\end{table*}

\begin{table*}[ht]
    \centering
    \footnotesize{
    \begin{tabular}{ccccccccc}
        \toprule
        
        \multicolumn{1}{l}{\phantom} &
        \multicolumn{4}{c}{$\sigma = 30$}    &
        \multicolumn{4}{c}{$\sigma = 90$}   \\
        \cmidrule(lr){2-5}
        \cmidrule(lr){6-9}
        &
        \multicolumn{1}{c}{DAVIS} &
        \multicolumn{1}{c}{Set8} &
        \multicolumn{1}{c}{Derfs} &
        \multicolumn{1}{c}{Vid3oC} &
        \multicolumn{1}{c}{DAVIS} &
        \multicolumn{1}{c}{Set8} &
        \multicolumn{1}{c}{Derfs} &
        \multicolumn{1}{c}{Vid3oC} \\

        \midrule
        UDVD-S & 33.68 / 78.16 & \textbf{32.90} / 81.85 & \textbf{33.95} / 81.91 & 34.65 / 84.60 & \textbf{29.05} / \textbf{53.53} &\textbf{28.07} / \textbf{55.35} & \textbf{29.42} / \textbf{59.25} & \textbf{29.94} / \textbf{63.79} \\
        
        \midrule
        UDVD$^*$ & 33.78 / 79.88 & 31.90 / \textbf{82.53} & 32.58 / 81.44 & 34.24 / 83.96 & 28.87 / 51.22 & 27.25 / 51.84 & 28.26 / 52.44 & 29.23 / 60.08   \\
        FastDVDnet$^*$ & \textbf{33.91} / 76.99 & 31.81 / 80.21 & 32.45 / 81.64 & 35.05 / 84.44 & 28.01 / 47.53 & 26.54 / 50.16 & 27.36 / 52.87 & 28.42 / 55.99 \\
        
        \midrule
        MF2F  & \textbf{33.91} / \textbf{80.01} & 31.84 / 80.55 & 32.87 / \textbf{82.22} & \textbf{35.18} / \textbf{85.71} & 28.81 / 51.24 & 27.25 / 52.78 & 28.29 / 55.06 & 29.67 / 61.28 \\

        \bottomrule \\
    \end{tabular}
    }
    \caption{\textbf{Results for UDVD trained on individual noisy videos}. The top row shows \textbf{PSNR}/\textbf{VMAF}\cite{vmaf} values (averaged over the entire dataset) for UDVD trained on each individual video sequence with early stopping (labelled UDVD-S) using the last 5 frames of a video as a held-out set. We augmented the dataset with spatial flipping and time reversal (see Suppl. D for an ablation study). With the augmentations and early stopping, UDVD-S is comparable to (and often outperforms) UDVD or FastDVDnet trained on the full DAVIS dataset (indicated by $^*$) and MF2F, which fine-tunes a pre-trained CNN on each individual video. See Suppl. D for results on individual video sequences. 
        }
    \label{tab:single_video}
\end{table*}

\section{Datasets}
\label{sec:datasets}
We demonstrate the broad applicability of our approach by validating it on domains with different signal and noise structure: natural videos, raw videos, fluorescence microscopy, and electron microscopy. 

\noindent\textbf{Natural videos.} We perform controlled experiments on natural videos by adding iid Gaussian noise to the DAVIS dataset~\cite{davis}. The training/validation/test split is 60/30/30 videos, respectively. We use three additional datasets for testing - Set8~\cite{fastdvdnet} composed of 4 videos from the Derfs Test Media collection and 4 videos captured with a GoPro camera, Derfs~\cite{mf2f} with 7 videos, and the first 10 videos from Vid3oC~\cite{vid3oc} dataset (See Suppl. D for details).

\noindent\textbf{Raw videos.} We evaluate UDVD on a dataset of raw videos i.e with frame color channels interleaved according to the sensor mosaic containing real noise introduced in ~\cite{rawvideo}. The dataset contains 11 unique videos, each containing 7 frames, captured at five different ISO levels using a surveillance camera. Each video has 10 different noise realizations per frame, which are averaged to obtain an estimated clean version of the video.

\noindent\textbf{Fluorescence microscopy.} We apply our approach to fluorescence-microscopy recordings of live cells in ~\cite{celltrack}. We use two videos: Fluo-C2DL-MSC (CTC-MSC) depicting mesenchymal stem cells, and Fluo-N2DH-GOWT1 (CTC-N2DH) depicting GOWT1 cells. This dataset illustrates the challenges of applying supervised approaches to real data: there is no ground-truth clean data.

\noindent\textbf{Electron microscopy.} We also apply our methodology to a transmission electron microscopy dataset from~\cite{mohan2020deep}. The data consist of a 40-frame video depicting a platinum nanoparticle supported on a cerium oxide base. The average image intensity is 0.45 electrons/pixel, which results in an extremely low signal-to-noise ratio. As with the fluorescence-microscopy data, no ground-truth clean images are available.

\begin{figure*}[ht]
    \def\f1ht{1.15\linewidth}%
    \centering 
    \begin{tabular}{ >{\centering\arraybackslash}m{0.10\linewidth}
     >{\centering\arraybackslash}m{0.10\linewidth} 
     >{\centering\arraybackslash}m{0.10\linewidth}
     >{\centering\arraybackslash}m{0.10\linewidth}
     >{\centering\arraybackslash}m{0.10\linewidth}
     >{\centering\arraybackslash}m{0.10\linewidth}
     >{\centering\arraybackslash}m{0.10\linewidth}
     >{\centering\arraybackslash}m{0.10\linewidth}
     }
     \centering
     
     \footnotesize{Noisy} & \footnotesize{Denoised} & 
     \footnotesize{Noisy} & \footnotesize{Denoised} & 
     \footnotesize{Noisy} & \footnotesize{Denoised} & 
     \footnotesize{Noisy} & \footnotesize{Denoised} \\
     
     \includegraphics[width=\f1ht]{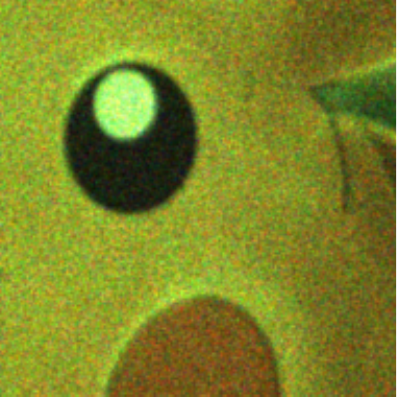} &
     \includegraphics[width=\f1ht]{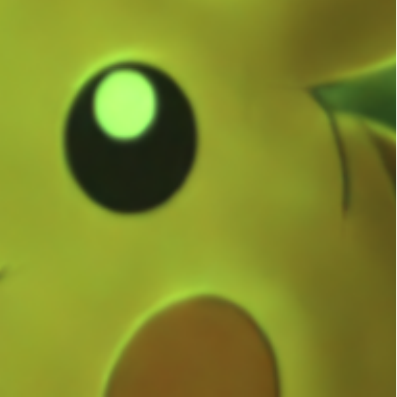} &
     \includegraphics[width=\f1ht]{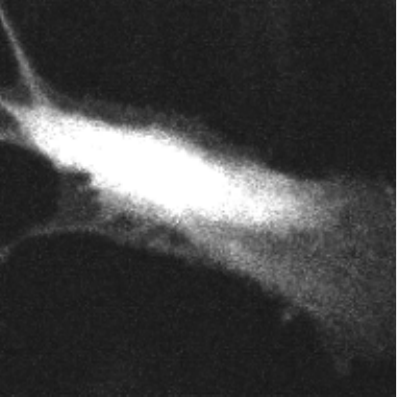} &
     \includegraphics[width=\f1ht]{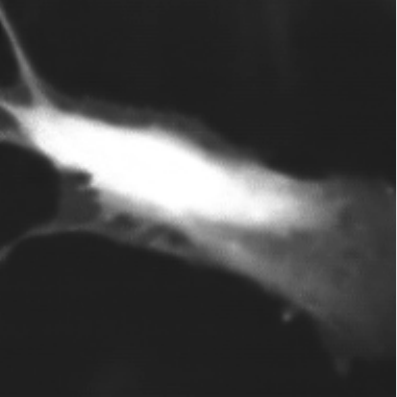} &
     \includegraphics[width=\f1ht]{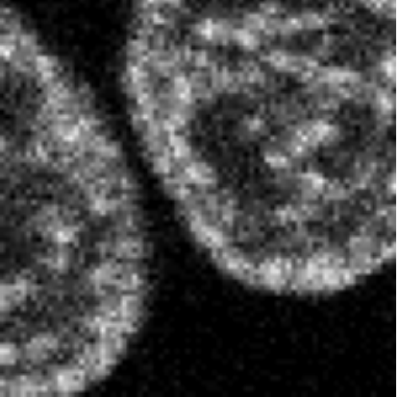} &
     \includegraphics[width=\f1ht]{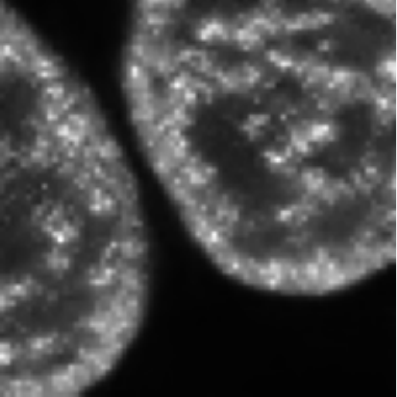} &
     \includegraphics[width=\f1ht]{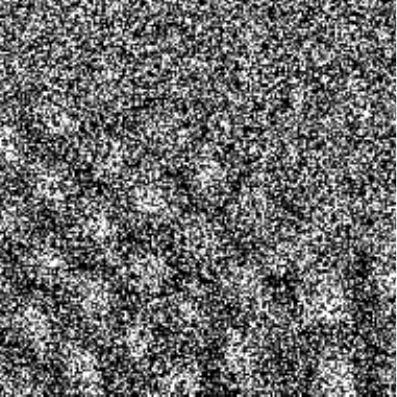} &
     \includegraphics[width=\f1ht]{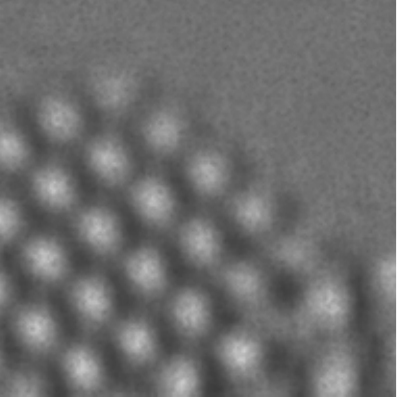} \\

     \includegraphics[width=\f1ht]{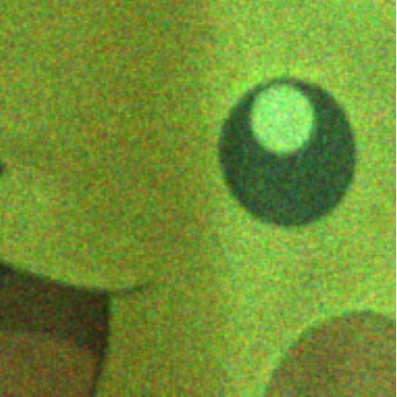} &
     \includegraphics[width=\f1ht]{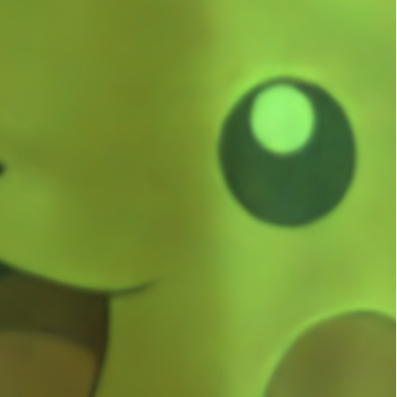} &
     \includegraphics[width=\f1ht]{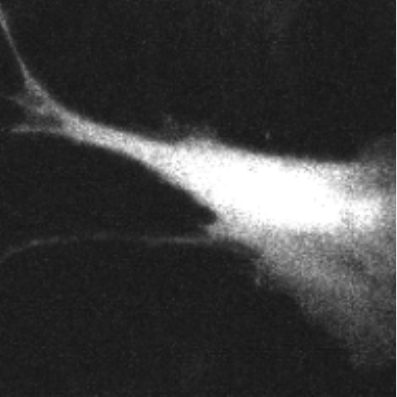} &
     \includegraphics[width=\f1ht]{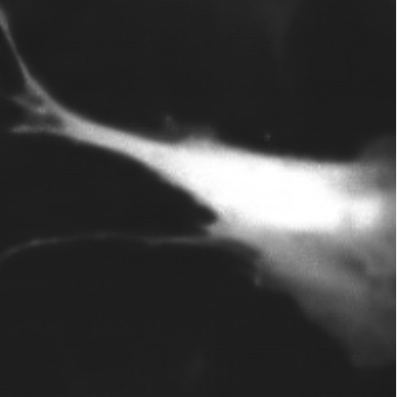} &
     \includegraphics[width=\f1ht]{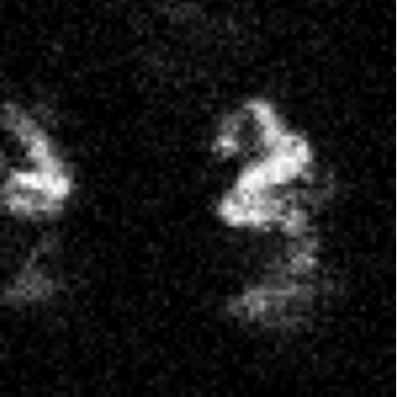} &
     \includegraphics[width=\f1ht]{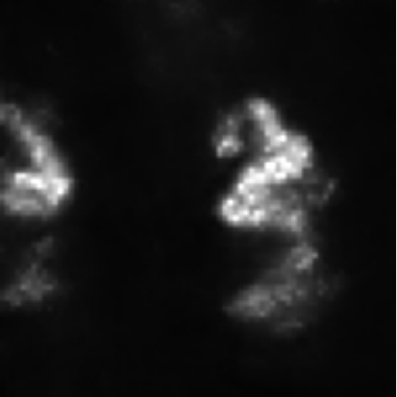} &
     \includegraphics[width=\f1ht]{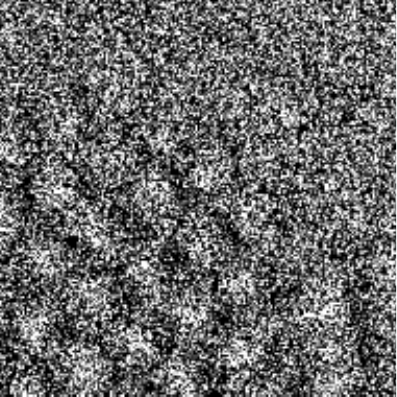} &
     \includegraphics[width=\f1ht]{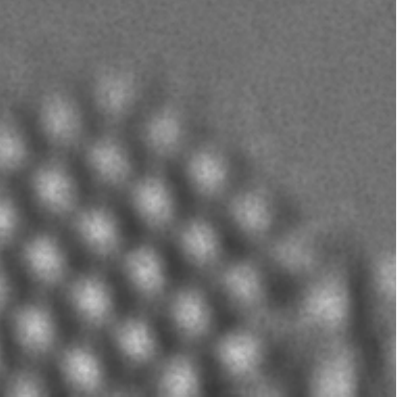} \\
     
     \multicolumn{2}{c}{ \makecell{\footnotesize{~(a) Raw video~(scene 11)} }} &
     \multicolumn{2}{c}{\makecell{\footnotesize{~(b) Fluor. micro.~(Mes. stem cells)}}} &
     \multicolumn{2}{c}{\makecell{\footnotesize{~(c) Fluor. micro.~(GOWT1 cells)}}} &
     \multicolumn{2}{c}{\makecell{\footnotesize{~(d) Electron micro.~(nanoparticle)}}}
     \end{tabular}
     \vspace{0.5mm}
\caption{\textbf{Denoising real-world data.} Results from applying UDVD to the raw video, fluorescence-microscopy and electron-microscopy datasets described in Section~\ref{sec:datasets}. Qualitatively, UDVD succeeds in removing noise while preserving the underlying signal structure, even for the highly noisy electron-microscopy data. Raw videos are converted to RGB for visualization. See Suppl.~D and F for denoised videos.}

\label{fig:real_data}
\end{figure*}

\section{Experiments and Results}
\label{sec:results}

\noindent\textbf{Comparison with other approaches on natural videos.} We train UDVD on the DAVIS training set (see Suppl. A for the training procedure). Following ~\cite{dncnn, biasfree, fastdvdnet, dvdnet,  blindspotnet, n2v, noise2self}, we add iid Gaussian noise with standard deviation $\sigma=30$ on the clean videos during training. UDVD is evaluated on the DAVIS test set and on Set8 by comparing to the clean ground-truth videos via PSNR. We compare UDVD with several popular methods: Bayesian processing of spatio-temporal patches (VNLB~\cite{nlb}), an extension of the popular image-denoising algorithm BM3D (VBM4D~\cite{vbm4d}) and supervised CNNs (VNLnet~\cite{vnlnet}, DVDnet~\cite{dvdnet}, FastDVDnet~\cite{fastdvdnet}). As shown in Table~\ref{tab:test_data}, UDVD achieves comparable performance to the supervised state-of-the-art on the DAVIS test set and slightly outperforms these methods on an independent test set (Set8) at multiple noise levels. It also outperforms traditional unsupervised techniques such as VNLB and VBM4D (see Figure~\ref{fig:comparison} and  Suppl.~C for visual examples). 

\noindent\textbf{Unsupervised denoising from limited data.} In order to validate our approach on a more challenging setting that is closer to the practical applications of unsupervised denoising, we trained and tested UDVD on individual videos from our test sets. As shown in Table~3 and 4 in Suppl. D, when combined with data augmentation and early stopping (using the last 5 frames of each video as a held-out validation set), this version of UDVD (called UDVD-S) achieves comparable results, or often outperforms supervised FastDVDnet and unsupervised UDVD trained on a large dataset~(DAVIS) (see Table~\ref{tab:single_video} for results on 4 different datasets).

To the best of our knowledge, all the existing unsupervised video denoising techniques are based on the F2F~\cite{f2f} framework, where a backbone CNN pre-trained with supervision is fine-tuned on the video to be denoised. We compared UDVD-S against the most recent such method -- MF2F~\cite{mf2f} which fine-tunes a FastDVDnet~\cite{fastdvdnet} trained with supervision on natural videos using an objective involving optical flow computed on consecutive noisy frames (see Section~\ref{sec:related_work}). Without any pre-training, UDVD-S outperforms MF2F in almost all videos in  Table~3 and 4 in Suppl. D, and datasets in Table~\ref{tab:single_video} (See Table 5 in Suppl. D.3 for measure of confidence). 
Note that (a) we trained MF2F using all the $5$ training schemes provided in the paper and reported the best results in Table~\ref{tab:single_video}, and (b) the metric we used to measure performance in  Table~\ref{tab:single_video} is the average PSNR of all denoised frames, unlike in Ref.~\cite{mf2f} where the first 10 frames of each video were excluded (see Suppl. D.3 for more details and results).

\begin{figure*}[ht]
    \def\f1ht{\linewidth}%

    \centering 
    \begin{tabular}{ >{\centering\arraybackslash}m{0.365\linewidth}
     >{\centering\arraybackslash}m{0.17\linewidth}
     >{\centering\arraybackslash}m{0.17\linewidth}
     >{\centering\arraybackslash}m{0.17\linewidth}
     >{\centering\arraybackslash}m{0.04\linewidth}
     }
     \centering

      {\scalebox{1.5}{ $ \underset{d_t(i)}{\includegraphics[width=0.1\linewidth]{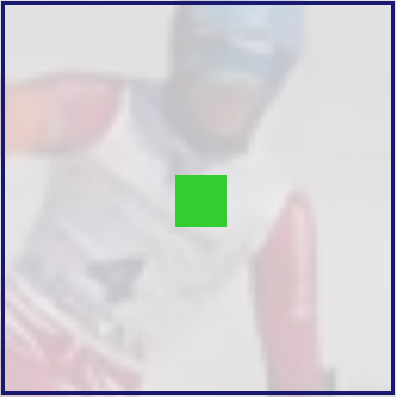}} = \sum\limits_{k=-2}^{2} \bigg\langle \footnotesize{ \underset{y_{t-k}}{\includegraphics[width=0.1\linewidth]{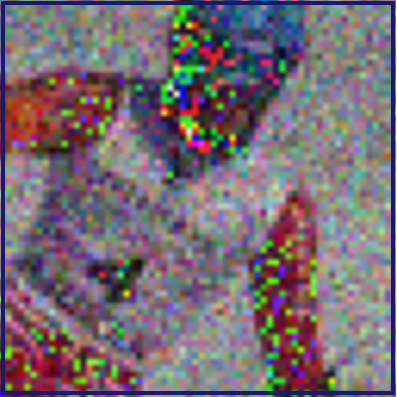}}, \underset{a(t-k, i)}{\includegraphics[width=0.1\linewidth]{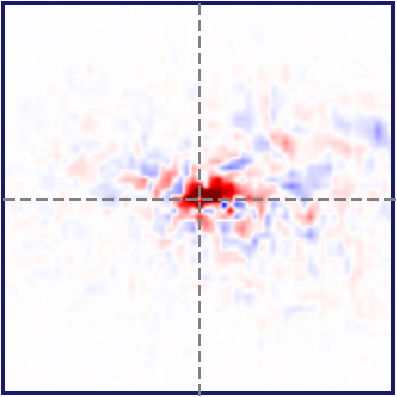}} }  \Bigg\rangle $ } } &
     \includegraphics[width=\f1ht]{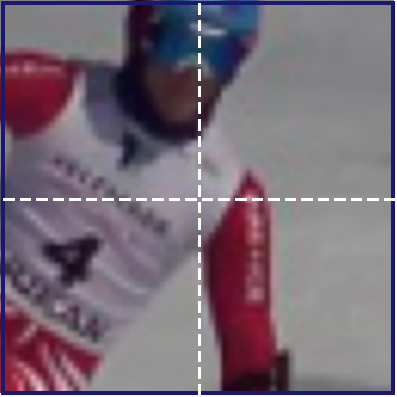} &
     \includegraphics[width=\f1ht]{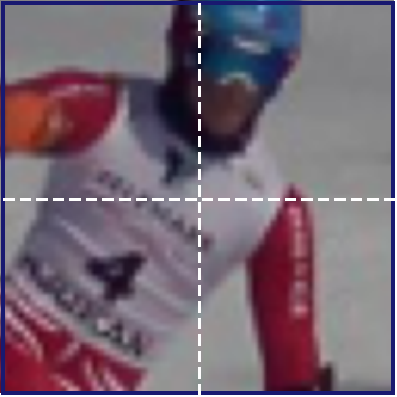} &
     \includegraphics[width=\f1ht]{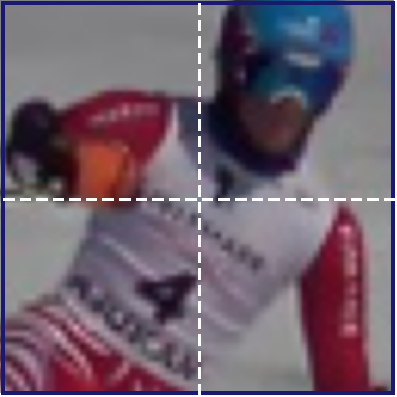} & \\%&
     
     \end{tabular}
     
     \centering 
     \begin{tabular}{ >{\centering\arraybackslash}m{0.17\linewidth}
     >{\centering\arraybackslash}m{0.17\linewidth}
     >{\centering\arraybackslash}m{0.17\linewidth}
     >{\centering\arraybackslash}m{0.17\linewidth}
     >{\centering\arraybackslash}m{0.17\linewidth}
     >{\centering\arraybackslash}m{0.04\linewidth}
     }
     \centering
     
     \includegraphics[width=\f1ht]{images/gradient_plots/noisy_DAVIS_giant-slalom_109_30.pdf} & 
     \includegraphics[width=\f1ht]{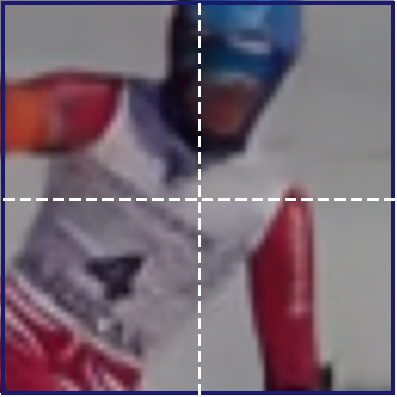} &
     \includegraphics[width=\f1ht]{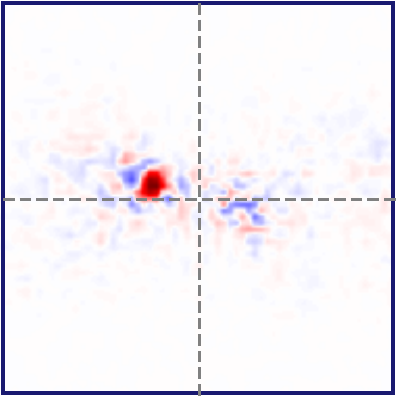} &
     \includegraphics[width=\f1ht]{images/gradient_plots/grad_DAVIS_giant-slalom_109_30_2.pdf} &
     \includegraphics[width=\f1ht]{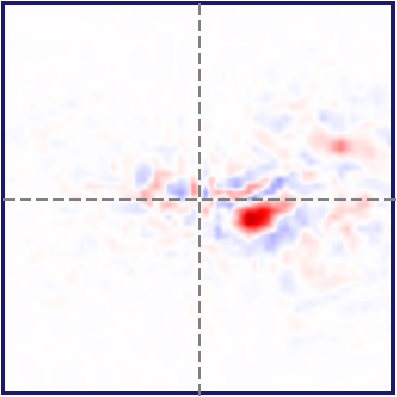} &
     \hskip-0.75cm \includegraphics[width=0.04\textwidth]{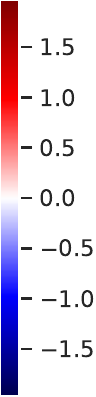}
     \\%&
     
     \includegraphics[width=\f1ht]{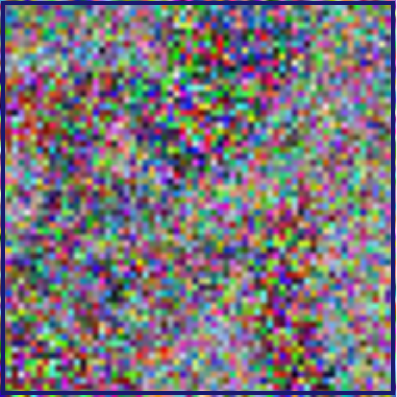} & 
     \includegraphics[width=\f1ht]{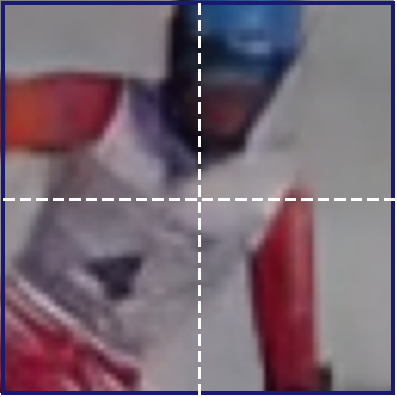} &
     \includegraphics[width=\f1ht]{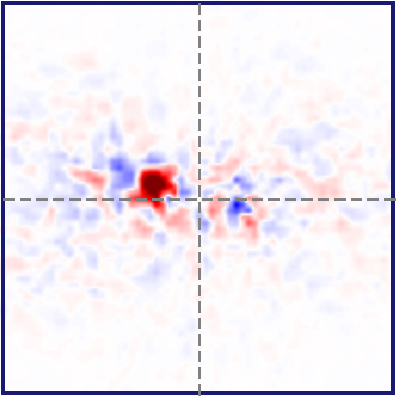} &
     \includegraphics[width=\f1ht]{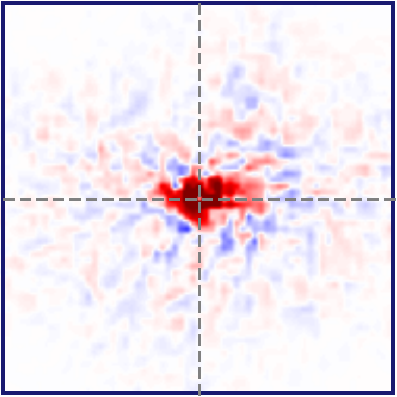} &
     \includegraphics[width=\f1ht]{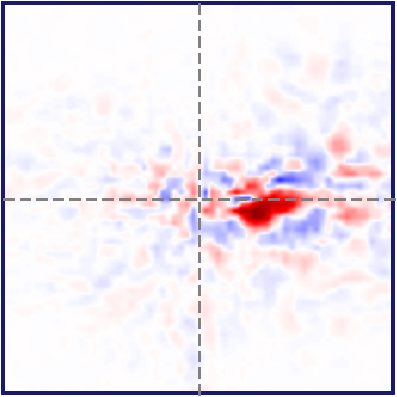} &
     \hskip-0.75cm \includegraphics[width=0.045\textwidth]{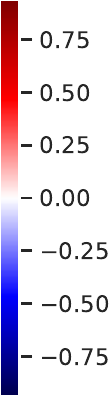}
     \\%&
     
     \hspace{4mm} $y_t$ &  $d_t$ & $a(t-1, i)$ & $a(t, i)$ & $a(t+1, i)$  \\
     \end{tabular}
     
     \vspace{0.2cm}
     
\caption{\textbf{Video denoising as spatiotemporal adaptive filtering}. Visualization of the equivalent linear weights ($a(k,i)$, Eq.~\ref{eq:local_linear_rep}) used to compute two example denoised pixels using UDVD. The left two columns show noisy frames $y_t$ at two noise levels, and the corresponding denoised frames, $d_t$. Three successive clean frames $\{x_{t-1}, x_t, x_{t+1}\}$ are shown in top row, for reference.  Corresponding weights $a(k, i)$ for pixel $i$ (intersection of the dashed white lines) in these three frames, are shown in the last three columns. The weights are seen to adapt to underlying video content, with their mode shifting to track the motion of the skier. As the noise level $\sigma$ increases (bottom row), their spatial extent grows, averaging out more of the noise while respecting object boundaries. For each denoised pixel, the sum of weights (over all pixel locations and frames) is approximately one, and thus can be interpreted as computing a local average (but note that some weights are negative, depicted in blue).}

\label{fig:jacobian}
\end{figure*}

\noindent\textbf{Use of temporal information.} UDVD estimates each frame from $k$ surrounding contiguous frames. To validate the effect of using more temporal information, we tested $k \in \{1,3,5\}$. As shown in Table~\ref{tab:test_data}, performance improves substantially and monotonically with $k$ (see Suppl. B for more noise levels ). This is in agreement with the literature on supervised learning \cite{fastdvdnet}. The performance gains arising from a longer temporal context are more substantial at higher noise levels (see Table~\ref{tab:test_data}). This is consistent with our analysis in Section~\ref{sec:analysis} which shows that, at low noise levels, UDVD($k=5$) tends to ignore the distant frames, but relies on them more at higher noise levels (see Figure~\ref{fig:jacobian} \& Suppl.~G).

\noindent \textbf{Generalization across noise levels.} UDVD generalizes strongly across noise levels not encountered during training. The results in Table~\ref{tab:test_data} are obtained with a network trained only at a fixed noise level of $\sigma=30$. This generalization ability is consistent with bias-free networks for image denoising~\cite{biasfree}. See Suppl.~F for more discussion and results.

\noindent \textbf{Raw videos with real noise.} We train UDVD on the first 9 realizations of the 5 videos from the test set of the raw video dataset (see Section~\ref{sec:datasets}), holding out the last realization for early stopping. We compare our performance with RViDeNet~\cite{rawvideo} which is pre-trained on a simulated dataset and then fine-tuned with supervision on 6 training videos from the raw video dataset. UDVD outperforms RViDeNet at all noise levels (see Table~\ref{tab:raw_video} and Fig~\ref{fig:real_data}). Note that UDVD was directly trained on the mosaiced raw videos. Existing unsupervised video denoising methods, like MF2F, cannot be applied directly on this dataset as their pre-trained backbone expects an input in the RGB domain (more details in Suppl. E).

\begin{table}[t]
    \centering
    \footnotesize{
    \begin{tabular}{ccccccc}
        \toprule
        \diagbox{CNN}{ISO} 
        & 1600 & 3200 & 6400 & 12800 & 25600 & \textbf{mean} \\
        \midrule
        UDVD & \textbf{48.04} & \textbf{46.24} & \textbf{44.70} & \textbf{42.19} & \textbf{42.11} & \textbf{44.69} \\
        RViDeNet~\cite{rawvideo} & 47.74 & 45.91 & 43.85 & 41.20 & 41.17 & 43.97 \\
        \bottomrule  
    \end{tabular}
    }
        \caption{\textbf{Raw video denoising}. PSNR values evaluated on the test set of the raw video dataset (Section~\ref{sec:datasets}) when denoised with (a) UDVD trained only the noisy test videos and (b) RViDeNet trained with supervision on a large dataset. The columns correspond to different ISO levels, with larger levels resulting in noisier data. 
        }
    \label{tab:raw_video}
\end{table}

\noindent \textbf{Real-world microscopy data.}  We train UDVD on the fluorescence-microscopy data described in Section~\ref{sec:datasets} following the same procedure as for the natural videos, including data augmentation. For the electron-microscopy data, we trained on the first 35 frames of the video, and used the remaining 5 as a validation set to perform early stopping based on mean-squared error. UDVD is able to effectively denoise the fluorescence-microscopy and the electron-microscopy datasets described in Section~\ref{sec:datasets}. This can be appreciated qualitatively in Figure~\ref{fig:real_data} and Suppl. E.

\section{Automatic Motion Compensation}
\label{sec:analysis}

Most previous approaches for video denoising rely on explicit motion compensation~\cite{liu2010high, arias2018video, buades2016patch, vbm4d}. This requires estimating the optical flow, which is the local translational motion of features in the image arising from the motion of objects and surfaces in a visual scene relative to the camera. Several CNN-based denoisers build motion estimation into the architecture~\cite{dvdnet, xue2019video}. In particular, motion compensation is critical to the F2F and MF2F frameworks for unsupervised denoising, which use motion compensation to register contiguous images~\cite{f2f, F2F_extension, mf2f}. In contrast, recent supervised video denoising networks like FastDVDnet \cite{fastdvdnet} and ViDeNN \cite{videnn}, as well as our unsupervised UDVD, do not perform any explicit motion compensation. Despite this, they achieve state-of-the-art results.  The empirical performance of these approaches suggests that the networks must somehow be exploiting temporal information successfully. Here, we study this phenomenon through an analysis of the denoising mapping, which reveals that these networks perform an implicit form of motion compensation.

\begin{figure*}[ht]
    \def\f1ht{\linewidth}%
    \centering 
    \begin{tabular}{ >{\centering\arraybackslash}m{0.3\linewidth}
     >{\centering\arraybackslash}m{0.3\linewidth}
     >{\centering\arraybackslash}m{0.3\linewidth} }
     \centering
     
      \hspace{4mm}  \footnotesize{(a) Noisy frame ($\sigma=30$)} &   \footnotesize{(b) Motion estimate from clean video%
      } 
      &  \footnotesize{(c) Motion estimate from UDVD gradients}  \\
      
      \includegraphics[width=\f1ht]{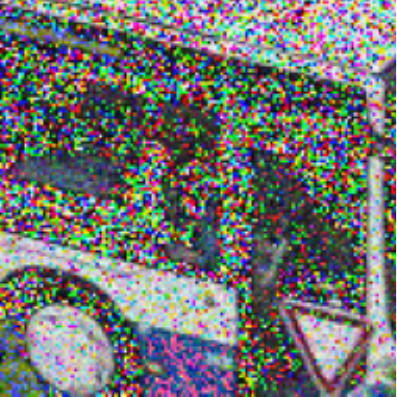} &
      \includegraphics[width=\f1ht]{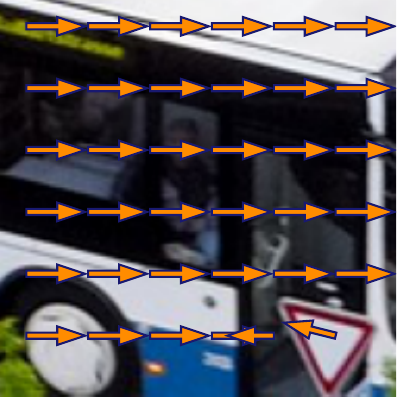} &
      \includegraphics[width=\f1ht]{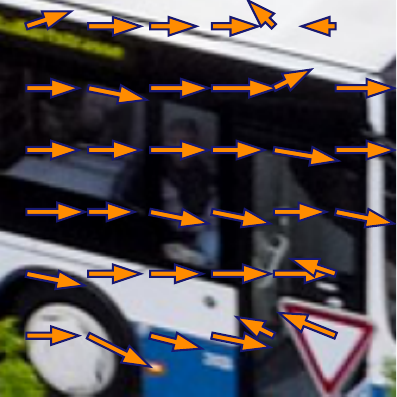} \\
    
     \end{tabular}
     \vspace{0.3cm}
     \caption{\textbf{CNNs trained for denoising automatically learn to perform motion estimation}. (a) Noisy frame from a video in the DAVIS dataset. (b) Optical flow direction at multiple locations of the image obtained using a state-of-the-art algorithm applied \emph{to the clean video}. (c) Optical flow direction estimated from the shift of the adaptive filter obtained by differentiating the network, which is trained exclusively with noisy videos and no optical flow information. Optical flow estimates are well-matched to those in (b), but deviate according to the aperture problem at oriented features (see black vertical edge of bus door), and in homogeneous regions (see bus roof, top right). }
\label{fig:optical_flow}

\end{figure*}

\noindent \textbf{Gradient-based analysis.} We use the approach of~\cite{biasfree} to analyze CNNs trained for image denoising. Let $y \in \mathbb{R}^{nT}$ be a flattened video sequence containing $T$ noisy frames with $n$ pixels each, processed by a CNN. We define the denoising function $f_i: \mathbb{R}^{nT} \rightarrow \mathbb{R}$ as the map between the noisy video and the denoised value $d_i:=f_i(y)$ of the CNN output at the $i$th pixel. A first-order Taylor decomposition of the denoising function may be written as:
\begin{align}
\label{eq:linearization}
    d_i:=f_i(y) &= %
    \langle \nabla f_i (y), y \rangle + b, %
\end{align}
where $\nabla f_i (y) \in \mathbb{R}^{nT}$ denotes the gradient of $f_i$ at $y$. The constant $b:=f_i(y)- \langle \nabla f_i (y), y \rangle$ is the net bias of the network, a combined function of all additive constants in the convolutional and batch-normalization layers of the CNN.

Our proposed architecture is bias-free (i.e., all additive constants are removed from the architecture, as proposed in~\cite{biasfree}), and thus $b=0$. As a result, the denoised value at the $i$th pixel may be written as:
\begin{align}
d(i) &= %
    \langle \nabla f_i (y), y \rangle 
    = \sum_{k=1}^{T} \langle a(k,i) , y_k \rangle, \label{eq:local_linear_rep}
\end{align}
where $y_k$ denotes each of the $T$ flattened frames that compose the noisy video, and the weights $a(k,i)$ correspond to the gradient of $f_i$ with respect to $y$. 
Each vector $a(k,i)$ can be interpreted as an \emph{equivalent filter} that produces an estimate of the denoised video at pixel $i$ via a weighted average of the noisy observations over space and time.

\noindent \textbf{Interpreting equivalent filters.} Visualizing these equivalent filters reveals that UDVD learns to denoise by performing  averaging over an adaptive spatiotemporal neighborhood of each pixel. As illustrated in Figure~\ref{fig:jacobian} (and Suppl.~G), when the noise level increases, the averaging is carried out over larger regions. This intuitive behavior is also seen in classical linear Wiener filters~\cite{wiener1950extrapolation}, where the filters are larger for higher levels of noise. The crucial difference is that in the case of CNNs, the equivalent filters are \emph{adapted} to the local video content: they respect object boundaries in space and time, taking into account their motion. This is apparent in Figure~\ref{fig:jacobian}: equivalent filters in adjoining frames are automatically shifted spatially to compensate for the movement of the skier (additional examples in Suppl.~G). We find that this implicit motion compensation is not unique to UDVD: CNNs trained in a supervised fashion have the same property (see also Suppl.~G). 

\noindent \textbf{Optical-flow estimation.} In order to validate our observation that CNNs exclusively trained for denoising implicitly detect and exploit video motion, we use the equivalent filters of the networks to estimate the optical flow. To estimate the optical flow from the $t^{th}$ frame to the $(t+1)^{th}$ frame at the $i$th pixel, we compute the difference between the position of the centroid of the equivalent filter corresponding to the pixel at times $t$, $a(t,i)$, and time $t+1$, $a(t+1,i)$. To increase the stability of the estimated flow, we compute the filter centroid through a robust weighted average that only includes entries with relatively large values (within 20\% of maximum value in the filter). 

The optical-flow estimates obtained from the gradients of the trained UDVD network are surprisingly precise, even at very high noise levels. Figure~\ref{fig:optical_flow}, and additional figures in Suppl.~G, show that the results are similar to those obtained by applying an algorithm for optical-flow estimation (DeepFlow \cite{deepflow}) on the corresponding \emph{clean} video. 
This demonstrates that the CNNs are able to implicitly estimate motion from data, despite the fact that they were not trained on that problem, and \emph{even in the presence of substantial noise corruption}, a setting that is quite challenging for optical-flow estimation techniques. 
We also observe that the optical-flow estimates obtained from UDVD gradients tend to be less accurate for pixels near strongly oriented features where local motion is only partially constrained (known as the \emph{aperture problem}) or in homogeneous regions, where the local motion is unconstrained (the \emph{blank wall problem}). 

\section{Conclusion}

In this work we propose a method for unsupervised deep video denoising that achieves comparable performance to state-of-the-art supervised approaches. Combined with data-augmentation techniques and early stopping, the method achieves effective denoising even when trained exclusively on short individual noisy sequences, which enables its application to real-world noisy data. In addition, we perform a gradient-based analysis of denoising CNNs, which reveals that they learn to perform implicit adaptive motion compensation. This suggests several interesting research directions. For example, denoising may be a useful pretraining task for optical-flow estimation or other computer-vision tasks requiring motion estimation.
\\~\\
\noindent \textbf{Acknowledgements.} This work was supported by HHMI, NSF NRT HDR Award 1922658, CBET 1604971, OAC-1940263 and OAC-1940097. We thank the HPC staff at NYU, ASU and RBCDSAI, IIT Madras for their support.

\appendix
\onecolumn

\section{Implementation Details of Unsupervised Deep Video Denoising}

\subsection{Restricting field of view}
\label{sec:blind-spot-layers}

In UDVD, we rotate the input frames by multiples of $90^{\circ}$ and process them through four separate branches (with shared parameters) containing asymmetric convolutional filters that are \emph{vertically causal}. As a result, the branches produce outputs that only depend on the pixels above ($0^{\circ}$ rotation), to the left ($90^{\circ}$), below ($180^{\circ}$) or to the right ($270^{\circ}$) of the output pixel. We use a UNet \cite{ronneberger2015u} style architecture for each branch of UDVD. The field of view of the UNet is constrained by restricting the field of view of the convolutional, downsampling and upsampling layers that are used to build the UNet. \\

\noindent \textbf{Convolutional Layers:} We restrict the receptive field of each convolutional layer to extend only upwards following the strategy proposed in \cite{blindspotnet}. Let the filter size be $h \times w$. We zero-pad the top region of the input tensor with $k = \lfloor h/2 \rfloor$ zero rows before convolution and remove the bottom $k$ rows after convolution. This is equivalent to convolving with a filter, where all weights below the center row are zero, so that the field of view only extends upwards. \\

\noindent \textbf{Downsampling and Upsampling Layers:} Following \cite{blindspotnet} we restrict the receptive field of the downsampling layer by creating an offset of one pixel (zero-pad with a row of zeros on the top and remove a row of pixels from below) before performing max-pooling using a $2 \times 2$ kernel. This operation restricts the field of view of the downsampling and upsampling operation pair.

\noindent Note that we do not use BatchNorm \cite{batchnorm} layers in UDVD as computing the spatial mean and variance would modify the field of view to include the center pixel. 

\subsection{Adding the Noisy Pixel Back}
\label{sec:add_noisy_back}

The denoised generated by the proposed architecture at each pixel is computed without using the noisy observation at that location. This avoids overfitting -- i.e. learning the trivial identity map that minimizes the mean-squared error cost function -- but ignores important information provided by the noisy pixel. In the case of Gaussian additive noise, we can use this information via a precision-weighted average between the network output and the noisy pixel value. Following~\cite{blindspotnet}, the weights in the average are derived by assuming a Gaussian distribution for the error in the blind-spot estimates of the (color) pixel values. The CNN architecture is trained to estimate the mean and covariance of this distribution at each pixel by maximizing the log likelihood of the noisy data. We explain this in detail in the following paragraphs. 

UDVD estimates the value of a pixel based on the noisy pixels in its neighbourhood. We model the distribution of the three color channels of a pixel $x \in \mathcal{R}^3$ given the noisy neighbourhood $\Omega_y$ as $p(x | \Omega_y) = \mathcal{N}(\mu_x, \Sigma_x)$, where $\mu_x \in \mathcal{R}^3$ and $\Sigma_x \in \mathcal{R}^3$ represent the mean vector and covariance matrix. Let $y = x + \eta$, $\eta \sim \mathcal{N}(0, \sigma^{2} I_{3})$ be the observed noisy pixel. We integrate the information in the noisy pixel with the UDVD output by computing the mean of the posterior $p(x | y, \Omega_y)$, given by 
\begin{equation}
    p(x | y, \Omega_y) \propto p(y | x) ~ p(x | \Omega_y)
\end{equation}
where $p(x | \Omega_y)$ is the prior and $p(y | x)$ is the noise model. Since both the prior and the noise model are Gaussian, we can write the optimal posterior estimate as, 
\begin{equation}
E[x|y] = (\Sigma_x^{-1} + \sigma^{-2} I )^{-1} (\Sigma_x^{-1} \mu_x + \sigma^{-2} y ).    
\end{equation}

Note that the posterior mean has a very intuitive interpretation. When the signal variance is high compared to noise variance (i.e. the uncertainty in our estimation is high) the posterior mean favours noisy pixel value. We estimate $\mu_x$ and $\Sigma_x$ as a function of the neighbourhood $\Omega_y$ using the network architecture discussed earlier. If $x$ is a grayscale image, then the output of the network consists of two channels - one for $\mu_x$ and one for $\sigma_x$. When the input image has $k$ channels, the output consists of $k$ channels for $\mu_x$ and $k(k-1)/2$ for the upper-triangular entries of $\Sigma_x$

One can estimate $\mu_x$ and $\Sigma_x$ directly from the noisy data by maximizing the likelihood. Using our distributional assumptions, the noisy pixels $y$ follows a Gaussian distribution, $y \sim \mathcal{N}(\mu_y, \Sigma_y)$, where $\mu_y = \mu_x$ and $\Sigma_y = \Sigma_x + \sigma^2 I$. Therefore, the loss function or the negative log likelihood is:
\begin{equation}
    \mathcal{L}(\mu_x, \Sigma_x) = \frac{1}{2}[(y - \mu_x)^T(\Sigma_x + \sigma^2 I)^{-1}(y - \mu_x)] + \frac{1}{2}\log|\Sigma_x + \sigma^2 I |.
\end{equation}
If $\sigma$ is unknown during training and has to be estimated, we use a separate neural network with the same architecture to do so. In such cases, we add a small regularization term equal to $-0.1\sigma$ for numerical stability, following \cite{blindspotnet}. 

For the experiments with real data, the noise distribution is unknown, so we simply ignore the central pixel.

\subsection{Architecture and Training}
\label{sec:arch}

\begin{table}[]
    \centering
    \begin{tabular}{lcr}
        \toprule
        Name & $N_{out}$ & Function \\
        \midrule
        Input & $k_1$ & \\
        enc\_conv\_0 & 48 & Convolution $3 \times 3$ \\
        enc\_conv\_1 & 48 & Convolution $3 \times 3$ \\
        enc\_conv\_2 & 48 & Convolution $3 \times 3$ \\
        pool\_1 & 48 & MaxPool $2 \times 2$ \\
        enc\_conv\_3 & 48 & Convolution $3 \times 3$ \\
        enc\_conv\_4 & 48 & Convolution $3 \times 3$ \\
        enc\_conv\_5 & 48 & Convolution $3 \times 3$ \\
        pool\_2 & 48 & MaxPool $2 \times 2$ \\
        enc\_conv\_6 & 96 & Convolution $3 \times 3$ \\
        enc\_conv\_7 & 96 & Convolution $3 \times 3$ \\
        enc\_conv\_8 & 48 & Convolution $3 \times 3$ \\
        upsample\_1 & 48 & NearestUpsample $2 \times 2$ \\
        concat\_1 & 96 & Concatenate output of pool\_1 \\
        dec\_conv\_0 & 96 & Convolution $3 \times 3$ \\
        dec\_conv\_1 & 96 & Convolution $3 \times 3$ \\
        dec\_conv\_2 & 96 & Convolution $3 \times 3$ \\
        dec\_conv\_3 & 96 & Convolution $3 \times 3$ \\
        upsample\_2 & 96 & NearestUpsample $2 \times 2$ \\
        concat\_2 & 96+$k_1$ & Concatenate output of Input \\
        dec\_conv\_4 & 96 & Convolution $3 \times 3$ \\
        dec\_conv\_5 & 96 & Convolution $3 \times 3$ \\
        dec\_conv\_6 & 96 & Convolution $3 \times 3$ \\
        dec\_conv\_7 & $k_2$ & Convolution $3 \times 3$ \\ 
        \bottomrule \\
    \end{tabular}
    \caption{\textbf{Network architecture used for UDVD}. The convolution and pooling layers are the blind-spot variants described in Section~\ref{sec:blind-spot-layers}. $k_1$ and $k_2$ represent the number of input and output channels of the base network respectively.}
    \label{tab:net_arch}
\end{table}

\textbf{Architecture:} The overall architecture is explained in Section 3 of the paper. The network architecture for the D1 and D2 blocks is described in Table \ref{tab:net_arch}. D1 has $k_1 = 9$ input channels and $k_2 = 32$ output channels. D2 has $k_1 = 96$ input channels and $k_2 = 96$ output channels. The architecture of D1 and D2 are analogous to the blocks in FastDVDnet \cite{fastdvdnet} to facilitate fair comparison with the supervised models. As described in Fig. 2 of the paper, D2 is followed by a derotation and the output is passed to a series of three cascaded $1 \times 1$ convolutions and non-linearity for reconstruction with 4 and 96 intermediate output channels, as in \cite{blindspotnet}. The final convolutional layer is linear and has 9 output channels, 3 representing the RGB value of the denoised image and 6 representing its covariance matrix. We use the same architecture for fluorescence microscopy and electron microscopy with the number of input channels to UDVD modified to 5 and number of output channels modified to 1. \\ 

\noindent \textbf{Training Details:} Following the convention in image and video denoising, we train UDVD on $128 \times 128$ patches extracted from our dataset~\cite{dncnn, biasfree, blindspotnet, fastdvdnet, dvdnet} (this is also consistent with the training methodology of the supervised baselines). For the natural video and fluorescence microscopy datasets, no data augmentation was applied. For electron microscopy dataset, we applied spatial flipping, time reversal and time subsampling (i.e. skipping frames). \\

\noindent \textbf{Optimization Details:} All networks were trained using Adam \cite{adam} optimizer with a starting learning of $10^{-4}$. The learning rate was decreased by a factor of $2$ at checkpoints $[20, 25, 30]$ during a total training of $40$ epochs. We did not experiment with other learning rate schedules such as cosine scheduling, which is a popular choice in unsupervised image denoising \cite{blindspotnet}.

\section{Ablation Study on Number of Input Frames}
\label{sec:ablation}

\begin{table*}
    \centering
    \begin{tabular}{c|cccc|cccc}
        \toprule
        
        \multicolumn{1}{l}{\phantom} &
        \multicolumn{4}{c}{DAVIS} &
        \multicolumn{4}{c}{Set8} \\
        
        \cmidrule(lr){2-5}
        \cmidrule(lr){6-9}
        \multicolumn{1}{l}{\phantom} &
        \multicolumn{1}{c}{Supervised CNN} &
        \multicolumn{3}{c}{Unsupervised CNN (UDVD)} &
        \multicolumn{1}{c}{Supervised CNN} &
        \multicolumn{3}{c}{Unsupervised CNN (UDVD)} \\
        
        \cmidrule(lr){2-2}
        \cmidrule(lr){3-5}
        \cmidrule(lr){6-6}
        \cmidrule(lr){7-9}
        \multicolumn{1}{c}{$\sigma$} &
        \multicolumn{1}{c}{5 frames} &
        \multicolumn{1}{c}{1 frame} &
        \multicolumn{1}{c}{3 frames} &
        \multicolumn{1}{c}{5 frames} &
        \multicolumn{1}{c}{5 frames} &
        \multicolumn{1}{c}{1 frame} & 
        \multicolumn{1}{c}{3 frames} & 
        \multicolumn{1}{c}{5 frames} \\
        
        \midrule
        
        20 & 35.86 & 34.13 & 34.91 & 35.16 & 33.37 & 32.39 & 33.09 & 33.36 \\
        30 & 34.06 & 32.80 & 33.48 & 33.92 & 31.60 & 30.91 & 31.62 & 32.01 \\
        40 & 32.80 & 31.48 & 32.20 & 32.68 & 30.37 & 29.63 & 30.42 & 30.82 \\
        50 & 31.83 & 30.47 & 31.20 & 31.70 & 29.42 & 28.65 & 29.47 & 29.89 \\
        60 & 31.01 & 29.65 & 30.39 & 30.90 & 29.08 & 27.86 & 28.70 & 29.13 \\
        70 & 30.21 & 28.96 & 29.70 & 30.22 & 28.37 & 27.20 & 28.06 & 28.49 \\
        80 & 29.28 & 28.37 & 29.10 & 29.63 & 27.60 & 26.65 & 27.50 & 27.94 \\
        
        \bottomrule
    \end{tabular}
        \caption{\textbf{Performance of UDVD}. Table shows the mean PSNR values of a state-of-the-art supervised video denoiser (FastDVDnet \cite{fastdvdnet} ) and UDVD with the denoised frame being predicted from $k \in \{1, 3, 5\}$ surrounding frames. The performance of UDVD monotonically increases with $k$ and is comparable for supervised denoising across all noise levels. All the three UDVD networks reported here are trained for only $\sigma=30$. FastDVDnet is trained for $\sigma \in [5, 55]$. }
    \label{tab:frame_ablation}
\end{table*}

We perform an ablation study on the number of frames $k$ UDVD uses as input, $k \in \{1, 3, 5\}$. UDVD with $k=1$ is equivalent to the blind-spot network proposed for image denoising in~\cite{blindspotnet}. In this section we describe the architectural and training details for UDVD with $k \in \{1, 3, 5\}$ and present some additional results. \\

\noindent \textbf{Architectural Details:} UDVD with $k=1$ contains only one UNet style network in each branch with architecture described in Table~\ref{tab:net_arch} and Section~\ref{sec:arch}. There are 3 input channels and 9 output channels (3 for the RGB channels in each denoised pixel and 6 for the corresponding covariance matrix). UDVD with $k=3$ has a similar architecture as for $k=1$ but has 9 input channels instead (3 channels for each frame). The architecture for $k=5$ is described in Section~\ref{sec:arch}.\\

\noindent \textbf{Training Details:} UDVD with $k \in \{1, 3, 5\}$ was trained on the DAVIS dataset with $\sigma=30$. The training details were as described in Section~\ref{sec:arch}. \\

\noindent \textbf{Results:} As shown in Table 1 of the paper and Table~\ref{tab:frame_ablation} performance improves substantially and monotonically with $k$ (the number of surrounding frames used to denoise each frame) across a wide range of noise levels. This difference in performance can also be observed visually. Fig~\ref{fig:img_vs_video} shows an example where the texture details of the brick wall and the fence are not well recovered when using only a single noisy frame. The texture is estimated better when using 5 noisy frames to predict the denoised output.

\def\nsp{\hspace*{-.07in}}%
\begin{figure*}

     \def\f1ht{\linewidth}%

     \begin{tabular}{>{\centering\arraybackslash}m{0.1\linewidth} 
     >{\centering\arraybackslash}m{0.27\linewidth}
     >{\centering\arraybackslash}m{0.27\linewidth}
     >{\centering\arraybackslash}m{0.27\linewidth}}
     
    \footnotesize{(a) Ground Truth} & 
    \includegraphics[width=\f1ht]{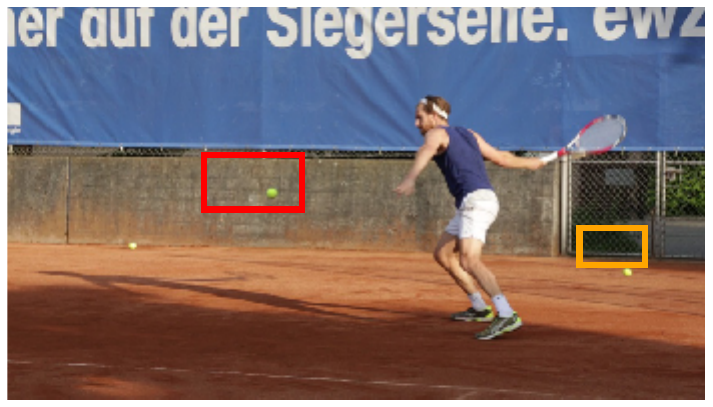}\nsp \nsp &
    \includegraphics[width=\f1ht]{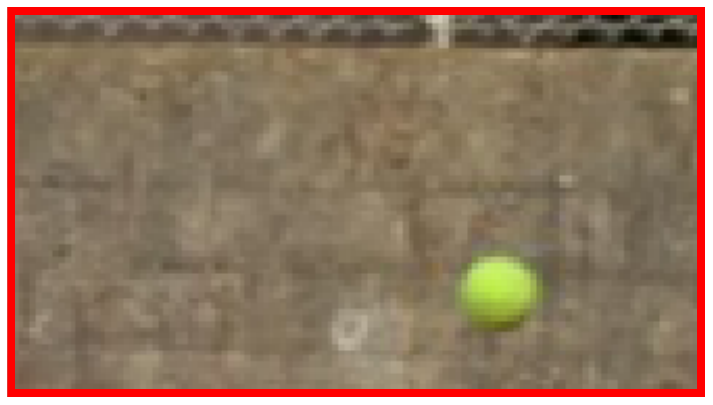}\nsp \nsp &
    \includegraphics[width=\f1ht]{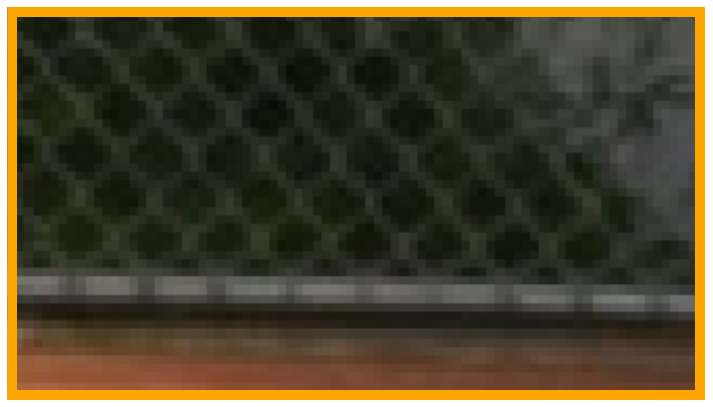} \\

    \footnotesize{(b) Noisy} & 
    \includegraphics[width=\f1ht]{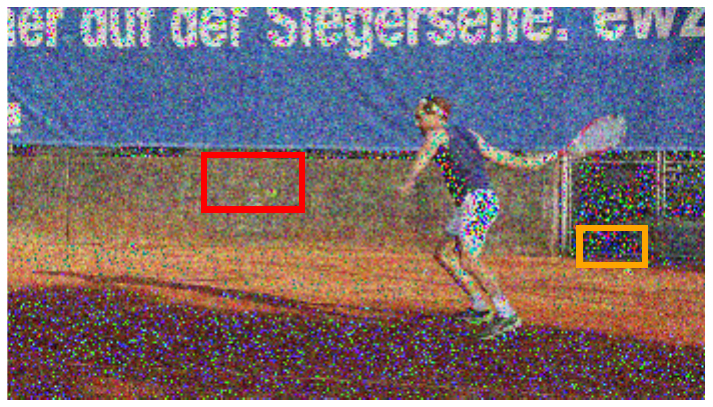}\nsp \nsp &
    \includegraphics[width=\f1ht]{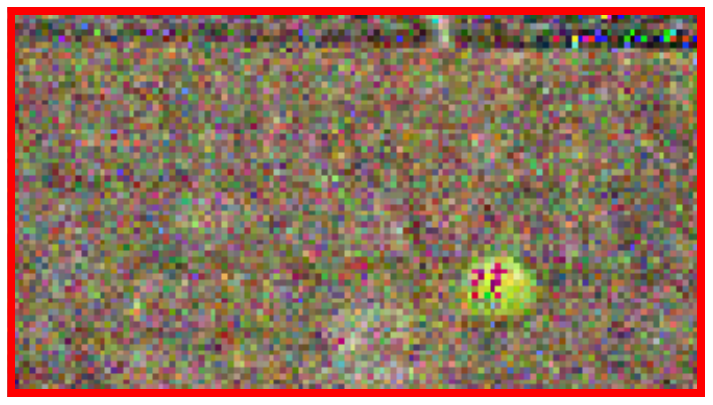}\nsp \nsp &
    \includegraphics[width=\f1ht]{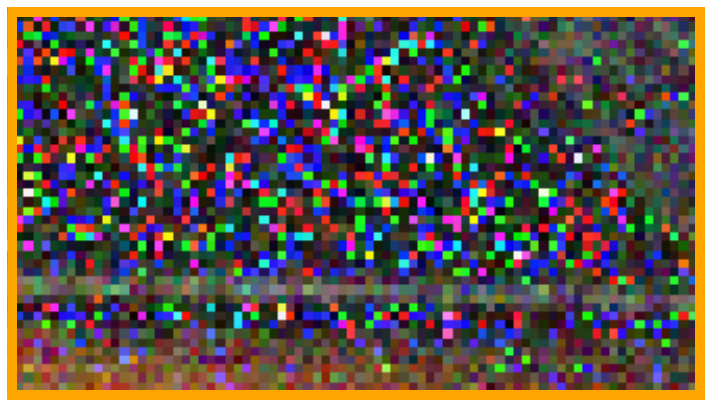} \\

    \footnotesize{(c) UDVD ($k=1$)} & 
    \includegraphics[width=\f1ht]{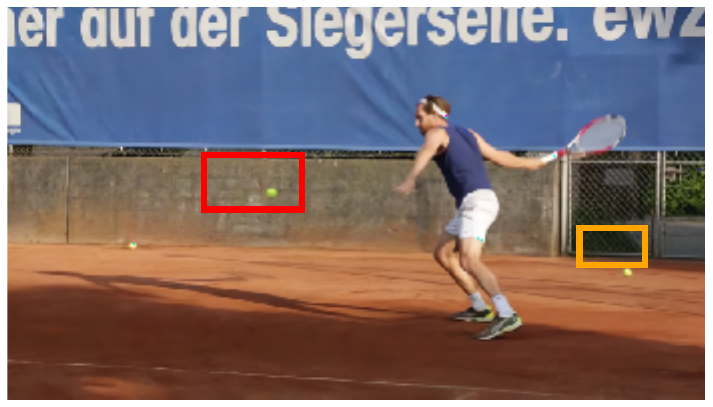}\nsp \nsp &
    \includegraphics[width=\f1ht]{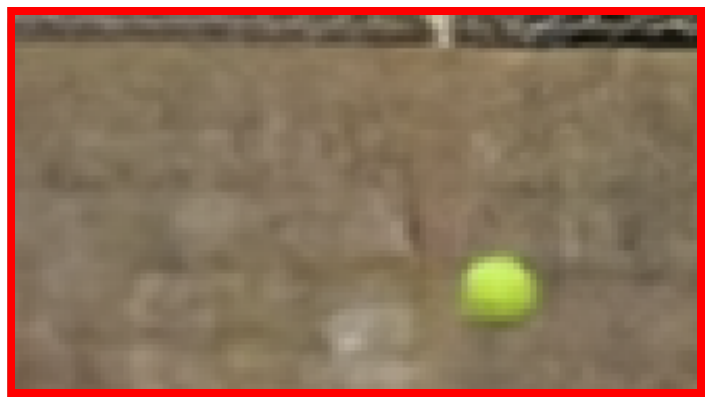}\nsp \nsp &
    \includegraphics[width=\f1ht]{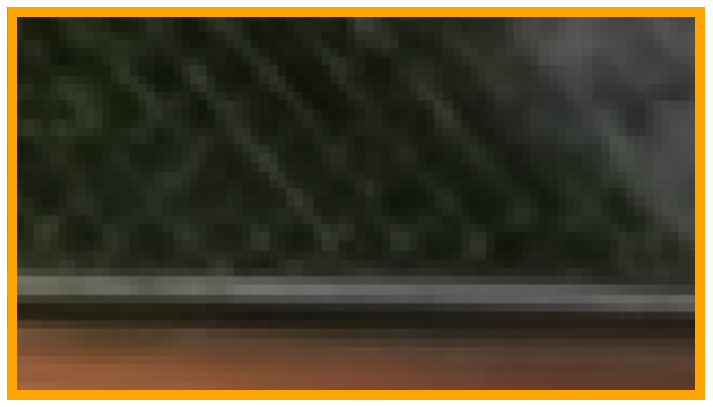} \\

    \footnotesize{(d) UDVD ($k=3$)} & 
    \includegraphics[width=\f1ht]{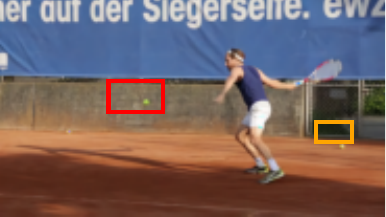}\nsp \nsp &
    \includegraphics[width=\f1ht]{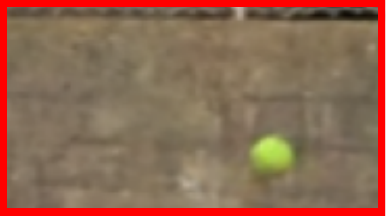}\nsp \nsp &
    \includegraphics[width=\f1ht]{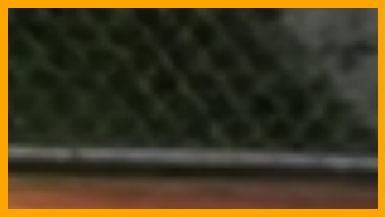} \\

    \footnotesize{(e) UDVD ($k=5$)} & 
    \includegraphics[width=\f1ht]{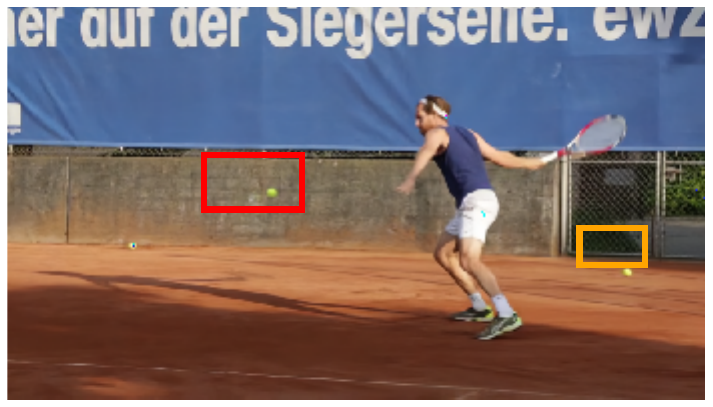}\nsp \nsp &
    \includegraphics[width=\f1ht]{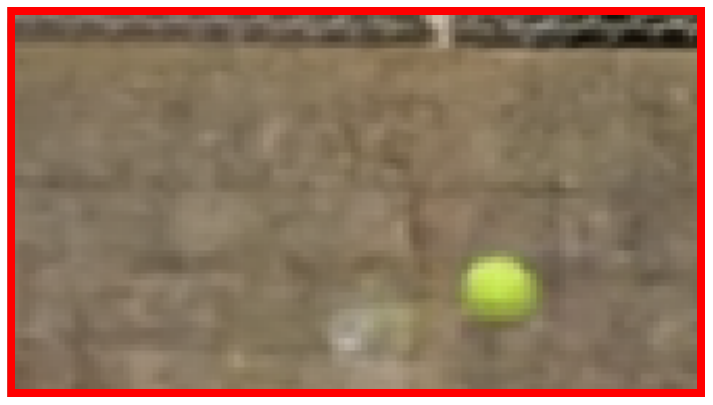}\nsp \nsp &
    \includegraphics[width=\f1ht]{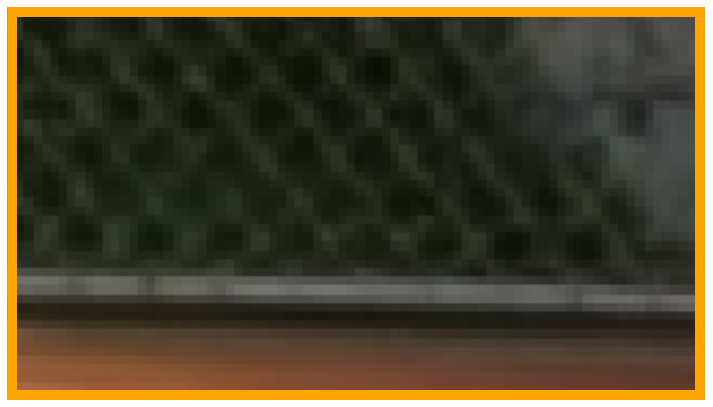} \\

     \end{tabular}
\caption{\textbf{Comparison of blind image and video denoising}. Example from the DAVIS dataset. (a) Ground truth frame. (b) Noisy frame. (c) Reconstruction using a single frame. The texture details of the brick wall and the fence are not recovered well. Reconstruction using (d) 3 and (e) 5 surrounding frames produces an improved texture estimate.}
 \label{fig:img_vs_video}
\end{figure*}

\section{Denoising Results on Natural Video Datasets}
\label{sec:denoise_natural}

\subsection{Comparison to Supervised Video Denoising}
In this section we provide additional comparisons between UDVD and supervised CNN-based methods.
\begin{enumerate}
    \item Table~\ref{tab:frame_ablation} shows the performance of UDVD trained at $\sigma=30$, and FastDVDnet trained for $\sigma \in [5, 55]$ when evaluated on the DAVIS test set and Set8 corrupted with $\sigma \in \{20, 40, \dots, 80\}$. UDVD achieves comparable performance to FastDVDnet on DAVIS test set and slightly outperforms it on Set8 at all noise levels.

    \item Examples of noisy videos, and denoised counterparts obtained using UDVD are included in the official github repository\footnote{\href{https://github.com/sreyas-mohan/udvd}{https://github.com/sreyas-mohan/udvd}} (\texttt{hypermooth.mp4, rafting.mp4, motorbike.mp4} and \texttt{snowboard.mp4}).
\end{enumerate}

\subsection{Comparison to Burst Denoising}
When several photographs are captured in quick succession to each other, the resulting set of images are often blurry or noisy (particularly when the object is in motion). Burst denoising aims to recover estimate the original scene from the set of burst photographs. Recent methods have solved burst denoising by applying deep neural network to map a stack of burst images to a single clean frame~\cite{Godard_2018_ECCV, Mildenhall_2018_CVPR, Xia_2020_CVPR}. A popular burst denoising method, KPN~\cite{Mildenhall_2018_CVPR} achieved a PSNR of 27.83 on the DAVIS dataset at $\sigma=30$ \footnote{Evaluated using the pre-trained model provided here: https://github.com/z-bingo/kernel-prediction-networks-PyTorch}, while UDVD achieves a PSNR of 33.92. UDVD is expected to outperform burst denoising methods as these methods (1) are trained for jittered motions, and cannot exploit systematic motion in natural videos like video denoising methods, and (2) often do not expect a motion change of more than 2 pixels from one frame to another~\cite{Mildenhall_2018_CVPR}, while the motion in natural videos is usually much larger (see Section 6 in main paper). 

\section{UDVD-S: Denoising Using Only a Single Video}
\label{sec:single_video}

UDVD, combined with aggressive data augmentation and early stopping, achieves state-of-the-art performance even when trained on only a single short video. In this section, we analyze the contribution of each of the data augmentation and early stopping scheme to the performance of UDVD-S through an ablation study. We also provide more details about our comparison to MF2F~\cite{mf2f}.

\begin{table*}[t]
    \centering
    \footnotesize{
    \begin{tabular}{lcccccccccc}
        \toprule
        
        \multicolumn{1}{l}{\phantom} &
        \multicolumn{10}{c}{$\sigma = 30$} \\
        \cmidrule(lr){2-11}
        &
        \multicolumn{1}{c}{ten-v} &
        \multicolumn{1}{c}{snow} & 
        \multicolumn{1}{c}{hyper} & 
        \multicolumn{1}{c}{raft} &
        \multicolumn{1}{c}{motor} &
        \multicolumn{1}{c}{trac} &
        \multicolumn{1}{c}{sunf} &
        \multicolumn{1}{c}{touch} &
        \multicolumn{1}{c}{park} &
        \multicolumn{1}{c}{\textbf{mean}} \\
        
        \midrule
        No. of frames &
        \multicolumn{1}{c}{75} &
        \multicolumn{1}{c}{59} &
        \multicolumn{1}{c}{37} & 
        \multicolumn{1}{c}{29} &
        \multicolumn{1}{c}{32} &
        \multicolumn{1}{c}{85} &
        \multicolumn{1}{c}{85} &
        \multicolumn{1}{c}{85} &
        \multicolumn{1}{c}{85} &
        \multicolumn{1}{c}{-} \\
        
        \midrule
        
        No Aug  \hfill(without ES)  & 33.37 & 29.10 & 29.72 & 27.26 & 27.28 & 32.52 & 35.07 & 32.65 & 30.20 & 30.80 \\
        No Aug \hfill(with ES) & 34.35 & 30.67 & 32.42 & 30.72 & 29.21 & 33.08 & 37.04 & 33.63 & 30.40 & 32.39 \\
        F  \hfill(without ES)       & 34.00 & 30.60 & 30.15 & 30.16 & 28.44 & 33.09 & 36.86 & 33.56 & 30.37 & 31.91 \\
        F   \hfill(with ES) & 34.68 & 30.76 & 32.41 & 30.76 & 29.33 & 33.35 & 37.13 & 33.74 & 30.53 & 32.52 \\
        F+TR  \hfill(without ES)    & 34.18 & 30.73 & 31.06 & 30.31 & 28.98 & 33.53 & \textbf{37.29} & 33.51 & \textbf{30.56} & 32.24 \\
        F+TR    \hfill(with ES)   & \textit{34.70} & \textit{30.78} & \textbf{32.60} & \textit{30.80} & \textbf{29.36} & \textbf{33.54} & \textbf{37.29} & \textbf{33.88} & \textbf{30.56} & \textbf{32.61} \\
        \midrule
        UDVD$^*$         & \textbf{34.82} & \textbf{30.83} & 32.34 & 30.82 & 29.24 & 31.73 & 35.33 & 33.48 & 28.98 & 31.95 \\
        FastDVDnet$^*$   & 34.58 & 30.78 & 32.48 & \textbf{30.94} & 29.35 & 31.39 & 35.06 & 33.71 & 28.73 & 31.89 \\
        \midrule
        MF2F - 8 sigmas               & 34.45 & 30.44 & 30.93 & 29.70 & 28.81 & 31.61 & 34.43 & 33.41 & 28.79 & 31.40 \\
        MF2F - online no teacher      & 34.50 & 30.42 & 30.54 & 29.45 & 28.40 & 32.11 & 35.19 & 33.47 & 28.89 & 31.44 \\
        MF2F - online with teacher    & 34.48 & 30.44 & 31.13 & \textit{29.91} & \textit{28.92} & 32.08 & 35.20 & 33.44 & 28.91 & 31.61 \\
        MF2F - offline no teacher     & \textit{34.66} & 30.49 & 30.20 & 29.38 & 28.36 & \textit{32.19} & 35.50 & 33.58 & 28.98 & 31.48 \\
        MF2F - offline with teacher   & 34.63 & \textit{30.52} & \textit{31.16} & 29.55 & \textit{28.92} & 31.93 & \textit{35.52} & \textit{33.61} & \textit{29.04} & \textit{31.65} \\
        \bottomrule \\
    \end{tabular}
    }
        \caption{\textbf{Results for UDVD and MF2F trained on individual noisy videos for $\sigma=30$}. The top block show PSNR values for UDVD trained on each individual video sequence with and without data augmentation (spatial flipping(F) and time-reversal(TR)) and early stopping (ES). Early stopping was performed using the last 5 frames of each video as the held-out set. The last block shows the result of running MF2F~\cite{mf2f} with all the 5 different fine-tuning scheme proposed in Ref.~\cite{mf2f}. With the augmentations and early stopping, UDVD-S, on average outperforms UDVD and FastDVDnet trained on the full DAVIS dataset (indicated by $^*$) and MF2F which fine-tunes a pre-trained FastDVDNet on each individual video. The best performing method for each video is highlighted in bold and the best performing method in each block is highlighted in italics. The tennis-vest video is from DAVIS and the rest of the 8 videos are from Set8. 
        }
    \label{tab:single_video_30}
\end{table*}

\begin{table*}[ht]
    \centering
    \footnotesize{
    \begin{tabular}{lcccccccccc}
        \toprule
        
        \multicolumn{1}{l}{\phantom} &
        \multicolumn{10}{c}{$\sigma = 90$} \\
        \cmidrule(lr){2-11}
        &
        \multicolumn{1}{c}{ten-v} &
        \multicolumn{1}{c}{snow} & 
        \multicolumn{1}{c}{hyper} & 
        \multicolumn{1}{c}{raft} &
        \multicolumn{1}{c}{motor} &
        \multicolumn{1}{c}{trac} &
        \multicolumn{1}{c}{sunf} &
        \multicolumn{1}{c}{touch} &
        \multicolumn{1}{c}{park} &
        \multicolumn{1}{c}{\textbf{mean}} \\
        
        \midrule
        No. of frames &
        \multicolumn{1}{c}{75} &
        \multicolumn{1}{c}{59} &
        \multicolumn{1}{c}{37} & 
        \multicolumn{1}{c}{29} &
        \multicolumn{1}{c}{32} &
        \multicolumn{1}{c}{85} &
        \multicolumn{1}{c}{85} &
        \multicolumn{1}{c}{85} &
        \multicolumn{1}{c}{85} &
        \multicolumn{1}{c}{-} \\
        
        \midrule
        
        No Aug \hfill(without ES)  & 24.13 & 22.89 & 22.04 & 20.99 & 20.06 & 24.84 & 25.98 & 25.67 & 23.35 & 23.33 \\
        No Aug  \hfill(with ES)                & 30.15 & 25.49 & 27.48 & 26.05 & 23.79 & 28.18 & 31.91 & 29.87 & 25.46 & 27.60 \\
        F \hfill(without ES)       & 27.21 & 24.42 & 24.05 & 23.32 & 21.84 & 27.42 & 29.53 & 28.01 & 25.03 & 25.65 \\
        F  \hfill(with ES)   & 30.35 & \textbf{25.60} & 27.72 & \textit{26.16} & 23.89 & \textbf{28.71} & 32.17 & 29.93 & 25.59 & 27.79 \\
        F+TR \hfill(without ES)    & 27.11 & 24.77 & 24.25 & 23.55 & 21.98 & 27.80 & 30.22 & 28.56 & 25.44 & 25.96 \\
        F+TR  \hfill(with ES)   & \textbf{30.40} & 25.59 & \textbf{27.75} & \textit{26.16} & \textbf{23.92} & 28.63 & \textbf{32.18} & \textbf{29.96} & \textbf{25.62} & \textbf{27.80} \\
        \midrule
        UDVD$^*$        & 28.78 & 25.16 & 26.78 & 25.81 & 23.57 & 26.42 & 29.04 & 28.71 & 24.23 & 26.50 \\
        FastDVDnet$^*$  & 29.44 & 25.25 & 27.30 & \textbf{26.35} & 23.68 & 27.42 & 30.29 & 29.61 & 24.72 & 27.12 \\
        \midrule
        MF2F - 8 sigmas               & 28.79 & 25.04 & 27.14 & 26.21 & 23.56 & 26.89 & 29.19 & 29.04 & 24.35 & 26.69 \\
        MF2F - online no teacher      & 28.35 & 25.12 & 26.67 & 26.07 & 23.39 & 27.28 & 30.01 & 29.49 & 24.64 & 26.78 \\
        MF2F - online with teacher    & \textit{29.44} & \textit{25.25} & \textit{27.30} & \textbf{26.35} & \textit{23.68} & \textit{27.42} & 30.09 & 29.53 & 24.71 & \textit{27.08} \\
        MF2F - offline no teacher     & 28.70 & 25.17 & 26.64 & 26.02 & 23.41 & 27.42 & \textit{30.29} & 29.60 & \textit{24.72} & 26.89 \\
        MF2F - offline with teacher   & 28.79 & \textit{25.25} & 27.22 & 26.31 & 23.62 & 27.34 & \textit{30.29} & \textit{29.61} & 24.69 & 27.01 \\
        \bottomrule \\
    \end{tabular}
    }
        \caption{\textbf{Results for UDVD and MF2F trained on individual noisy videos for $\sigma=90$}. The top block show PSNR values for UDVD trained on each individual video sequence with and without data augmentation (spatial flipping(F) and time-reversal(TR)) and early stopping (ES). Early stopping was performed using the last 5 frames of each video as the held-out set. The last block shows the result of running MF2F~\cite{mf2f} with all the 5 different fine-tuning scheme proposed in Ref.~\cite{mf2f}. With the augmentations and early stopping, UDVD-S, on average outperforms, UDVD or FastDVDnet trained on the full DAVIS dataset (indicated by $^*$) and MF2F which fine-tunes on a pre-trained FastDVDNet on each individual video. The best performing method for each video is highlighted in bold and the best performing method in each block is highlighted in italics. The tennis-vest video is from DAVIS and the rest of the 8 videos are from Set8. 
        }
    \label{tab:single_video_90}
\end{table*}

\subsection{Details of test sets.}

We evaluate UDVD-S and baselines on the following four datasets:

\begin{enumerate}
    \item \textbf{DAVIS~\cite{davis}}: We take all the 30 sequences from the test set of the DAVIS Challenge 2017.
    \item \textbf{Set8~\cite{fastdvdnet}}: Following FastDVDNet~\cite{fastdvdnet}, we use 4 sequences from the GoPro set (\textit{hypersmooth, motorbike, rafting, snowboard}) and 4 sequences from the Derfs Test Media Collection (\textit{park\_joy, sunflower, touchdown, tractor}).
    \item \textbf{Derfs}: Following \cite{mf2f}, we use 7 sequences from the Derfs Test Media Collection, which are  \textit{park\_joy, sunflower, touchdown, tractor} (shared with Set8), and \textit{blue\_sky, old\_town\_cross, pedestrian\_area}. We use the first 85 frames from each sequences with a spatial-resolution of $960 \times 540$~\cite{fastdvdnet}. 
    \item \textbf{Vid3oC~\cite{vid3oc}}: We use the first 10 sequences (\textit{000 to 009}) out of the 50 available sequences.
\end{enumerate}

\subsection{Ablation study}
\label{sec:augmentations}

We train UDVD-S on $128 \times 128$ patches extracted from the noisy video. (see Section~\ref{sec:arch}) for more details). For each patch, we apply each of the following data augmentations at random:
\begin{enumerate}
    \item \textbf{Spatial flipping}: We flip all the $5$ input patches vertically or horizontally. This operation only rearranges the pixel location and does not combine the pixel together in anyway, making sure that the noise statistics is still preserved after the augmentation. 
    \item \textbf{Time reversal}: We reverse the order of frames in the input to generate a new video. Like spatial flipping, this operation also preserves the noise statistics. 
\end{enumerate}

In addition to data augmentation, we employ early stopping by choosing the model parameters which produced the best error on a a held-out set of frames. We pick the last 5 frames of each video as our held out set. Tables~\ref{tab:single_video_30} and ~\ref{tab:single_video_90} show an ablation study over data augmentations and early stopping for $9$ different videos at two different noise levels, $\sigma=30$ and $\sigma=90$. Across videos and noise levels, data augmentation and early stopping significantly increase the performance of our method.  

\subsection{Comparison with MF2F}
\label{sec:mf2f}

\begin{figure*}[ht]
    \def\f1ht{0.20\textwidth}%
    \centering 
    \begin{tabular}{ >{\centering\arraybackslash}m{0.20\textwidth}
     >{\centering\arraybackslash}m{0.20\textwidth} 
     >{\centering\arraybackslash}m{0.20\textwidth}}
     \centering
     
     \hspace{2mm}  \footnotesize{(a) Noisy Input} &   \footnotesize{(b) UDVD-S} &  \footnotesize{(c) MF2F}  \\

     \includegraphics[width=\f1ht]{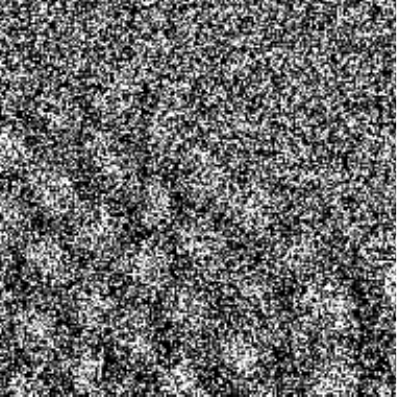} &
     \includegraphics[width=\f1ht]{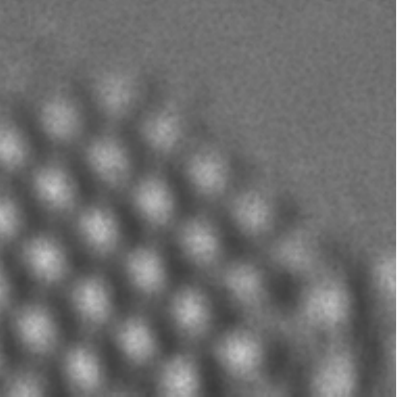} &
     \includegraphics[width=\f1ht]{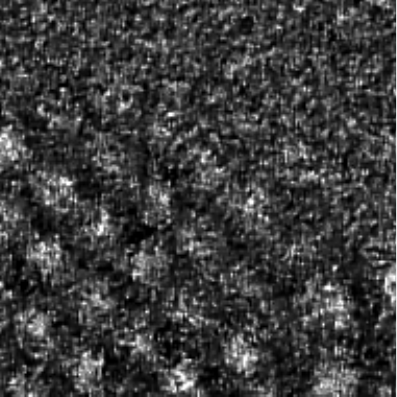} \\
    
    \end{tabular}
     
\caption{\textbf{UDVD-S outperforms MF2F on electron microscopy data}. UDVD-S is able to effectively denoise real-world data acquired from an electron-microscopy, but MF2F \emph{fails}.  
}

\label{fig:comparison_nano_mf2f}
\end{figure*}

We compare the performance of UDVD-S to an unsupervised denoising method MF2F~\cite{mf2f}. MF2F fine-tunes a pre-trained CNN on the noisy video using an objective function involving optical flow. The pre-trained CNN used in MF2F is FastDVDNet~\cite{fastdvdnet}, which is trained with supervised on a large dataset of natural videos(DAVIS~\cite{davis}). The authors of MF2F provide five different schemes for fine-tuning: 8 sigmas, online no teacher, online with teacher, offline no teacher and offline with teacher. Tables~\ref{tab:single_video_30} and \ref{tab:single_video_90} show the denoising results using each of these five training schemes. The best result (on 4 different dataets) is reported in Table~2 of the main paper. In addition to this, we also apply MF2F on real electron microscopy data (see Figure~\ref{fig:comparison_nano_mf2f}), where it \emph{fails}. We used the official implementation\footnote{https://github.com/cmla/mf2f} for all the training schemes.

\subsection{Measure of Confidence on Improvements}

We compute the mean and standard deviation of the improvement of UDVD-S with respect to MF2F in Table~\ref{tab:confidence}.

\begin{table*}[ht]
    \centering
    \begin{tabular}{ccccc}
        \toprule
        
        \multicolumn{1}{l}{\phantom} &
        \multicolumn{1}{c}{DAVIS} &
        \multicolumn{1}{c}{Set8} &
        \multicolumn{1}{c}{Derfs} & 
        \multicolumn{1}{c}{Vid3oC} \\

        \midrule
        
        $\sigma = 30$ & -0.33 $\pm$ 0.18 & 0.99 $\pm$ 0.21 & 1.09 $\pm$ 0.50 & -0.54 $\pm$ 0.52 \\
        $\sigma = 90$ & 0.15 $\pm$ 0.09 & 0.65 $\pm$ 0.23 & 1.13 $\pm$ 0.39 & 0.26 $\pm$ 0.28 \\
        
        \bottomrule \\
    \end{tabular}
    \caption{\textbf{Measure of confidence on improvement of UDVD-S with respect to MF2F}. We compute the mean and standard deviation of the difference between performance on UDVD-S and MF2F (in PSNR) on four different datasets and two different noise levels ($\sigma=30, 90$). UDVD-S outperforms MF2F with high certainty on two datasets at low noise level ($\sigma=30$) and all the four datasets at high noise level ($\sigma=90$).
        }
    \label{tab:confidence}
\end{table*}

\section{Denoising Results on Real-world Datasets}
\label{sec:denoise_micro}

\noindent \textbf{Raw videos}: The estimated ground truth, noisy raw data~\cite{rawvideo}, and the denoised videos obtained with UDVD can be found on the official github repository (\texttt{raw\_video.mp4}). The videos were converted to RGB for illustration. 

As discussed in the main paper, UDVD was directly trained on the mosaiced raw videos. Existing unsupervised video denoising methods, like MF2F~\cite{mf2f}, cannot be applied directly on this dataset as their pre-trained backbone expects an input in the RGB domain. In Ref.~\cite{mf2f}, the authors convert mosaiced videos into the RGB domain, apply MF2F~\cite{mf2f} and transform the denoised RGB videos back. \\

\noindent \textbf{Fluorescence and electron microscopy data}:
The noisy fluorescence microscopy and electron microscopy data, and the denoised videos obtained with UDVD can be found on the official github repository (\texttt{fluoro\_1.mp4, fluoro\_2.mp4} and \texttt{electron.mp4}).

\section{Generalization Across Noise and Frame Rate}
\label{sec:generalisation}

Ideally, a denoiser should be able to denoise videos corrupted at a wide range of noise levels. This is usually achieved by training the CNN on examples corrupted with a range of noise strength \cite{dncnn, fastdvdnet, dvdnet}. The range of noise levels on which the network is trained is called the \emph{training range} of the network.  \\

\begin{figure}
    \centering
    \includegraphics[width=0.49\textwidth]{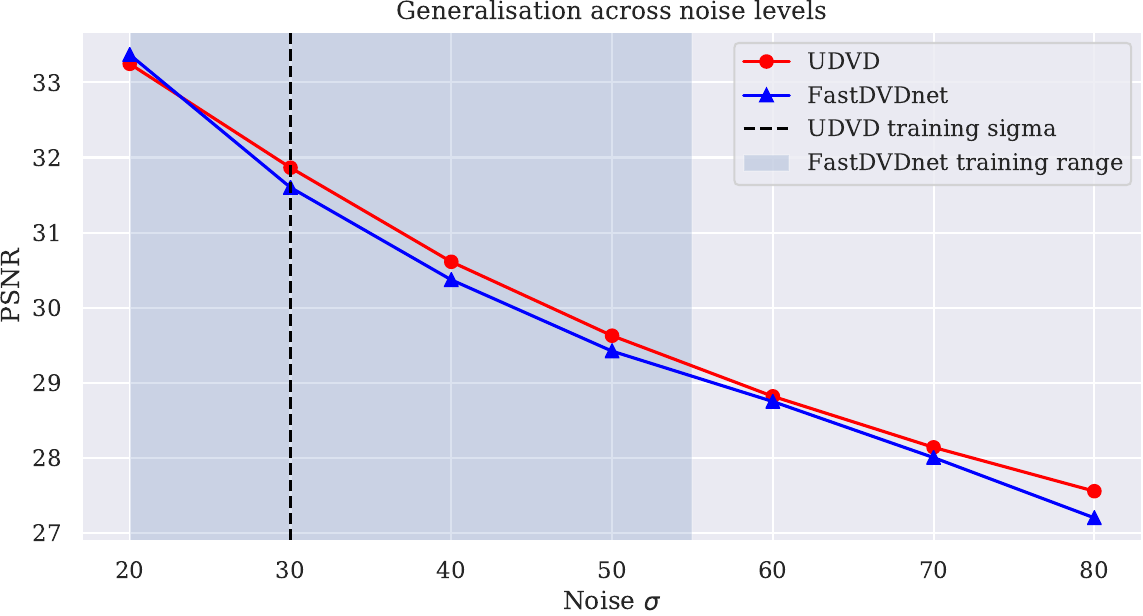}
    \includegraphics[width=0.49\textwidth]{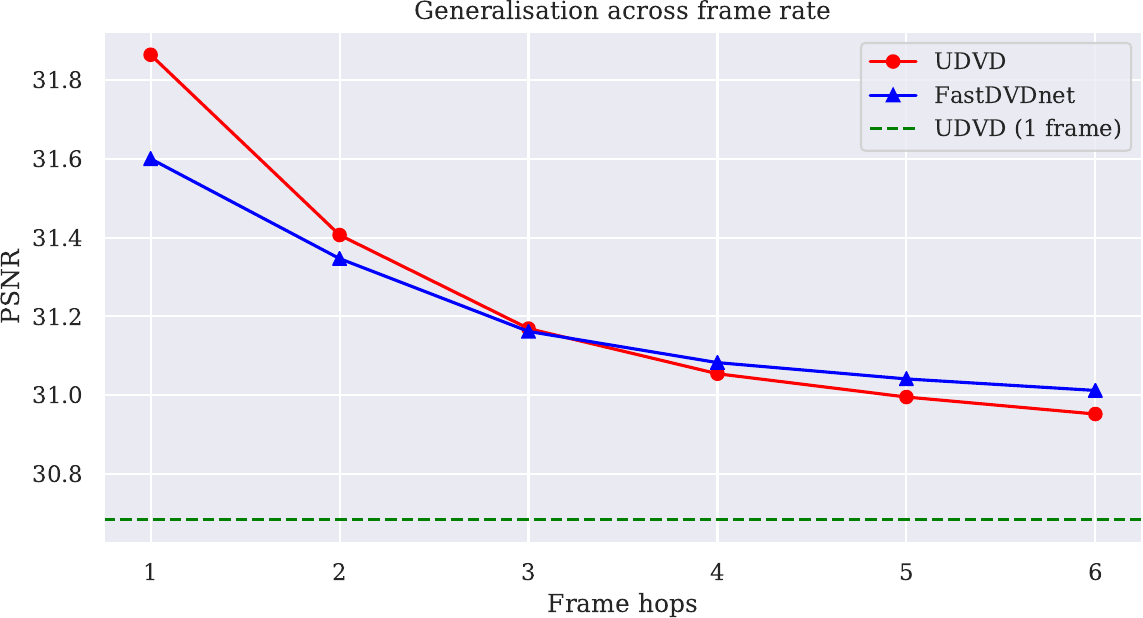}
    \caption{\textbf{Generalization across noise levels and frame rates.} (left) UDVD trained at only $\sigma=30$ generalizes well to noise levels not seen during training. The plotted points represent mean PSNR values evaluated on Set8. (right) UDVD generalizes well to faster videos (created by skipping frames) and consistently outperforms a baseline image denoiser (UDVD with a single input frame, shown as a green dashed line).}
    \label{fig:noise_frame_gen}
\end{figure}

\noindent \textbf{Generalization outside the training range:} The authors of \cite{biasfree} showed that CNNs trained for image denoising generalize well on test images corrupted with noise in the training range, but fails catastrophically when corrupted with noise strength outside the training range. The authors provided evidence that the overfitting is due to additive terms in the convolutional layers (and BatchNorm \cite{batchnorm} ) and showed that a CNN with no additive terms, called a \emph{bias-free} CNN generalizes well outside the training range. UDVD uses a bias-free architecture and generalizes well to noise levels outside its training range (Fig~\ref{fig:noise_frame_gen}). \\

\noindent \textbf{Generalization across frame rates:} To test generalization across frame rates, we simulated faster videos by skipping frames of videos in Set8. Fig~\ref{fig:noise_frame_gen} shows that UDVD generalizes robustly to faster videos and maintains a significant gain in performance over single-image denoising even when tested on videos where a large number of frames have been skipped (i.e. at a very low frame rate).

\section{Analysis of CNN-based Video Denoising}
\label{sec:thorough_analysis}

\subsection{Natural Videos}
\label{sec:filter_natural}
In Section~7 and Fig~4 of the paper we examined the equivalent filters and illustrated that UDVD learns to denoise by performing an average over a spatiotemporal neighbourhood of each pixel. Here we examine equivalent filters for more videos and a supervised CNN (FastDVDnet) and show that similar observations hold. \\

\noindent \textbf{Adaptive filtering:} Fig~\ref{fig:jacobian_1}, \ref{fig:jacobian_2}, \ref{fig:jacobian_3} and \ref{fig:jacobian_4} shows filters computed at a pixel for 4 different videos at 4 different noise levels. The filters adapt to the underlying signal content. They span larger areas as the noise level increases. These observations also holds for FastDVDnet, which is trained with supervision (Fig~\ref{fig:jacobian_fastfast})  \\

\noindent \textbf{Contribution of neighbouring frames for denoising:} UDVD tends to ignore temporally distant frames at lower noise levels as shown in Fig~\ref{fig:jacobian_1}, \ref{fig:jacobian_2}, \ref{fig:jacobian_3} and \ref{fig:jacobian_4}. This phenomenon is quantified in Fig~\ref{fig:filter_sum} by plotting the contribution of each frame to the denoised pixel by averaging over \textbf{5000} pixels from \textbf{250} random patches of size $128 \times 128$. At higher noise levels, UDVD seems to use distant frames more. This is consistent with the ablation study, which shows that for higher noise levels using more surrounding frames improves the denoising performance. 
Similar results hold for supervised CNN FastDVDnet, as shown in Fig~\ref{fig:jacobian_fastfast}. \\

\noindent \textbf{Local Averaging:} The weighting functions or equivalent filters perform an approximate averaging operation. They are mostly non-negative (although they do have some negative entries as depicted in blue in Fig~\ref{fig:jacobian_1}, \ref{fig:jacobian_2}, \ref{fig:jacobian_3} and \ref{fig:jacobian_4}) and they approximately sum up to one (see Fig~\ref{fig:filter_sum}).

\subsection{Real-world Data}

\begin{figure}
    \centering
    \includegraphics[width=0.49\textwidth]{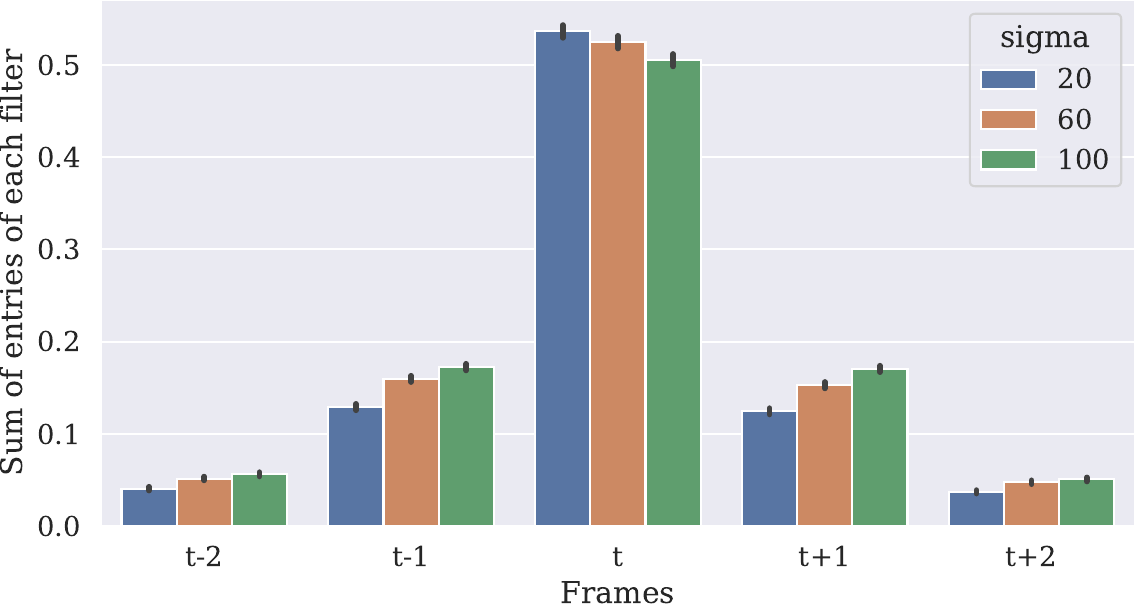}
    \includegraphics[width=0.49\textwidth]{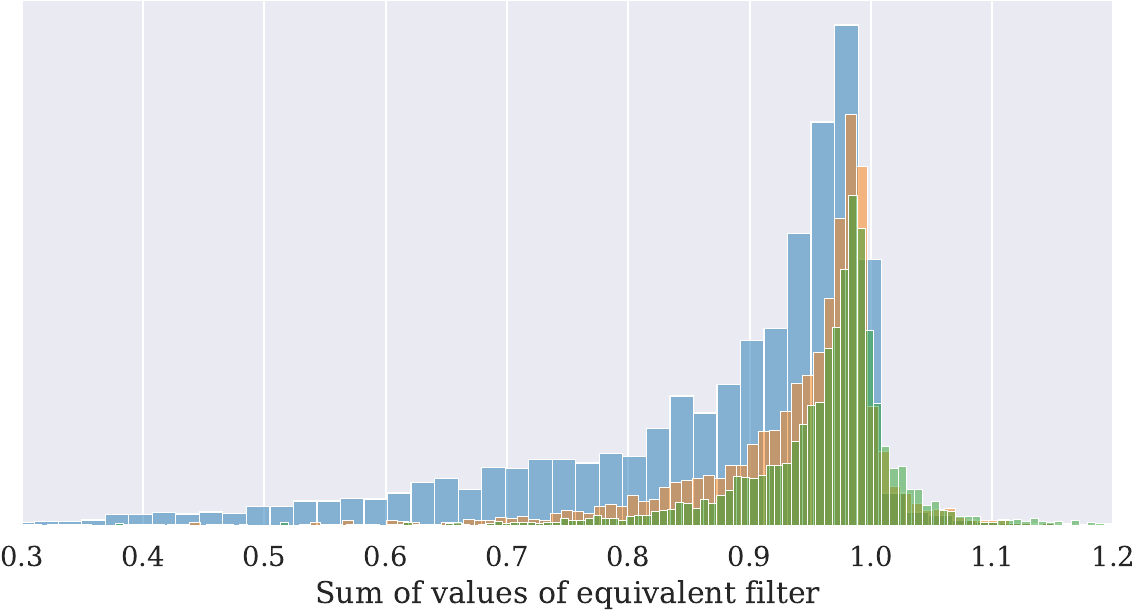}
    
    \includegraphics[width=0.49\textwidth]{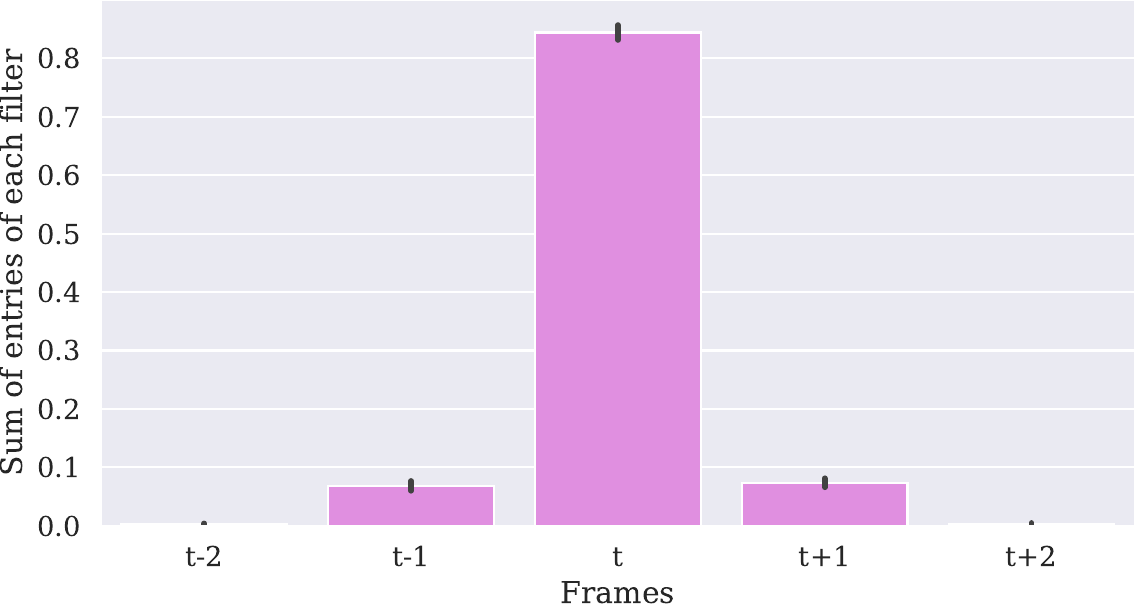}
    \includegraphics[width=0.49\textwidth]{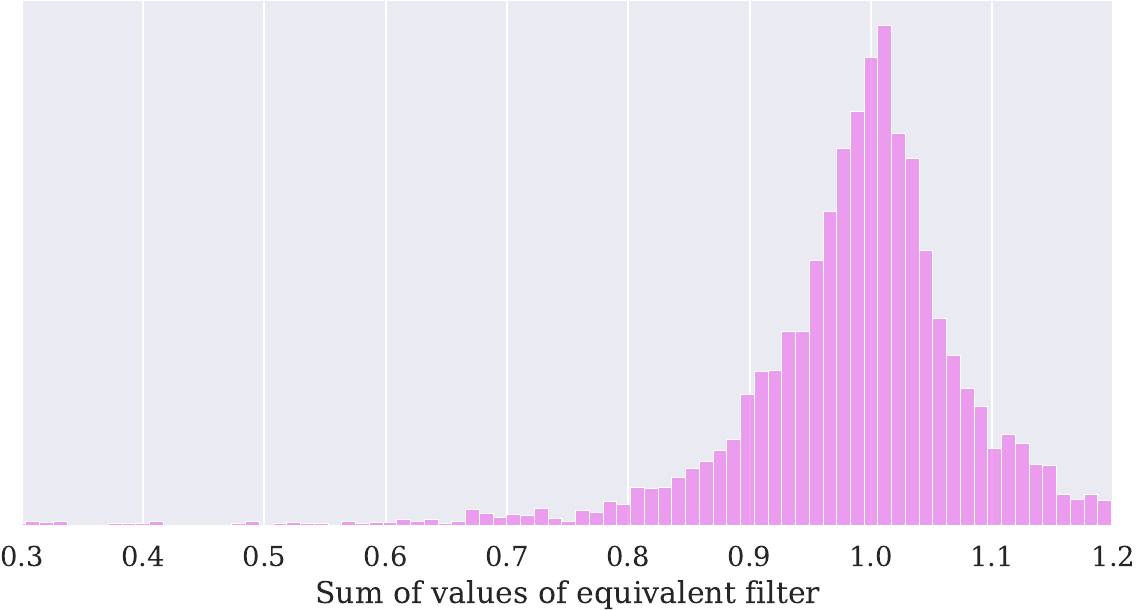}
    
    \includegraphics[width=0.49\textwidth]{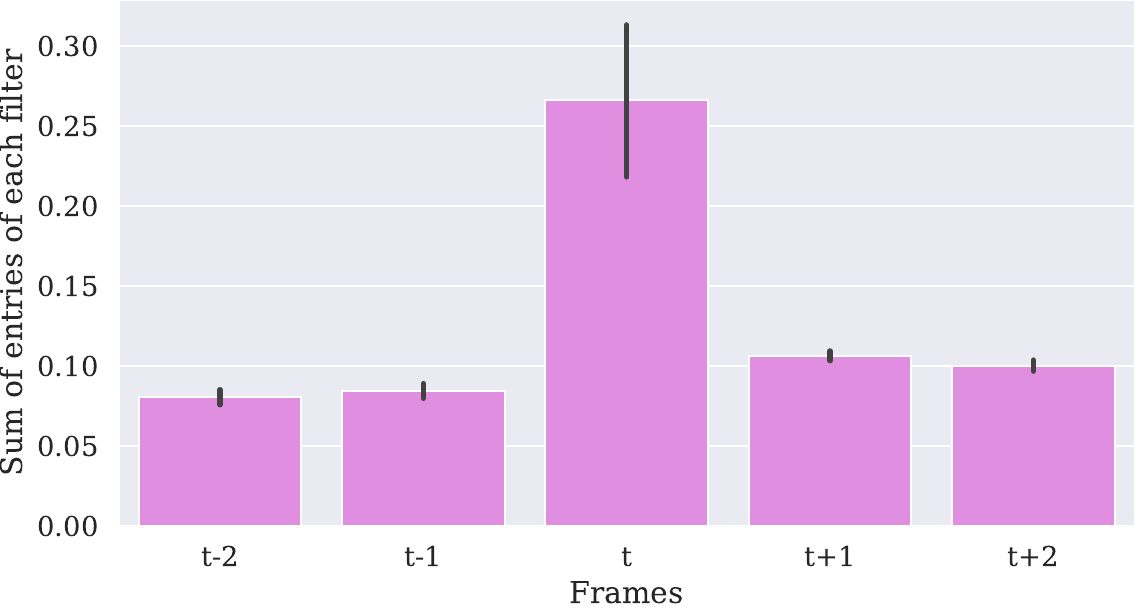}
    \includegraphics[width=0.49\textwidth]{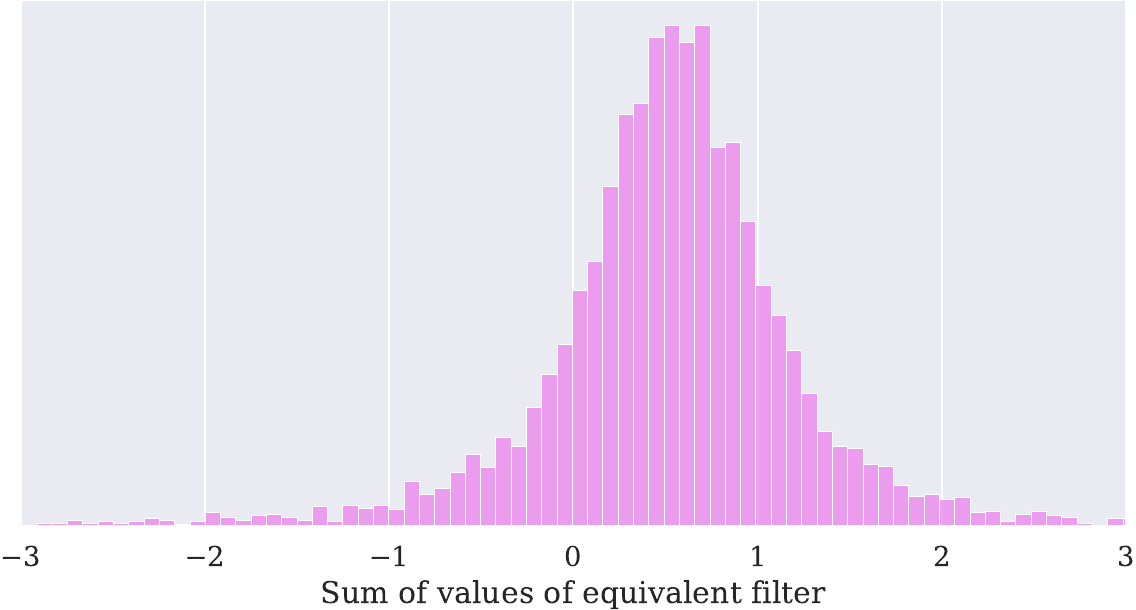}
    
    \caption{\textbf{Quantitative analysis of equivalent filters}. \emph{Left column:} The graphs show the sum of the entries of the equivalent filters in each frame, averaged over 5000 pixels from 250 random patches of size $128 \times 128$. For all datasets, the central frame dominates. For the DAVIS dataset (top), the contribution from the other frames increases with the noise level.
    For the fluorescence-microscopy data (mid) the contribution of the other frames is rather low, due to the high signal-to-noise ratio. For the electron-microscopy dataset the contribution of the other frames is larger (bottom). \emph{Right column:} Histogram of the sum of all entries in the equivalent filters (over all 5 frames) for 5000 pixels from 250 random patches of size $128 \times 128$ from the DAVIS test set (top), the fluorescence-microscopy dataset (mid) and the electron-microscopy dataset (bottom). For the DAVIS and fluorescence-microscopy datasets, the filters sum to $1$ in most cases. The peak of electron microscopy deviates significantly from $1$. This could be due to the noise model, which has non-Gaussian characteristics (it is Poisson with low counts).}
    \label{fig:filter_sum}
\end{figure}

Equivalent filters for the raw video, the fluorescence-microscopy and the electron-microscopy data are shown in Fig~\ref{fig:jacobian_micro}. The fluorescence -microscopy data have a low noise level. As expected from the results on natural videos (see Section~\ref{sec:denoise_natural}), the weighting functions are mostly confined to the middle frame (as quantified in Fig \ref{fig:filter_sum}). In the electron-microscopy dataset the weighting functions shows that the network relies on adjacent frames to estimate the denoised (as quantified in Fig \ref{fig:filter_sum}). %

\subsection{Motion Estimation}
\begin{figure*}
    \def\f1ht{\linewidth}%
     
     \centering 
     \begin{tabular}{ >{\centering\arraybackslash}m{0.02\linewidth}
     >{\centering\arraybackslash}m{0.11\linewidth}
     >{\centering\arraybackslash}m{0.11\linewidth}
     >{\centering\arraybackslash}m{0.11\linewidth}
     >{\centering\arraybackslash}m{0.11\linewidth}
     >{\centering\arraybackslash}m{0.11\linewidth}
     >{\centering\arraybackslash}m{0.11\linewidth}
     >{\centering\arraybackslash}m{0.11\linewidth}
     >{\centering\arraybackslash}m{0.03\linewidth}
     }
     \centering
     
     & & &
     \includegraphics[width=\f1ht]{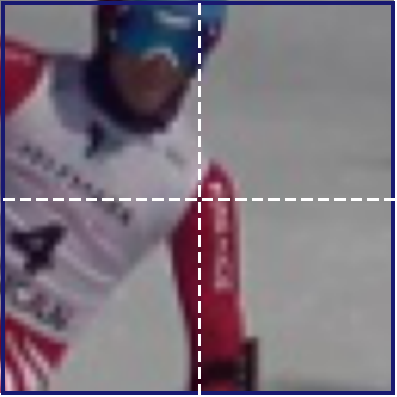} & 
     \includegraphics[width=\f1ht]{images/gradient_plots/truth_DAVIS_giant-slalom_109_1.pdf} &
     \includegraphics[width=\f1ht]{images/gradient_plots/truth_DAVIS_giant-slalom_109_2.pdf} &
     \includegraphics[width=\f1ht]{images/gradient_plots/truth_DAVIS_giant-slalom_109_3.pdf} & 
     \includegraphics[width=\f1ht]{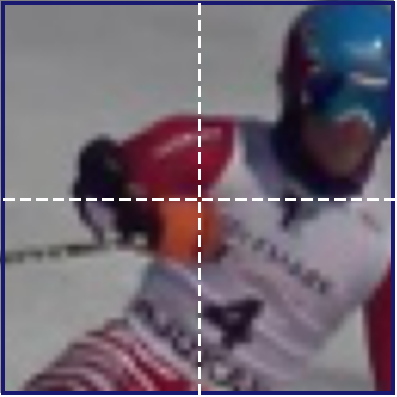} \\
     
     $30$ & \includegraphics[width=\f1ht]{images/gradient_plots/noisy_DAVIS_giant-slalom_109_30.pdf} & 
     \includegraphics[width=\f1ht]{images/gradient_plots/denoised_DAVIS_giant-slalom_109_30.pdf} &
     \includegraphics[width=\f1ht]{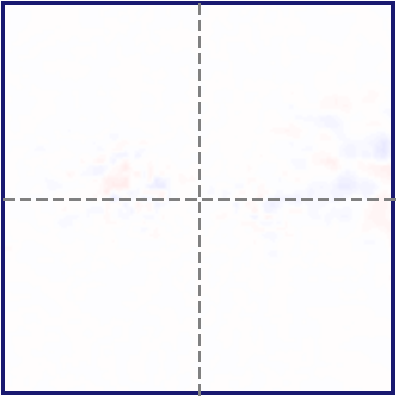} &
     \includegraphics[width=\f1ht]{images/gradient_plots/grad_DAVIS_giant-slalom_109_30_1.pdf} &
     \includegraphics[width=\f1ht]{images/gradient_plots/grad_DAVIS_giant-slalom_109_30_2.pdf} &
     \includegraphics[width=\f1ht]{images/gradient_plots/grad_DAVIS_giant-slalom_109_30_3.pdf} &
     \includegraphics[width=\f1ht]{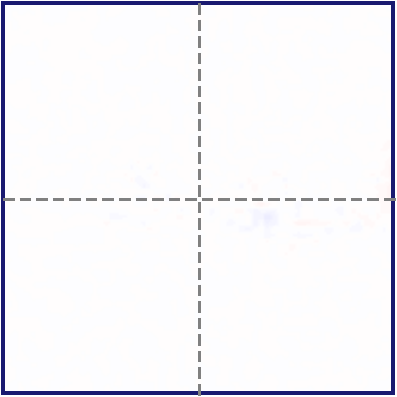} &
     \includegraphics[width=0.03\textwidth]{images/gradient_plots/cbar_DAVIS_giant-slalom_109_30.pdf} \\
     
     $45$ & \includegraphics[width=\f1ht]{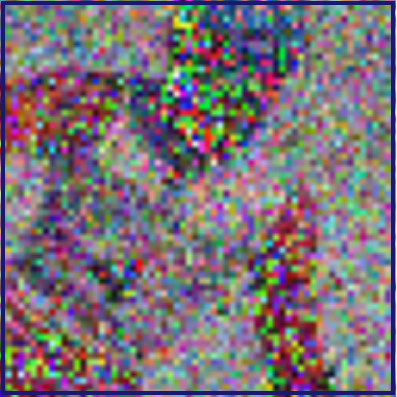} & 
     \includegraphics[width=\f1ht]{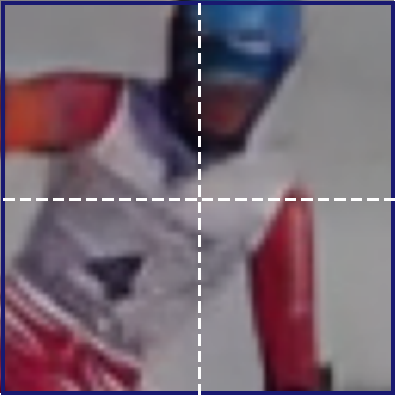} &
     \includegraphics[width=\f1ht]{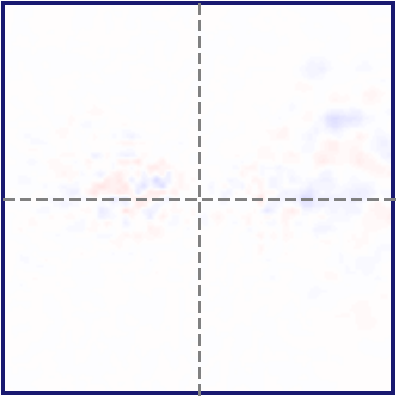} &
     \includegraphics[width=\f1ht]{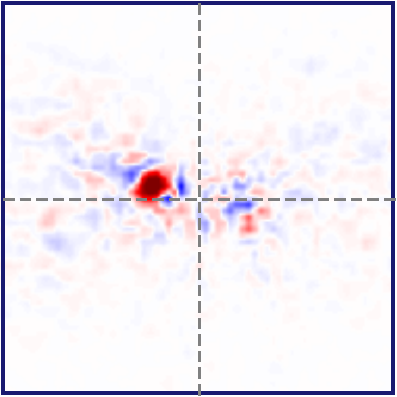} &
     \includegraphics[width=\f1ht]{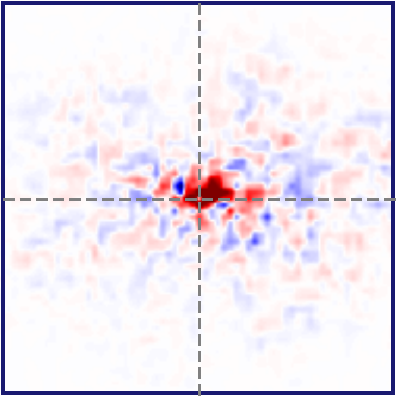} &
     \includegraphics[width=\f1ht]{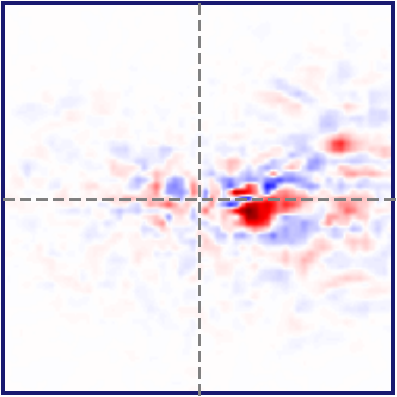} &
     \includegraphics[width=\f1ht]{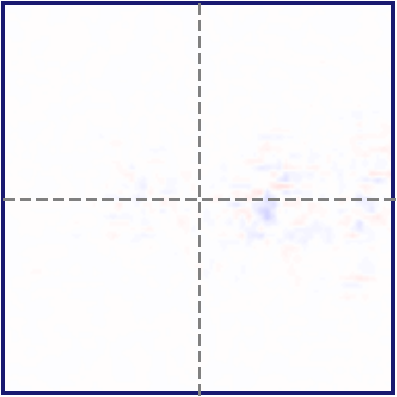} &
     \includegraphics[width=0.03\textwidth]{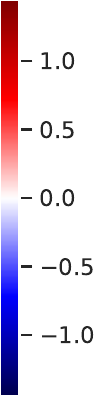} \\
     
     $60$ & \includegraphics[width=\f1ht]{images/gradient_plots/noisy_DAVIS_giant-slalom_109_60.pdf} & 
     \includegraphics[width=\f1ht]{images/gradient_plots/denoised_DAVIS_giant-slalom_109_60.pdf} &
     \includegraphics[width=\f1ht]{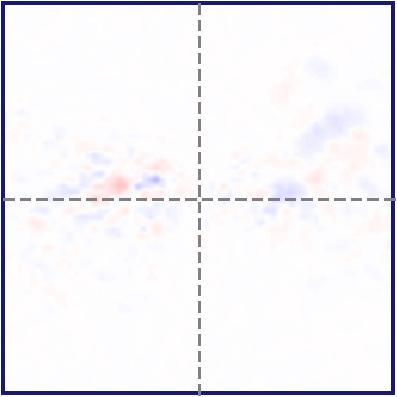} &
     \includegraphics[width=\f1ht]{images/gradient_plots/grad_DAVIS_giant-slalom_109_60_1.pdf} &
     \includegraphics[width=\f1ht]{images/gradient_plots/grad_DAVIS_giant-slalom_109_60_2.pdf} &
     \includegraphics[width=\f1ht]{images/gradient_plots/grad_DAVIS_giant-slalom_109_60_3.pdf} &
     \includegraphics[width=\f1ht]{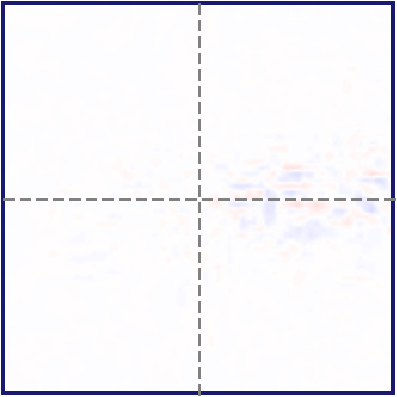} &
     \includegraphics[width=0.03\textwidth]{images/gradient_plots/cbar_DAVIS_giant-slalom_109_60.pdf} \\
     
     $75$ & \includegraphics[width=\f1ht]{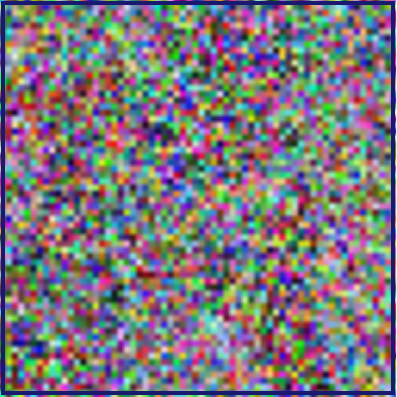} & 
     \includegraphics[width=\f1ht]{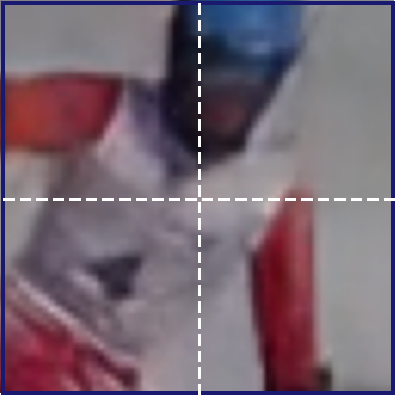} &
     \includegraphics[width=\f1ht]{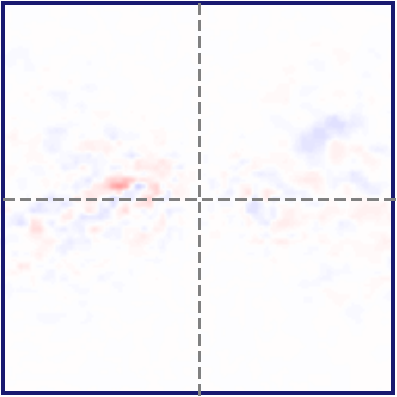} &
     \includegraphics[width=\f1ht]{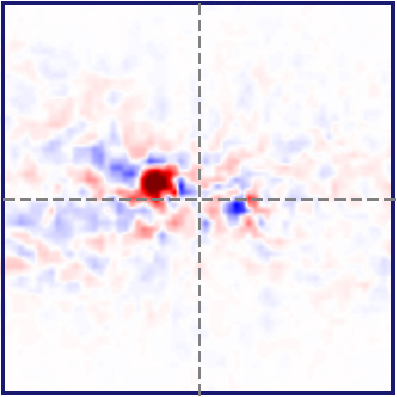} &
     \includegraphics[width=\f1ht]{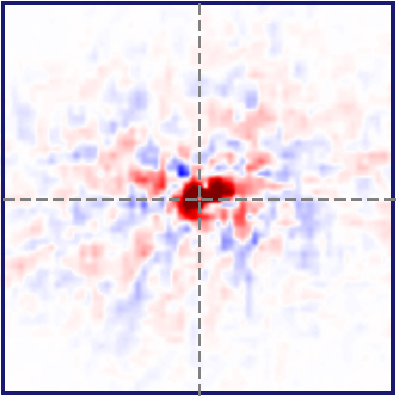} &
     \includegraphics[width=\f1ht]{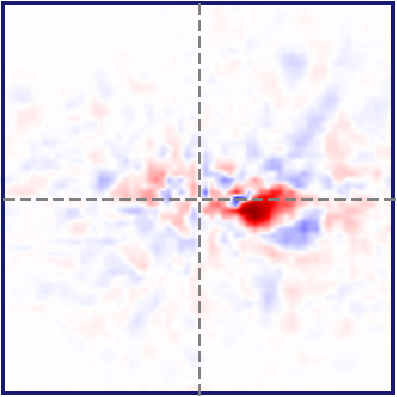} &
     \includegraphics[width=\f1ht]{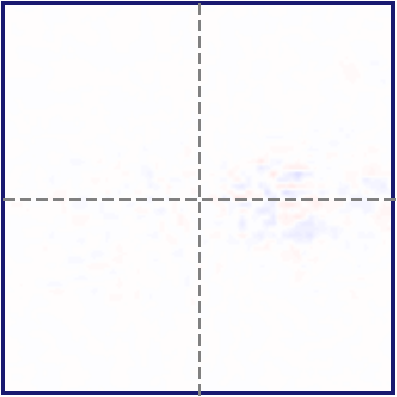} &
     \includegraphics[width=0.03\textwidth]{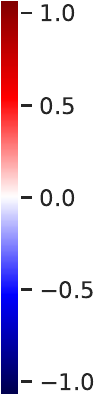} \\
     
      $\sigma$ & $y_t$ &  $d_t$ & $a(t-2, i)$ &  $a(t-1, i)$ & $a(t, i)$ & $a(t+1, i)$ & $a(t+2, i)$ &  \\
     \end{tabular}
     
     \vspace{0.2cm}
     
\caption{\textbf{Video denoising as spatiotemporal adaptive filtering; \texttt{giant-slalom} video from the DAVIS dataset}. Visualization of the linear weighting functions ($a(k,i)$, Section~6 of paper) of UDVD. The left two columns show the noisy frame $y_t$ at four levels of noise, and the corresponding denoised frame, $d_t$.  Weighting functions $a(k, i)$ corresponding to the pixel $i$ (at the intersection of the dashed white lines), for five successive frames, are shown in the last five columns. The weighting functions adapt to underlying image content, and are shifted to track the motion of the skier. As the noise level $\sigma$ increases, their spatial extent grows, averaging out more of the noise while respecting object boundaries.  
The weighting functions corresponding to the five frames approximately sum to one, and thus compute a local average (although some weights are negative, depicted in blue) as explained in Section~\ref{sec:filter_natural}.}

\label{fig:jacobian_1}
\end{figure*}

\begin{figure*}[ht]
    \def\f1ht{\linewidth}%
     
     \centering 
     \begin{tabular}{ >{\centering\arraybackslash}m{0.02\linewidth}
     >{\centering\arraybackslash}m{0.11\linewidth}
     >{\centering\arraybackslash}m{0.11\linewidth}
     >{\centering\arraybackslash}m{0.11\linewidth}
     >{\centering\arraybackslash}m{0.11\linewidth}
     >{\centering\arraybackslash}m{0.11\linewidth}
     >{\centering\arraybackslash}m{0.11\linewidth}
     >{\centering\arraybackslash}m{0.11\linewidth}
     >{\centering\arraybackslash}m{0.03\linewidth}
     }
     \centering
     
     & & &
     \includegraphics[width=\f1ht]{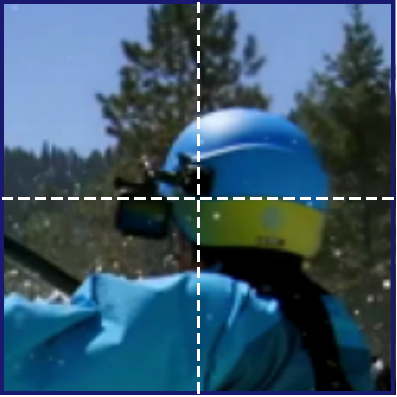} & 
     \includegraphics[width=\f1ht]{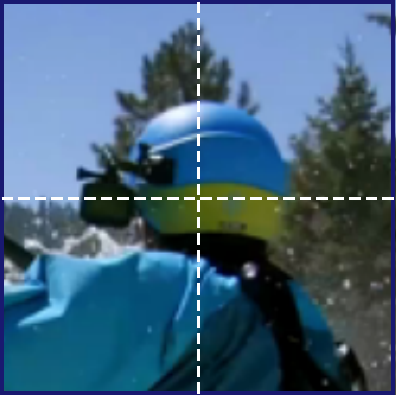} &
     \includegraphics[width=\f1ht]{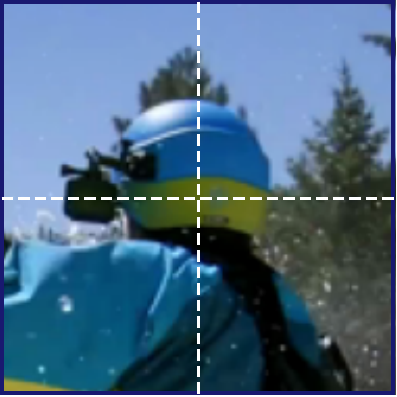} &
     \includegraphics[width=\f1ht]{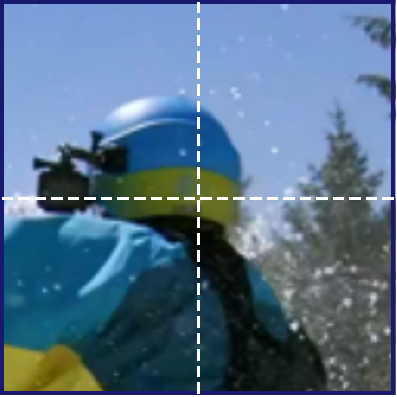} & 
     \includegraphics[width=\f1ht]{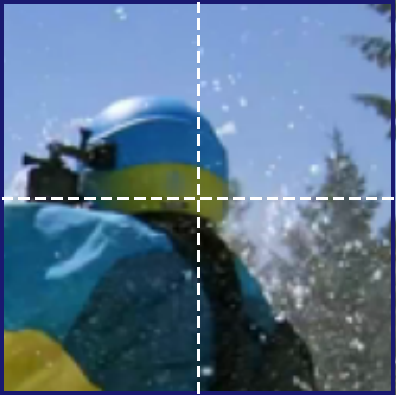} \\
     
     30 & \includegraphics[width=\f1ht]{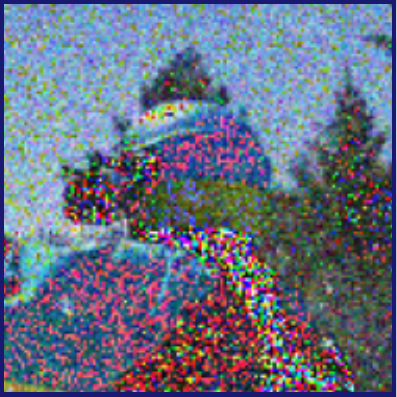} & 
     \includegraphics[width=\f1ht]{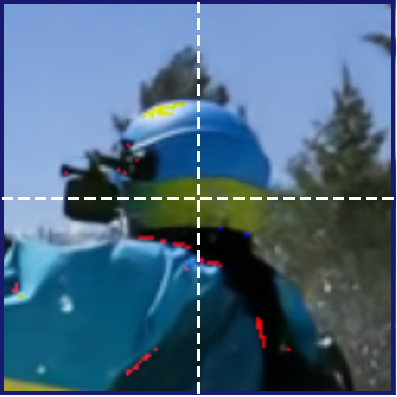} &
     \includegraphics[width=\f1ht]{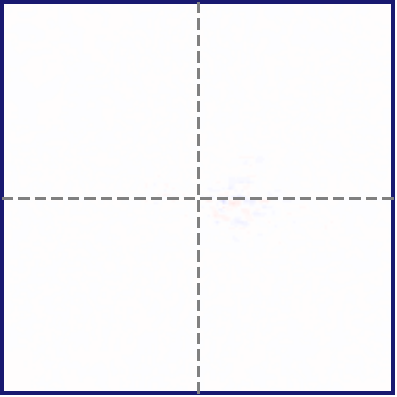} &
     \includegraphics[width=\f1ht]{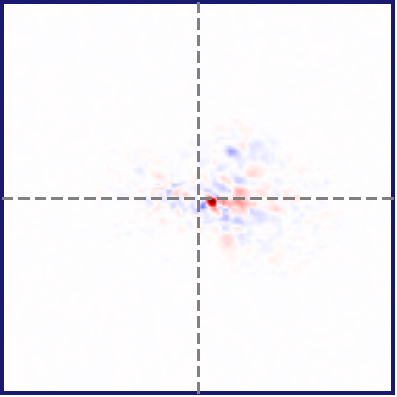} &
     \includegraphics[width=\f1ht]{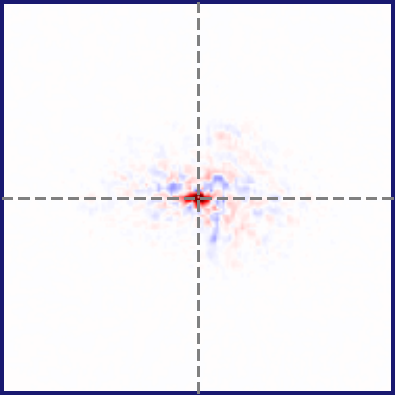} &
     \includegraphics[width=\f1ht]{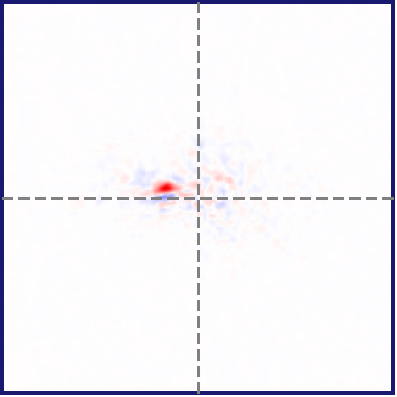} &
     \includegraphics[width=\f1ht]{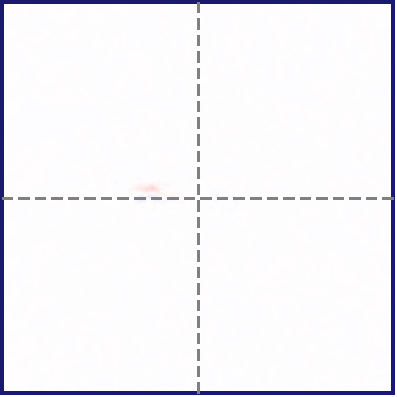} &
     \includegraphics[width=0.022\textwidth]{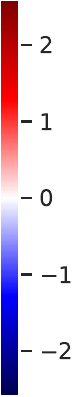} \\
     
     45 & \includegraphics[width=\f1ht]{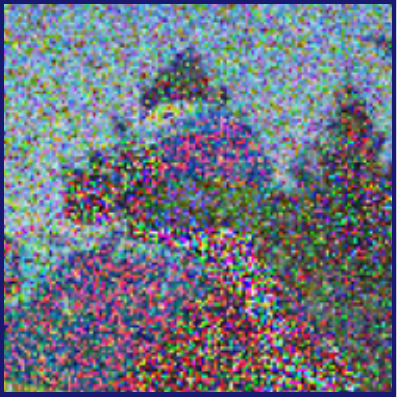} & 
     \includegraphics[width=\f1ht]{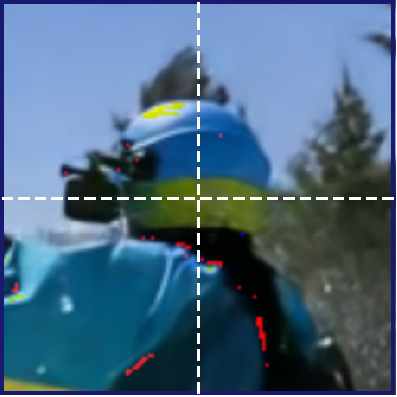} &
     \includegraphics[width=\f1ht]{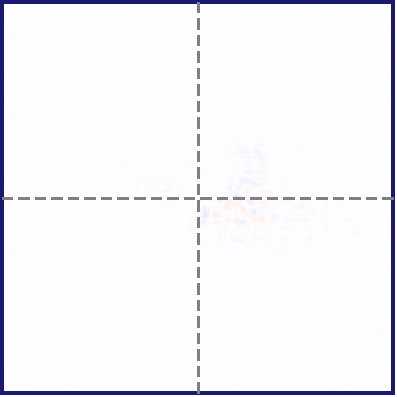} &
     \includegraphics[width=\f1ht]{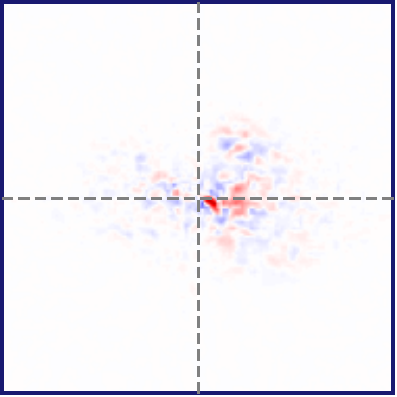} &
     \includegraphics[width=\f1ht]{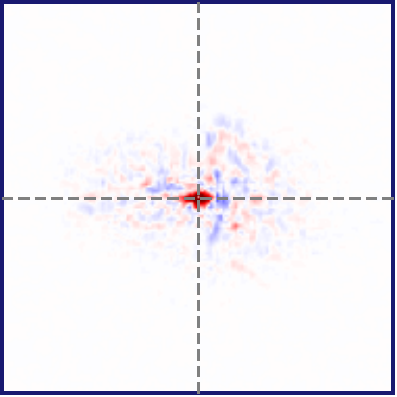} &
     \includegraphics[width=\f1ht]{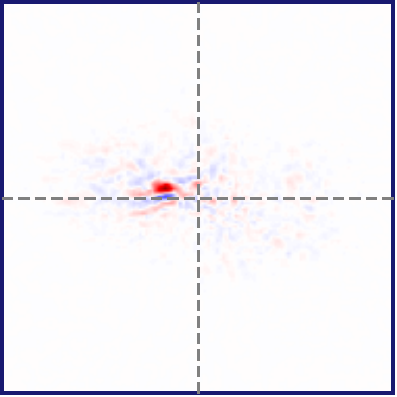} &
     \includegraphics[width=\f1ht]{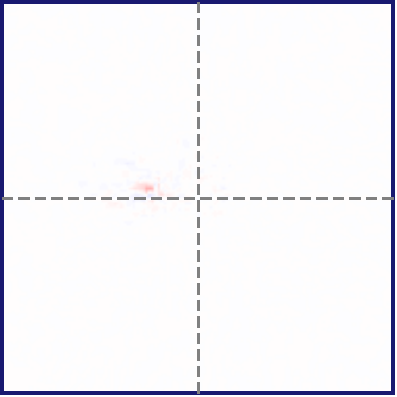} &
     \includegraphics[width=0.022\textwidth]{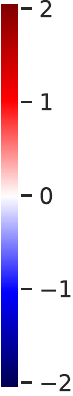} \\
     
     60 & \includegraphics[width=\f1ht]{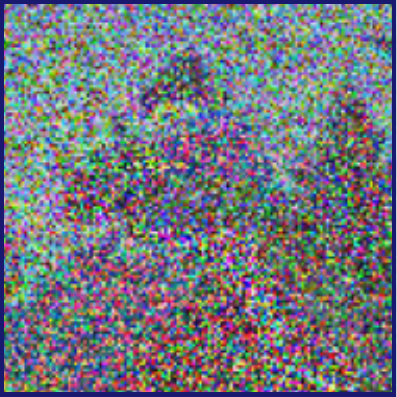} & 
     \includegraphics[width=\f1ht]{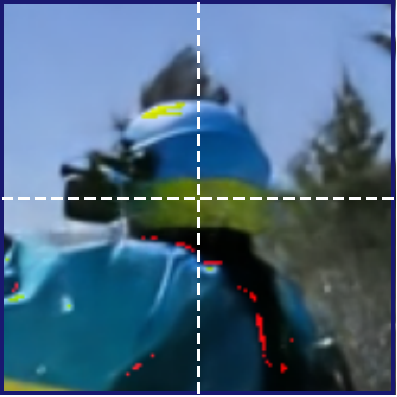} &
     \includegraphics[width=\f1ht]{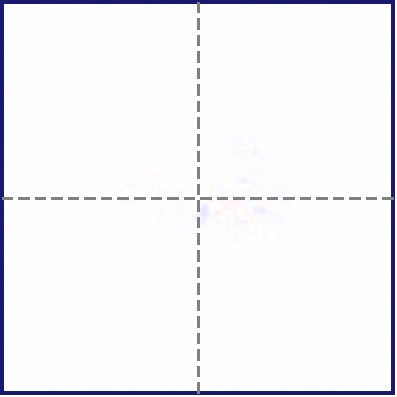} &
     \includegraphics[width=\f1ht]{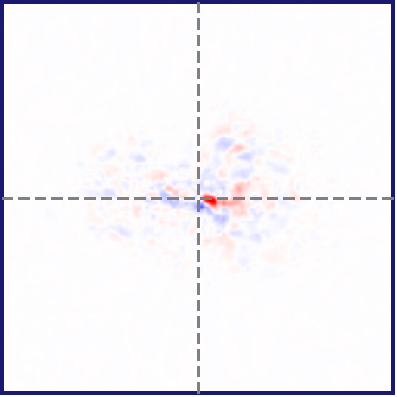} &
     \includegraphics[width=\f1ht]{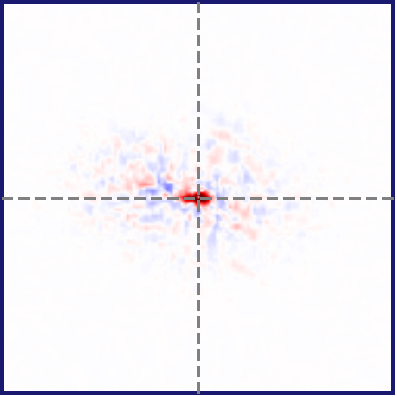} &
     \includegraphics[width=\f1ht]{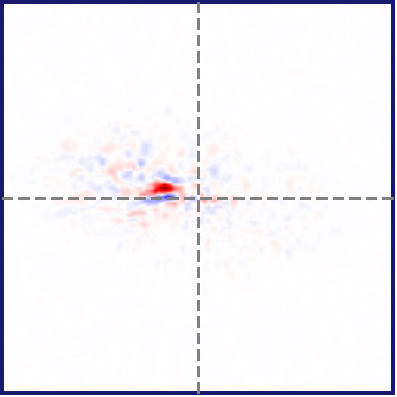} &
     \includegraphics[width=\f1ht]{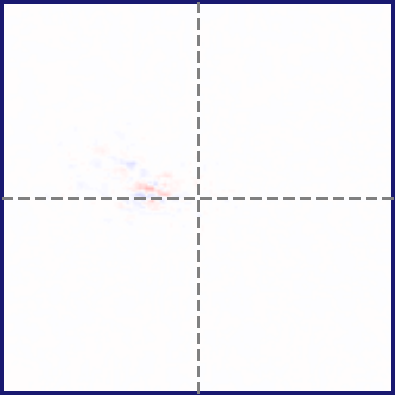} &
     \includegraphics[width=0.022\textwidth]{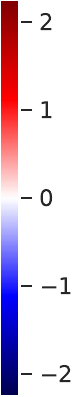} \\
     
     75 & \includegraphics[width=\f1ht]{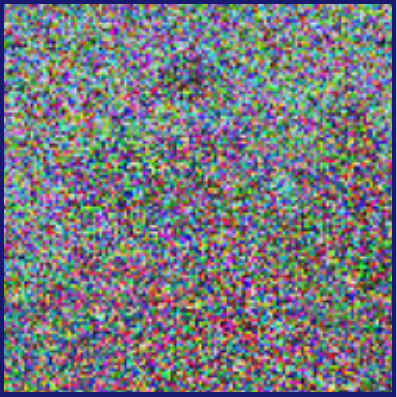} & 
     \includegraphics[width=\f1ht]{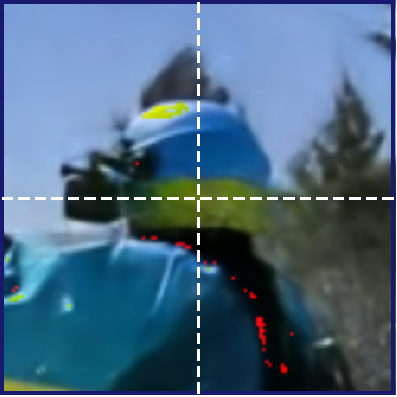} &
     \includegraphics[width=\f1ht]{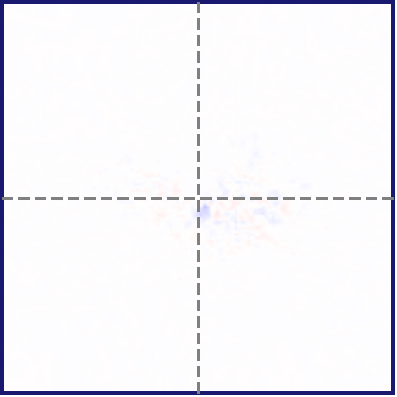} &
     \includegraphics[width=\f1ht]{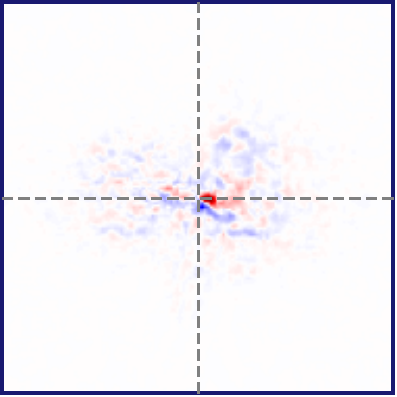} &
     \includegraphics[width=\f1ht]{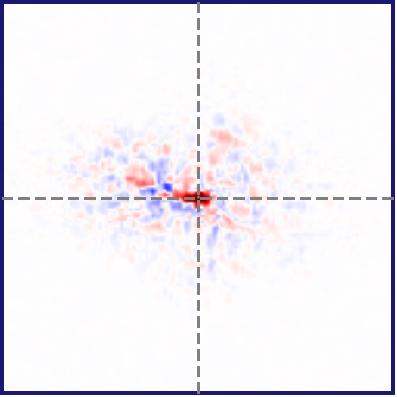} &
     \includegraphics[width=\f1ht]{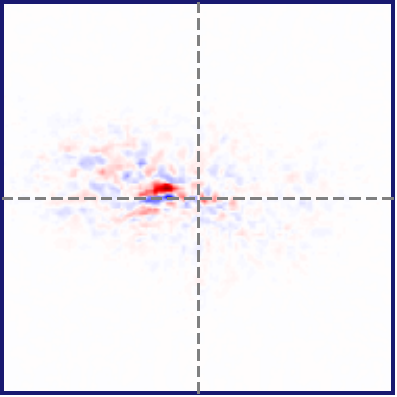} &
     \includegraphics[width=\f1ht]{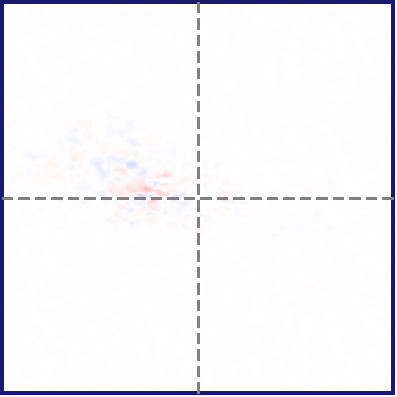} &
     \includegraphics[width=0.03\textwidth]{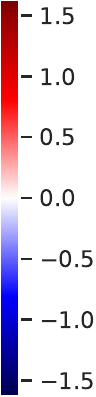} \\
     
     $\sigma$ & $y_t$ &  $d_t$ & $a(t-2, i)$ &  $a(t-1, i)$ & $a(t, i)$ & $a(t+1, i)$ & $a(t+2, i)$ &  \\
     \end{tabular}
     
     \vspace{0.2cm}
     
\caption{\textbf{Video denoising as spatiotemporal adaptive filtering; \texttt{rafting} video from the GoPro dataset}. Visualization of the equivalent filters, as described in Fig~\ref{fig:jacobian_1}.}

\label{fig:jacobian_2}
\end{figure*}

\begin{figure*}[ht]
    \def\f1ht{\linewidth}%
     
     \centering 
     \begin{tabular}{ >{\centering\arraybackslash}m{0.02\linewidth}
     >{\centering\arraybackslash}m{0.11\linewidth}
     >{\centering\arraybackslash}m{0.11\linewidth}
     >{\centering\arraybackslash}m{0.11\linewidth}
     >{\centering\arraybackslash}m{0.11\linewidth}
     >{\centering\arraybackslash}m{0.11\linewidth}
     >{\centering\arraybackslash}m{0.11\linewidth}
     >{\centering\arraybackslash}m{0.11\linewidth}
     >{\centering\arraybackslash}m{0.03\linewidth}
     }
     \centering
     
     & & &
     \includegraphics[width=\f1ht]{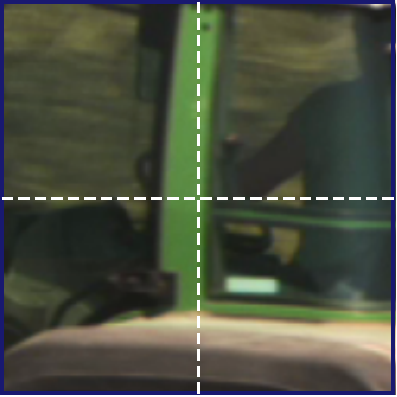} & 
     \includegraphics[width=\f1ht]{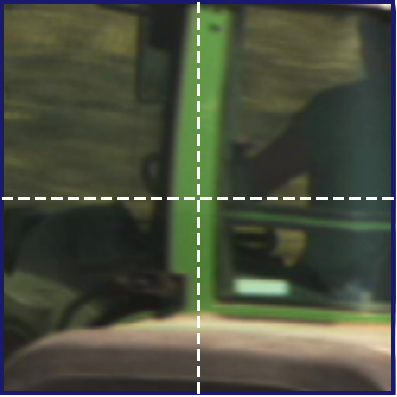} &
     \includegraphics[width=\f1ht]{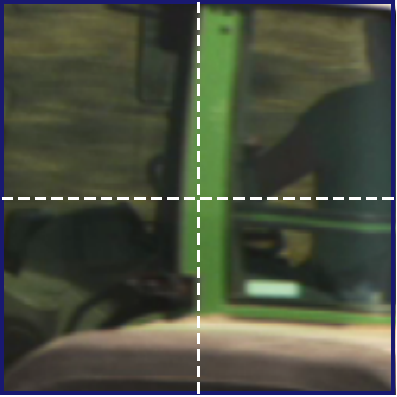} &
     \includegraphics[width=\f1ht]{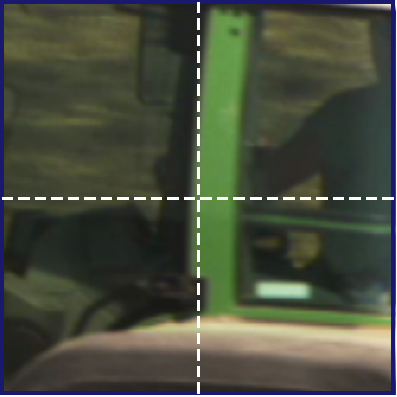} & 
     \includegraphics[width=\f1ht]{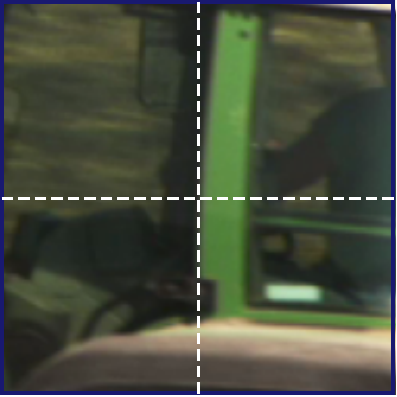} \\
     
     30 & \includegraphics[width=\f1ht]{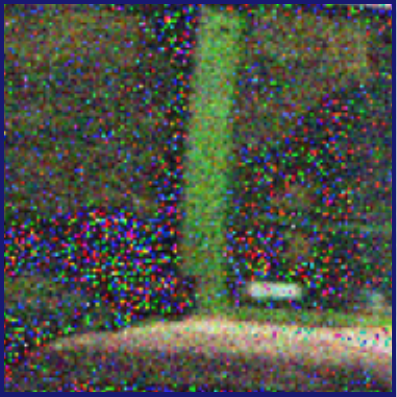} & 
     \includegraphics[width=\f1ht]{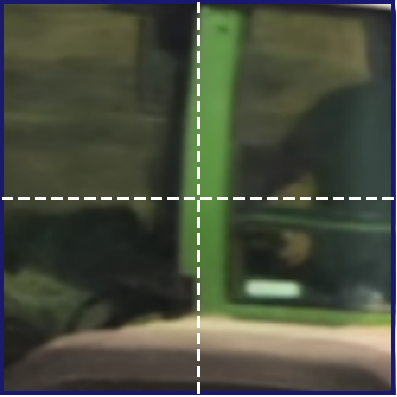} &
     \includegraphics[width=\f1ht]{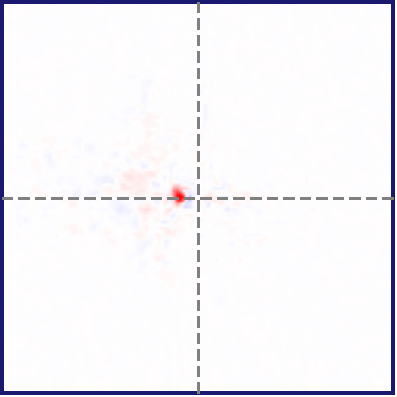} &
     \includegraphics[width=\f1ht]{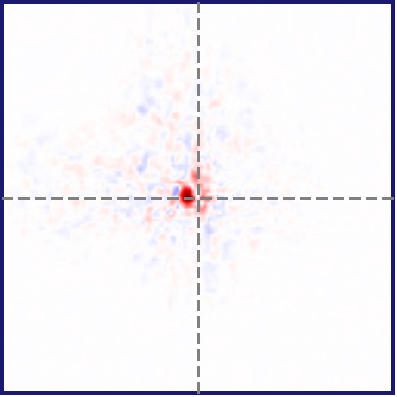} &
     \includegraphics[width=\f1ht]{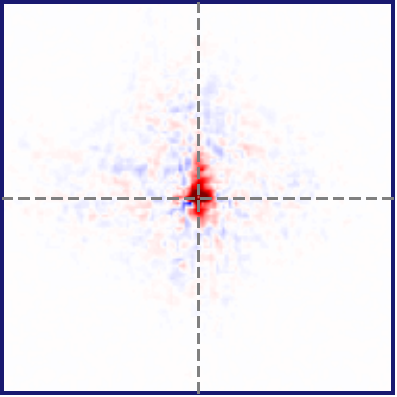} &
     \includegraphics[width=\f1ht]{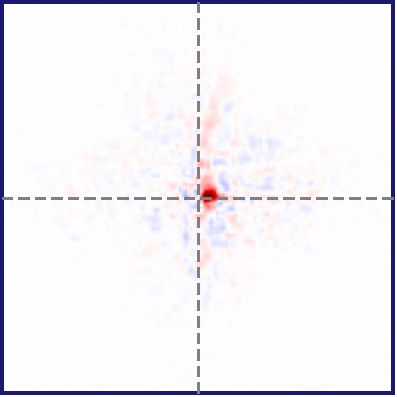} &
     \includegraphics[width=\f1ht]{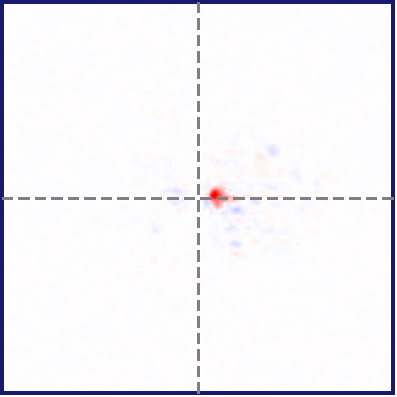} &
     \includegraphics[width=0.03\textwidth]{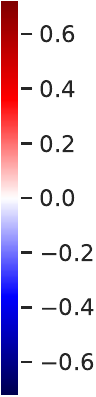} \\
     
     45 & \includegraphics[width=\f1ht]{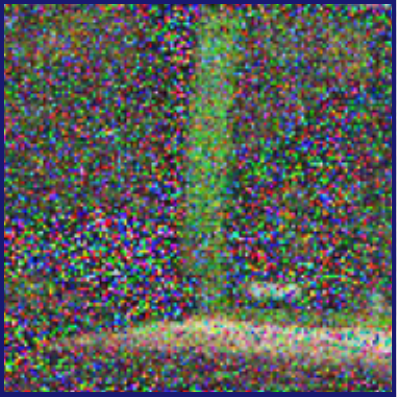} & 
     \includegraphics[width=\f1ht]{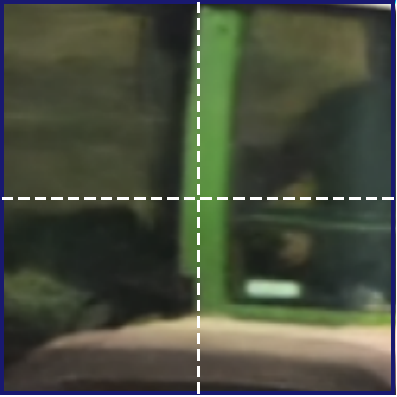} &
     \includegraphics[width=\f1ht]{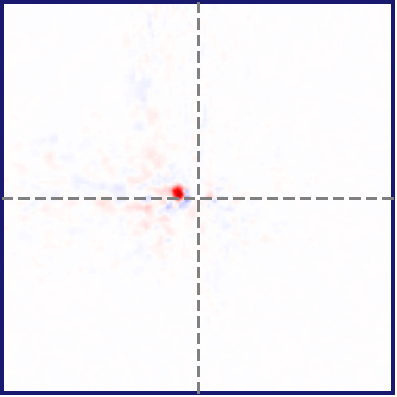} &
     \includegraphics[width=\f1ht]{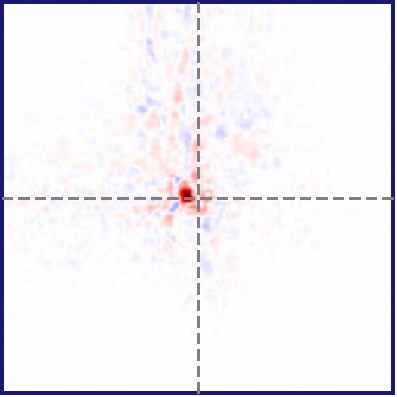} &
     \includegraphics[width=\f1ht]{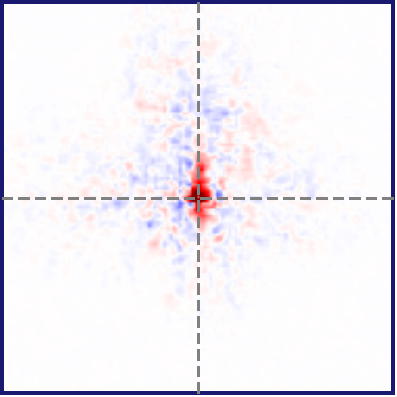} &
     \includegraphics[width=\f1ht]{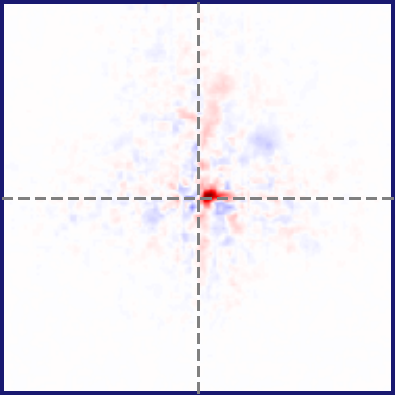} &
     \includegraphics[width=\f1ht]{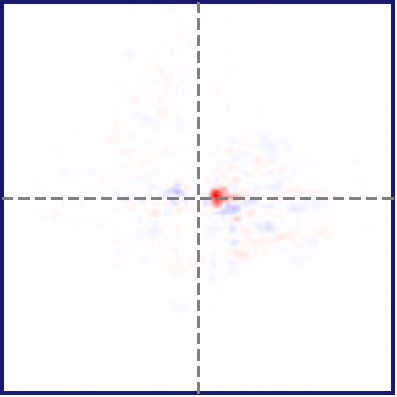} &
     \includegraphics[width=0.03\textwidth]{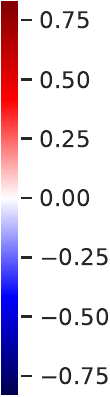} \\
     
     60 & \includegraphics[width=\f1ht]{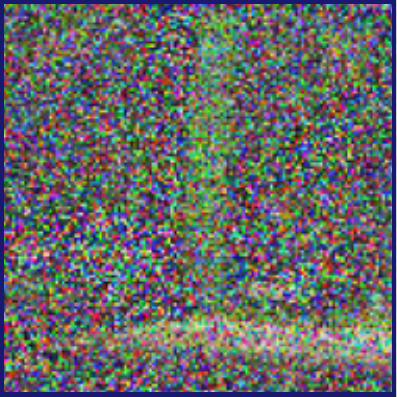} & 
     \includegraphics[width=\f1ht]{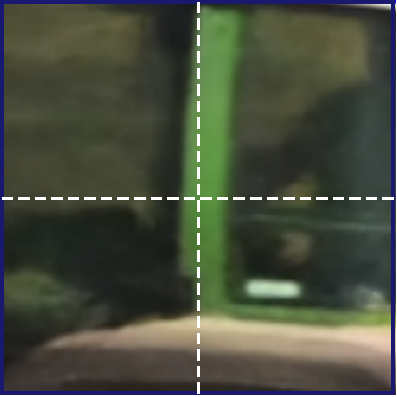} &
     \includegraphics[width=\f1ht]{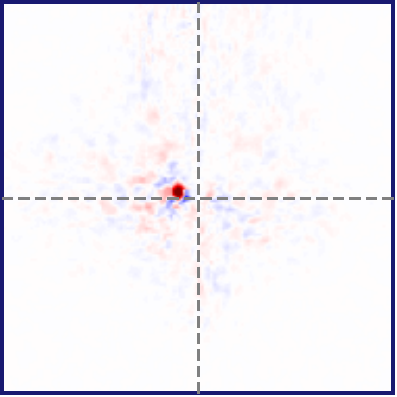} &
     \includegraphics[width=\f1ht]{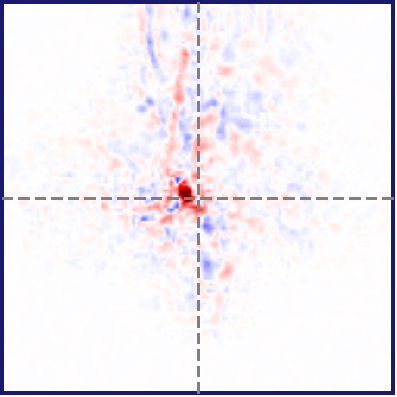} &
     \includegraphics[width=\f1ht]{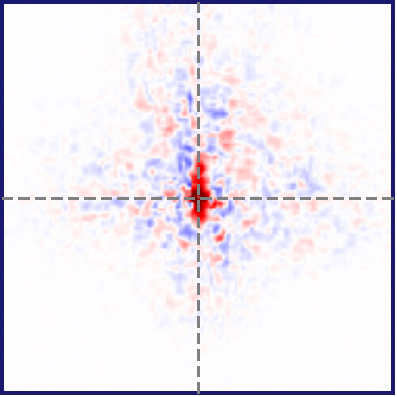} &
     \includegraphics[width=\f1ht]{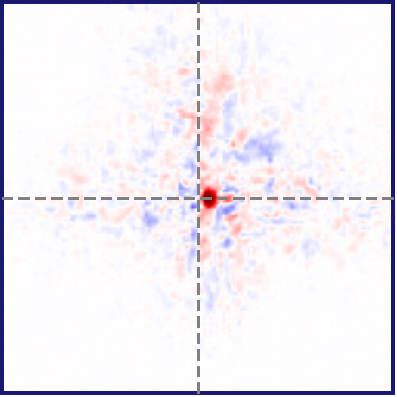} &
     \includegraphics[width=\f1ht]{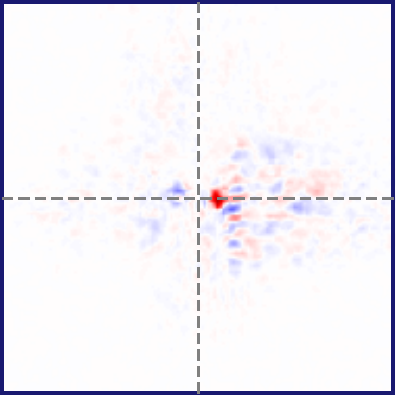} &
     \includegraphics[width=0.03\textwidth]{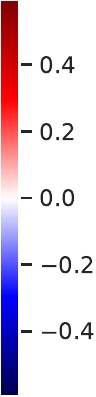} \\
     
     75 & \includegraphics[width=\f1ht]{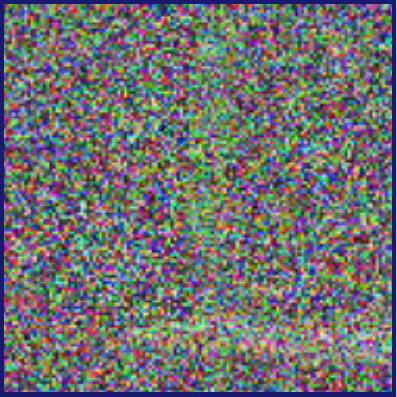} & 
     \includegraphics[width=\f1ht]{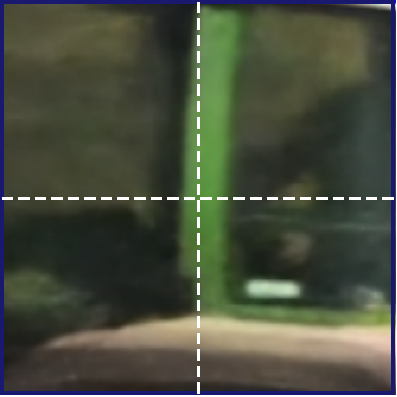} &
     \includegraphics[width=\f1ht]{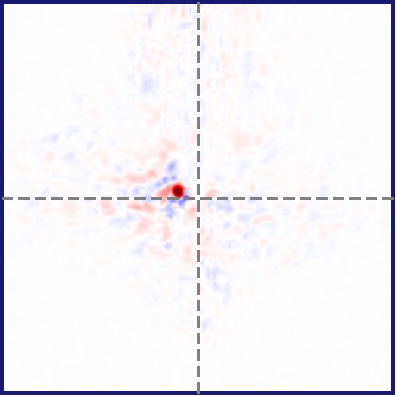} &
     \includegraphics[width=\f1ht]{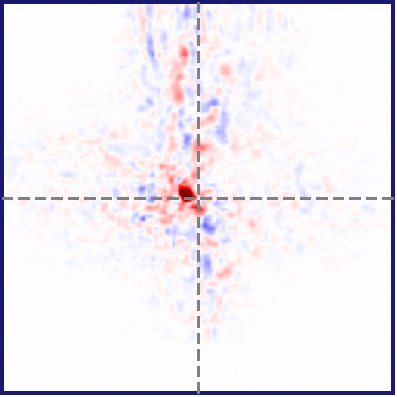} &
     \includegraphics[width=\f1ht]{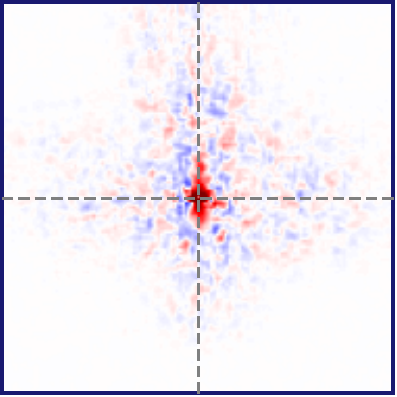} &
     \includegraphics[width=\f1ht]{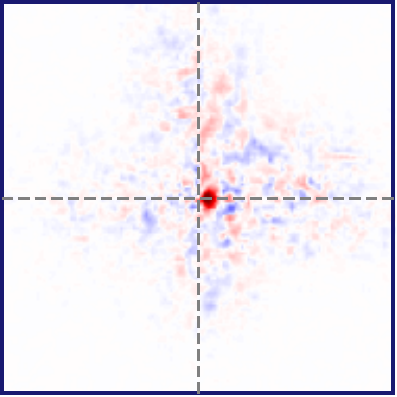} &
     \includegraphics[width=\f1ht]{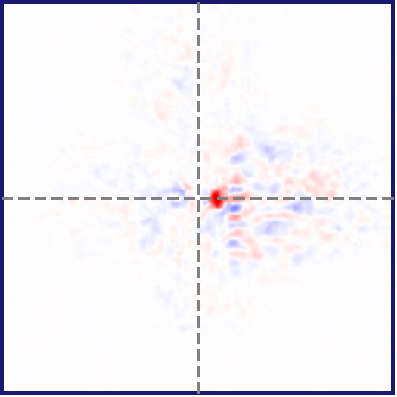} &
     \includegraphics[width=0.03\textwidth]{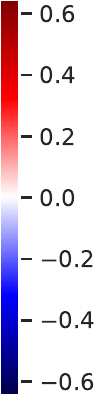} \\
     
        $\sigma$ & $y_t$ &  $d_t$ & $a(t-2, i)$ &  $a(t-1, i)$ & $a(t, i)$ & $a(t+1, i)$ & $a(t+2, i)$ &  \\
     \end{tabular}
     
     \vspace{0.2cm}
     
\caption{\textbf{Video denoising as spatiotemporal adaptive filtering; \texttt{tractor} video from Set8}. Visualization of the equivalent filters, as described in Fig~\ref{fig:jacobian_1}.}

\label{fig:jacobian_3}
\end{figure*}

\begin{figure*}[ht]
    \def\f1ht{\linewidth}%
     
     \centering 
    \begin{tabular}{ >{\centering\arraybackslash}m{0.02\linewidth}
     >{\centering\arraybackslash}m{0.11\linewidth}
     >{\centering\arraybackslash}m{0.11\linewidth}
     >{\centering\arraybackslash}m{0.11\linewidth}
     >{\centering\arraybackslash}m{0.11\linewidth}
     >{\centering\arraybackslash}m{0.11\linewidth}
     >{\centering\arraybackslash}m{0.11\linewidth}
     >{\centering\arraybackslash}m{0.11\linewidth}
     >{\centering\arraybackslash}m{0.03\linewidth}
     }
     \centering
     
     & & &
     \includegraphics[width=\f1ht]{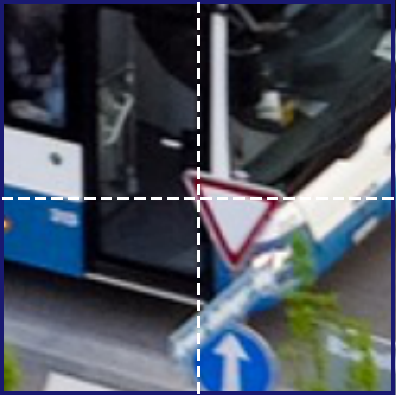} & 
     \includegraphics[width=\f1ht]{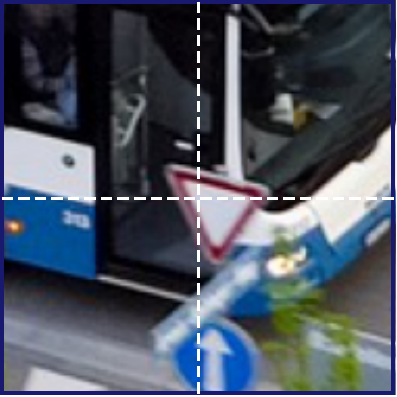} &
     \includegraphics[width=\f1ht]{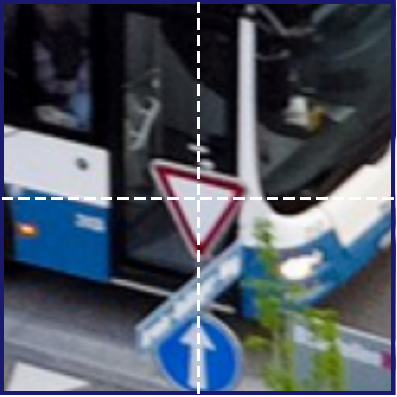} &
     \includegraphics[width=\f1ht]{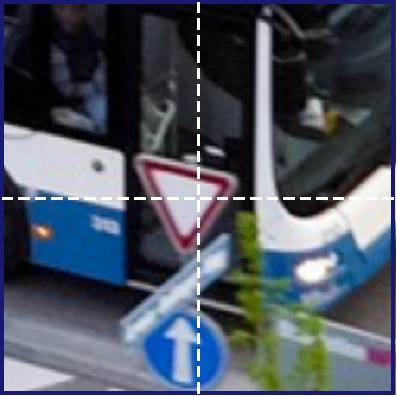} & 
     \includegraphics[width=\f1ht]{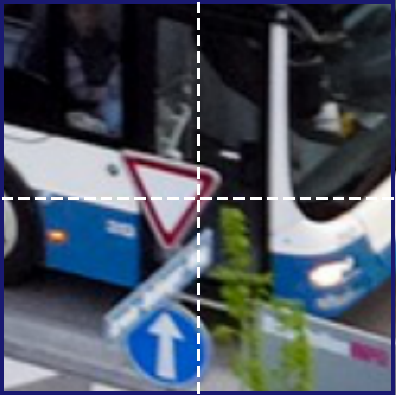} \\
     
     30 & \includegraphics[width=\f1ht]{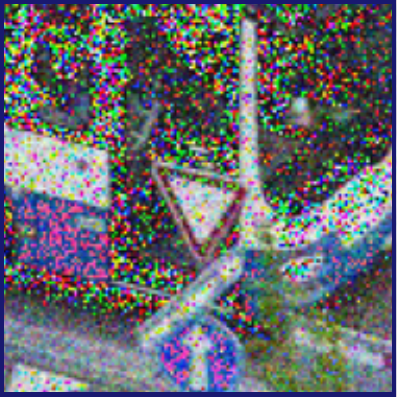} & 
     \includegraphics[width=\f1ht]{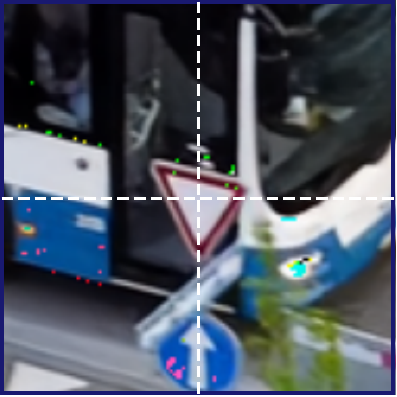} &
     \includegraphics[width=\f1ht]{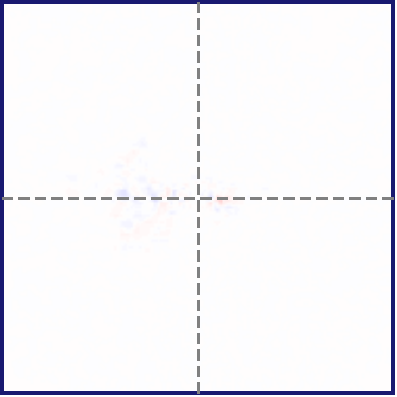} &
     \includegraphics[width=\f1ht]{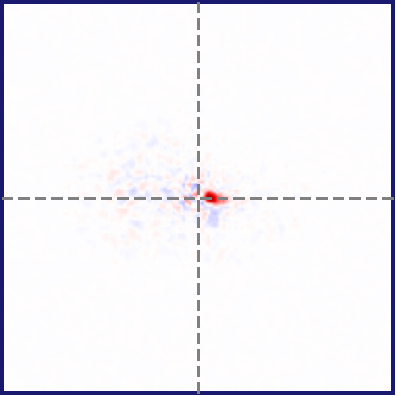} &
     \includegraphics[width=\f1ht]{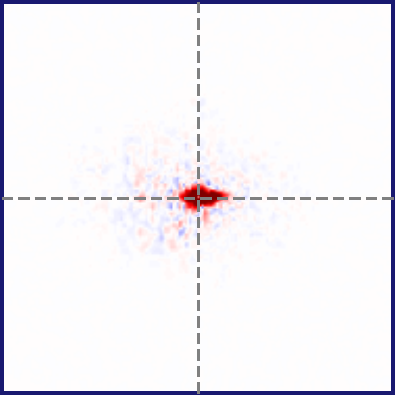} &
     \includegraphics[width=\f1ht]{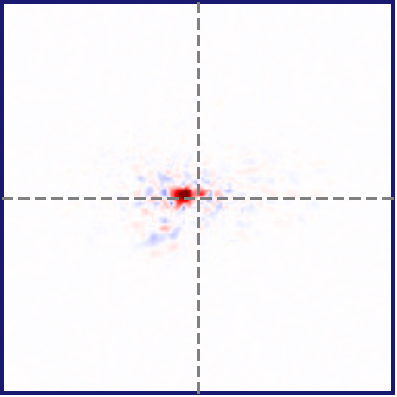} &
     \includegraphics[width=\f1ht]{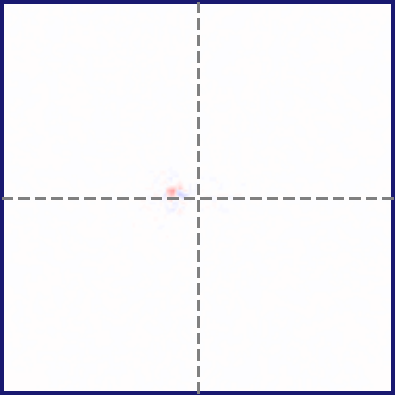} &
     \includegraphics[width=0.03\textwidth]{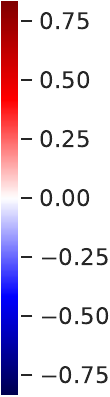} \\
     
     45 & \includegraphics[width=\f1ht]{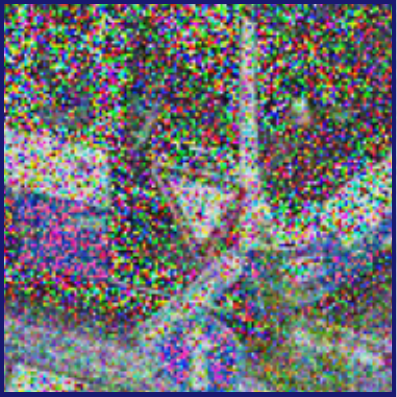} & 
     \includegraphics[width=\f1ht]{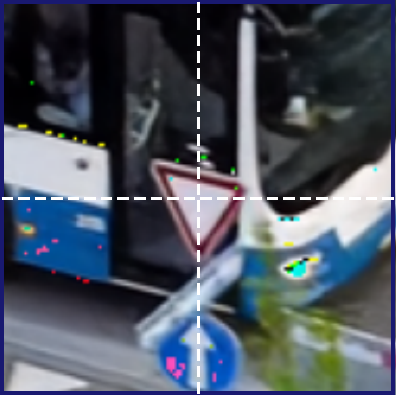} &
     \includegraphics[width=\f1ht]{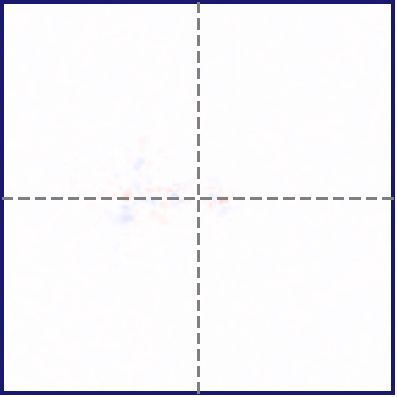} &
     \includegraphics[width=\f1ht]{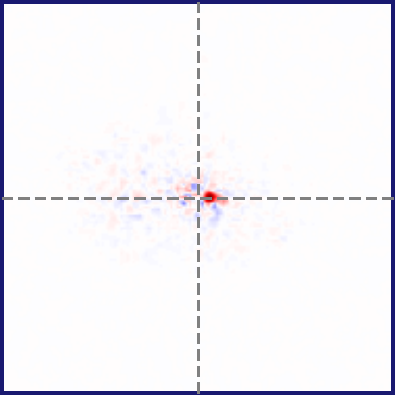} &
     \includegraphics[width=\f1ht]{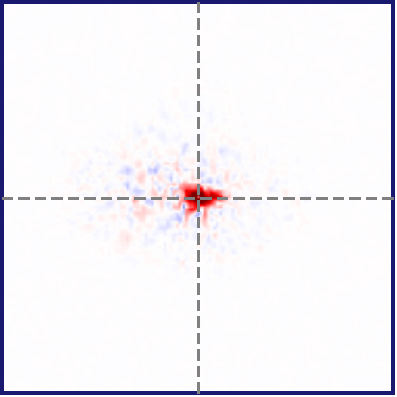} &
     \includegraphics[width=\f1ht]{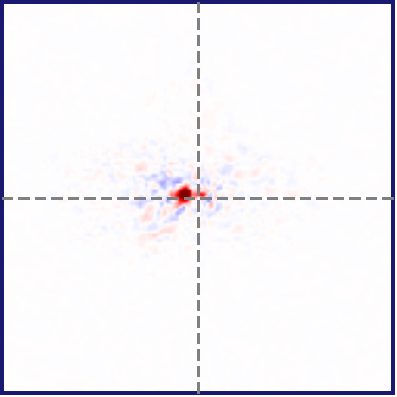} &
     \includegraphics[width=\f1ht]{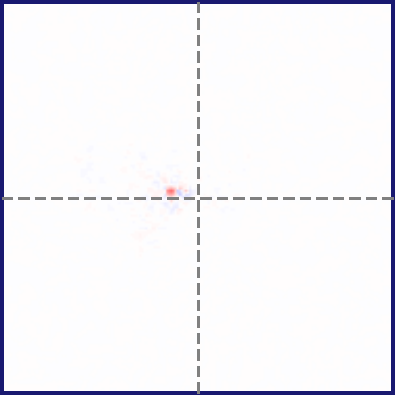} &
     \includegraphics[width=0.03\textwidth]{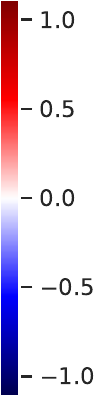} \\
     
     60 & \includegraphics[width=\f1ht]{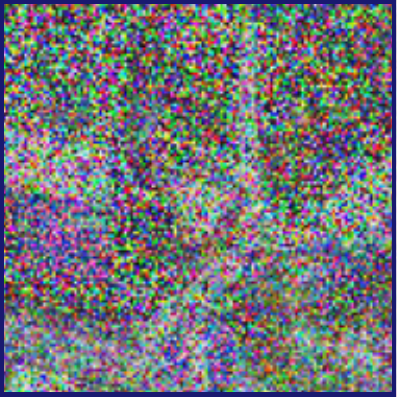} & 
     \includegraphics[width=\f1ht]{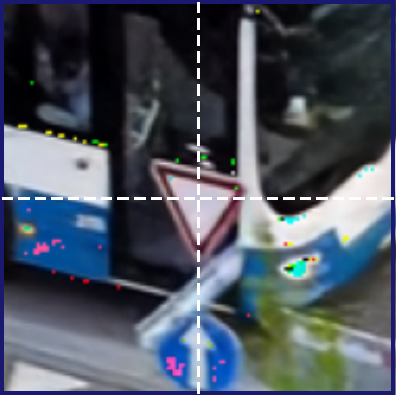} &
     \includegraphics[width=\f1ht]{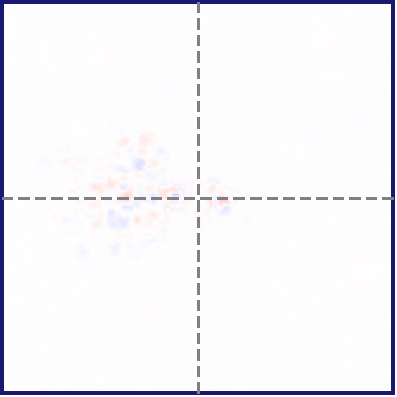} &
     \includegraphics[width=\f1ht]{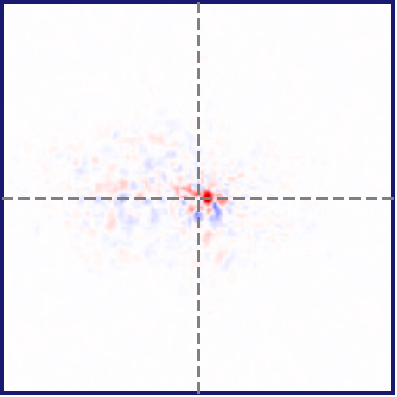} &
     \includegraphics[width=\f1ht]{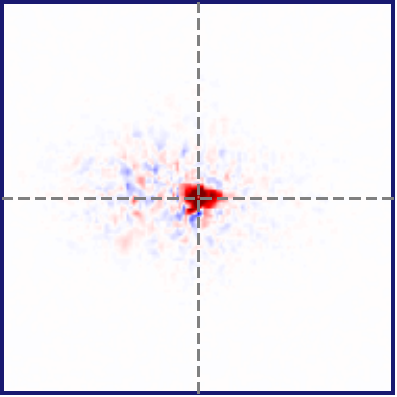} &
     \includegraphics[width=\f1ht]{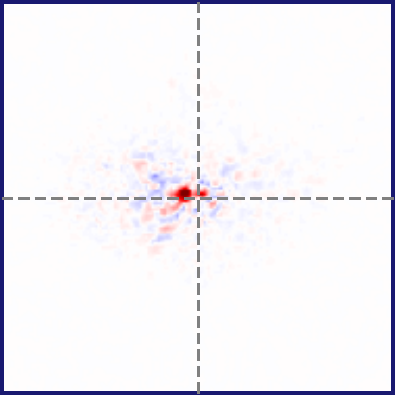} &
     \includegraphics[width=\f1ht]{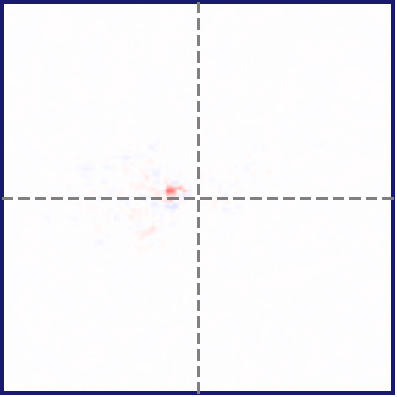} &
     \includegraphics[width=0.03\textwidth]{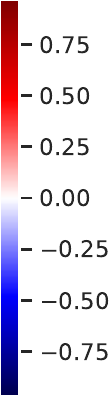} \\
     
     75 & \includegraphics[width=\f1ht]{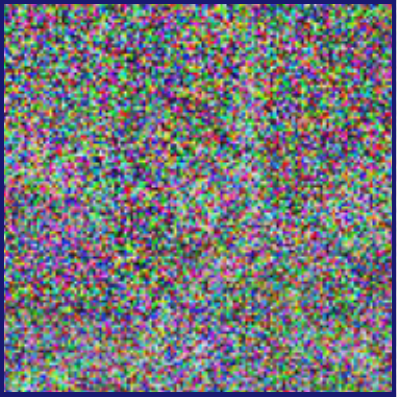} & 
     \includegraphics[width=\f1ht]{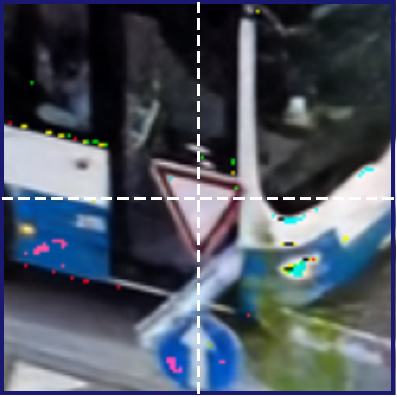} &
     \includegraphics[width=\f1ht]{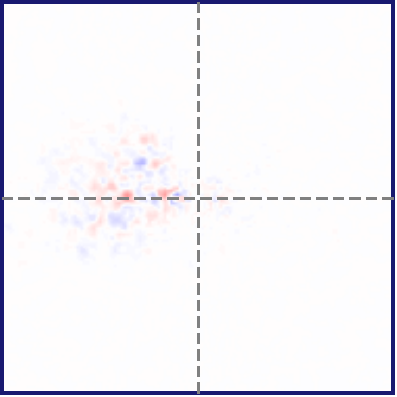} &
     \includegraphics[width=\f1ht]{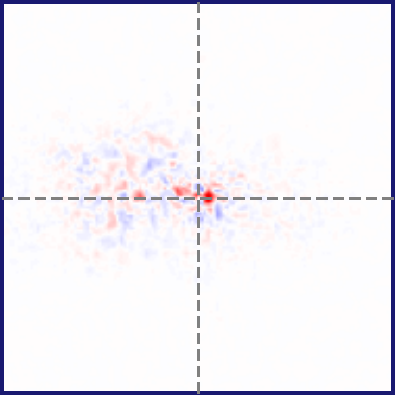} &
     \includegraphics[width=\f1ht]{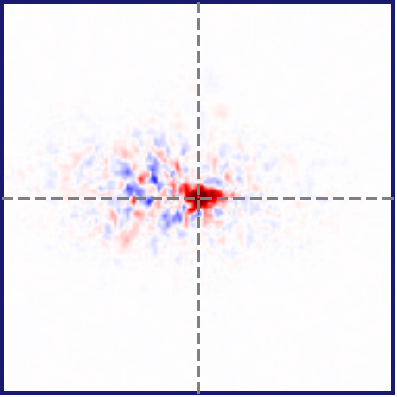} &
     \includegraphics[width=\f1ht]{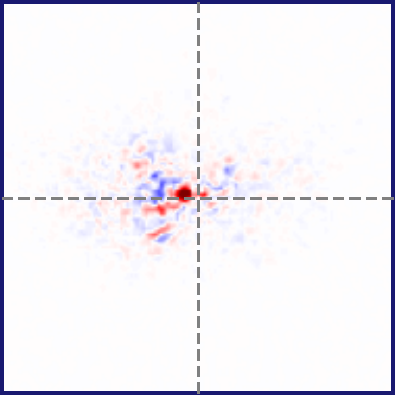} &
     \includegraphics[width=\f1ht]{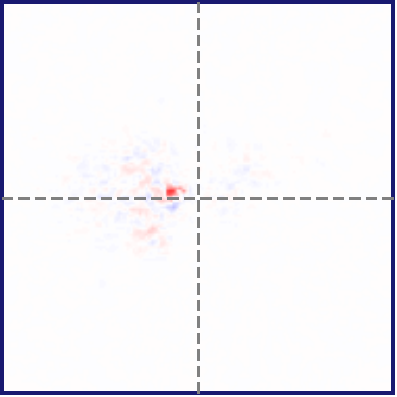} &
     \includegraphics[width=0.03\textwidth]{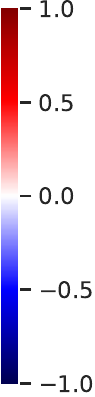} \\
     
     $\sigma$ & $y_t$ &  $d_t$ & $a(t-2, i)$ &  $a(t-1, i)$ & $a(t, i)$ & $a(t+1, i)$ & $a(t+2, i)$ & \\
     \end{tabular}
     
     \vspace{0.2cm}
     
\caption{\textbf{Video denoising as spatiotemporal adaptive filtering; \texttt{bus} video from the DAVIS dataset}. Visualization of the equivalent filters, as described in Fig~\ref{fig:jacobian_1}. }
\label{fig:jacobian_4}
\end{figure*}

\begin{figure*}[ht]
    \def\f1ht{\linewidth}%
     
     \centering 
     \begin{tabular}{ >{\centering\arraybackslash}m{0.02\linewidth}
     >{\centering\arraybackslash}m{0.11\linewidth}
     >{\centering\arraybackslash}m{0.11\linewidth}
     >{\centering\arraybackslash}m{0.11\linewidth}
     >{\centering\arraybackslash}m{0.11\linewidth}
     >{\centering\arraybackslash}m{0.11\linewidth}
     >{\centering\arraybackslash}m{0.11\linewidth}
     >{\centering\arraybackslash}m{0.11\linewidth}
     >{\centering\arraybackslash}m{0.03\linewidth}
     }
     \centering
     
     & & &
     \includegraphics[width=\f1ht]{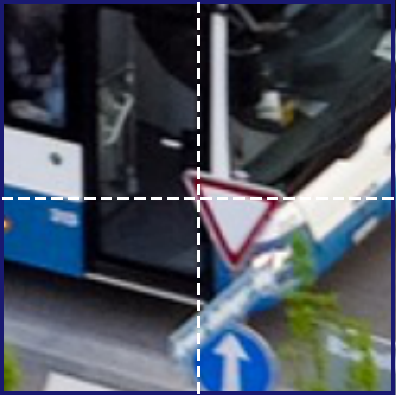} & 
     \includegraphics[width=\f1ht]{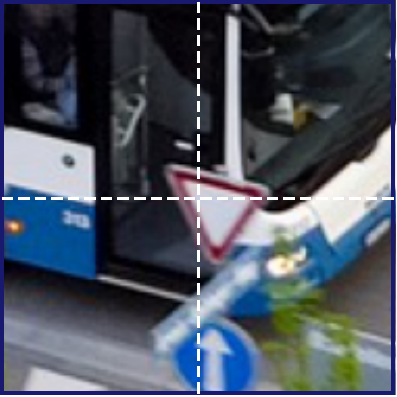} &
     \includegraphics[width=\f1ht]{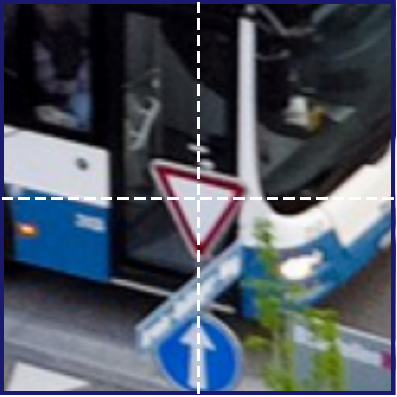} &
     \includegraphics[width=\f1ht]{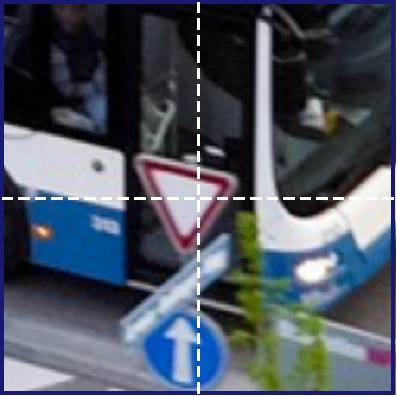} & 
     \includegraphics[width=\f1ht]{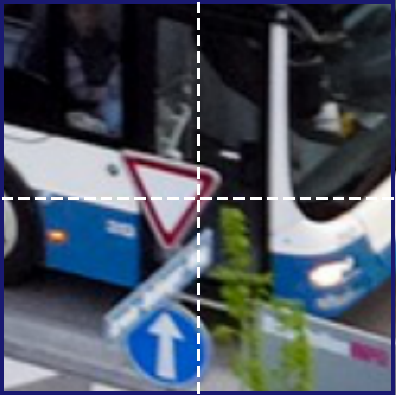} \\
     
     30 & \includegraphics[width=\f1ht]{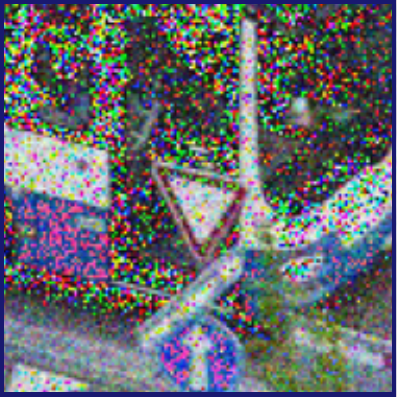} & 
     \includegraphics[width=\f1ht]{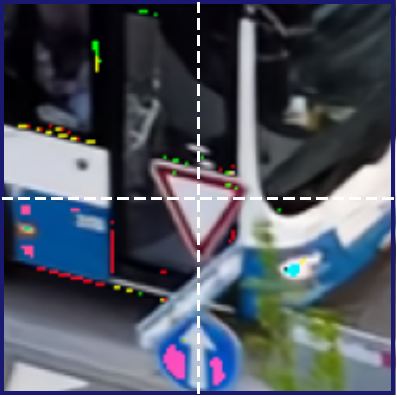} &
     \includegraphics[width=\f1ht]{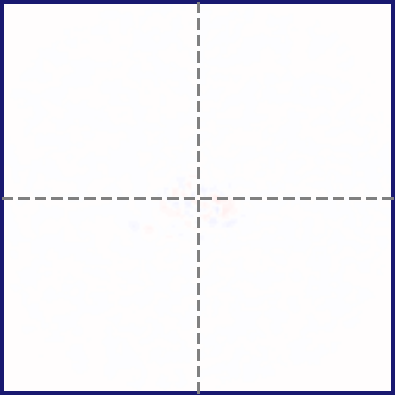} &
     \includegraphics[width=\f1ht]{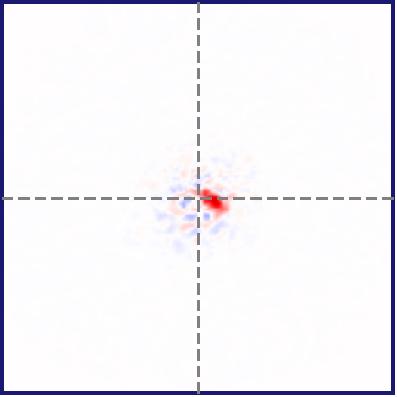} &
     \includegraphics[width=\f1ht]{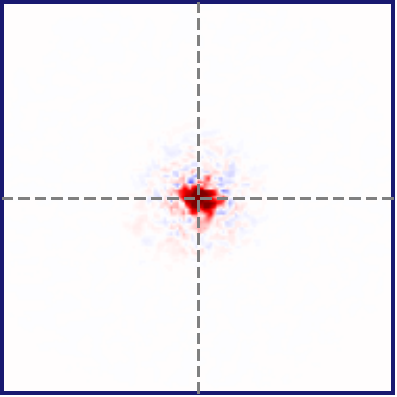} &
     \includegraphics[width=\f1ht]{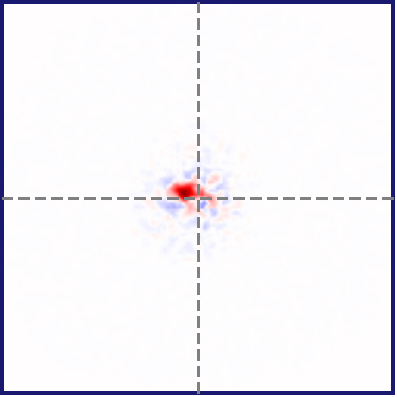} &
     \includegraphics[width=\f1ht]{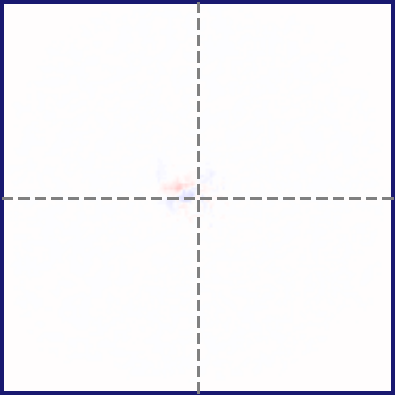} &
     \includegraphics[width=0.03\textwidth]{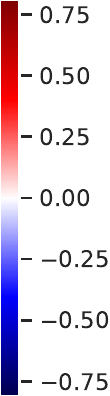} \\
     
     45 & \includegraphics[width=\f1ht]{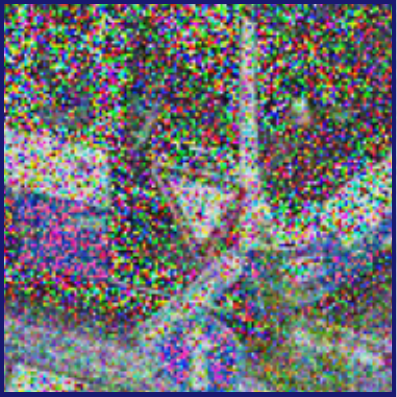} & 
     \includegraphics[width=\f1ht]{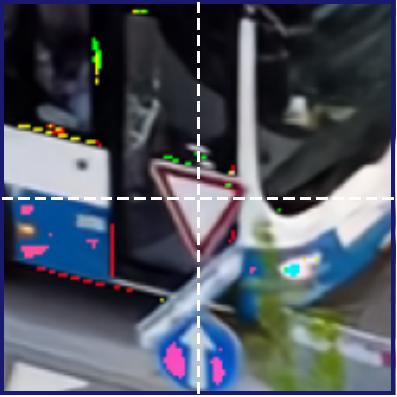} &
     \includegraphics[width=\f1ht]{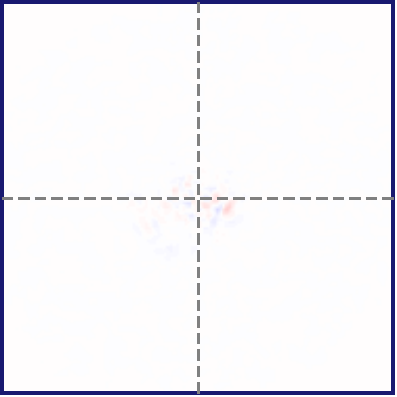} &
     \includegraphics[width=\f1ht]{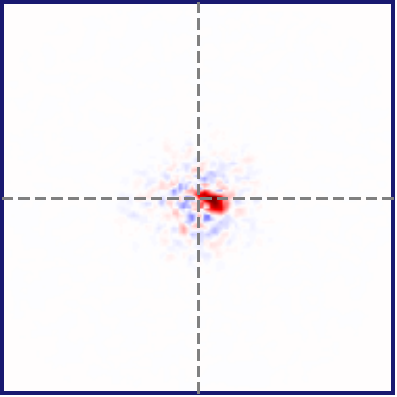} &
     \includegraphics[width=\f1ht]{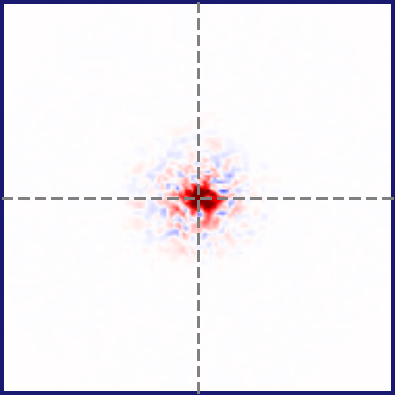} &
     \includegraphics[width=\f1ht]{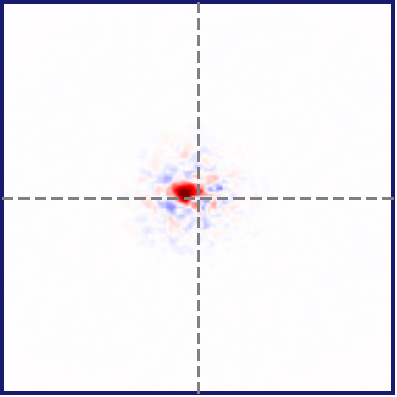} &
     \includegraphics[width=\f1ht]{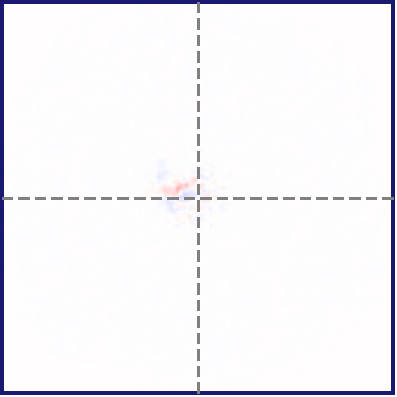} &
     \includegraphics[width=0.03\textwidth]{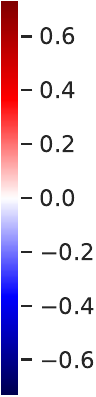} \\
     
     60 & \includegraphics[width=\f1ht]{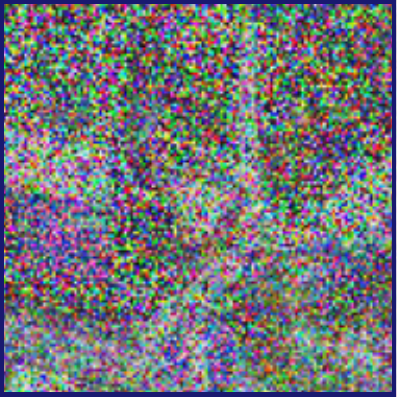} & 
     \includegraphics[width=\f1ht]{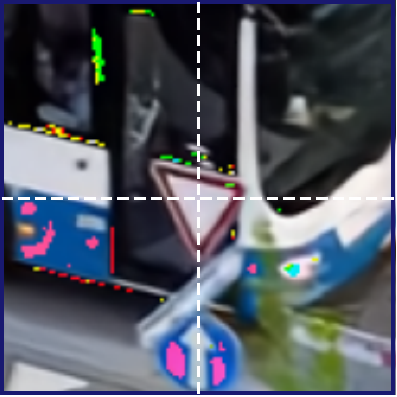} &
     \includegraphics[width=\f1ht]{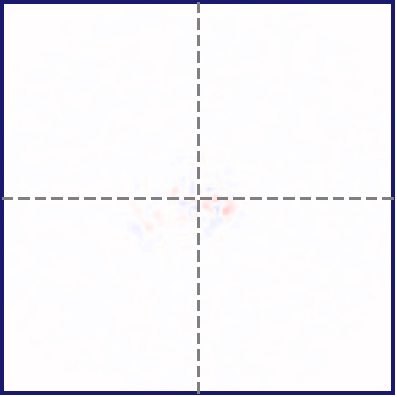} &
     \includegraphics[width=\f1ht]{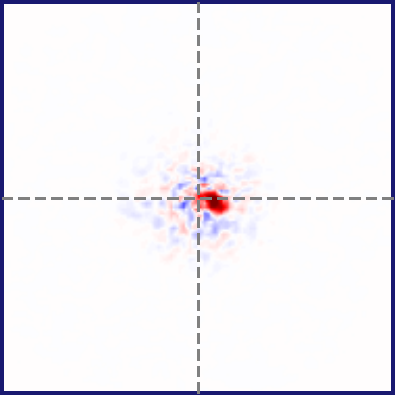} &
     \includegraphics[width=\f1ht]{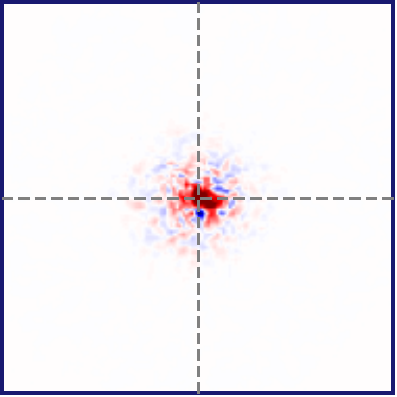} &
     \includegraphics[width=\f1ht]{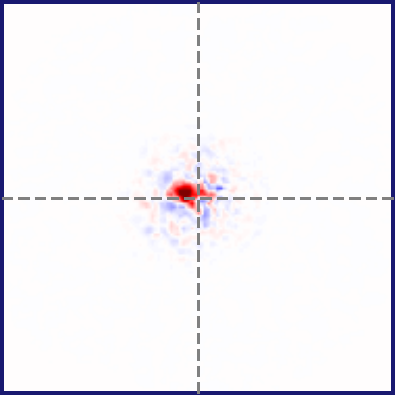} &
     \includegraphics[width=\f1ht]{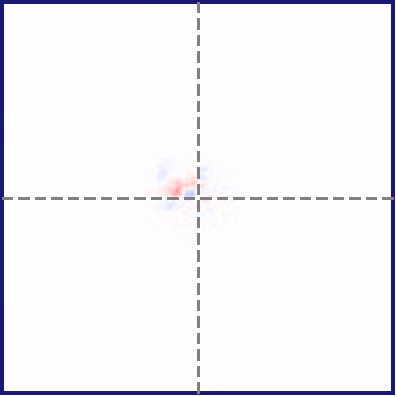} &
     \includegraphics[width=0.03\textwidth]{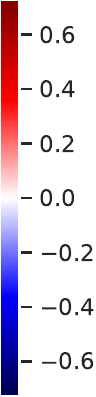} \\
     
     75 & \includegraphics[width=\f1ht]{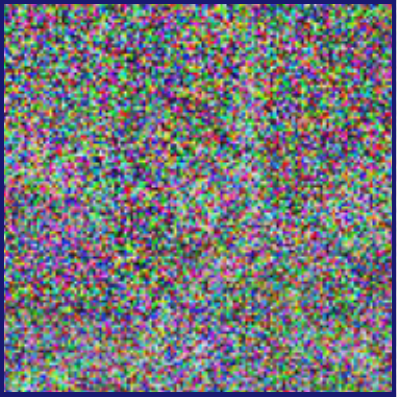} & 
     \includegraphics[width=\f1ht]{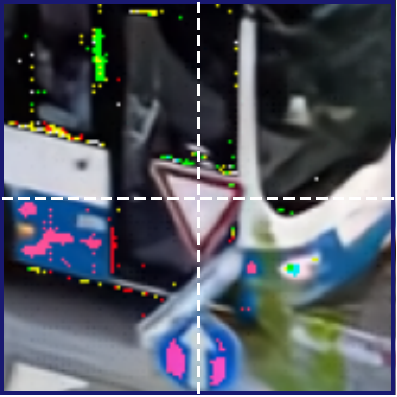} &
     \includegraphics[width=\f1ht]{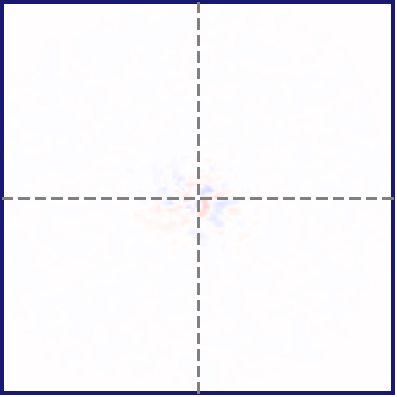} &
     \includegraphics[width=\f1ht]{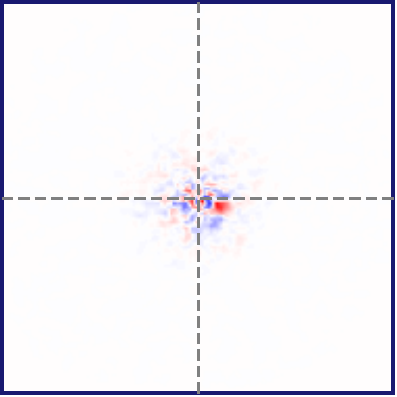} &
     \includegraphics[width=\f1ht]{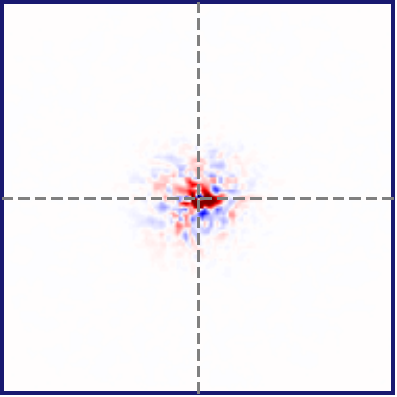} &
     \includegraphics[width=\f1ht]{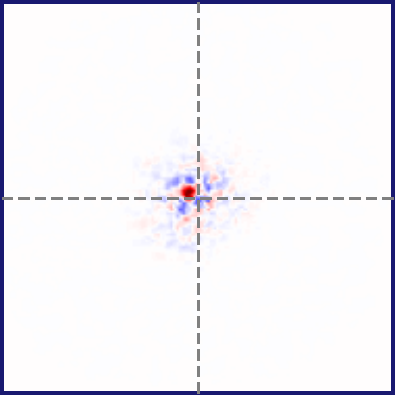} &
     \includegraphics[width=\f1ht]{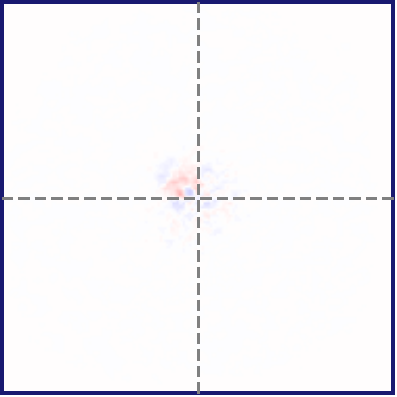} &
     \includegraphics[width=0.025\textwidth]{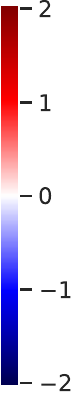} \\
     
     $\sigma$ & $y_t$ &  $d_t$ & $a(t-2, i)$ &  $a(t-1, i)$ & $a(t, i)$ & $a(t+1, i)$ & $a(t+2, i)$ & \\
     \end{tabular}
     
     \vspace{0.2cm}
     
\caption{ \textbf{Video denoising using FastDVDnet as spatiotemporal adaptive filtering; \texttt{bus} video from the DAVIS dataset}. Visualization of the linear weighting functions ($a(k,i)$, Section~6 of paper) of FastDVDnet which is trained with supervision. The left two columns show the noisy frame $y_t$ at four levels of noise, and the corresponding denoised frame, $d_t$.  Weighting functions $a(k, i)$ corresponding to the pixel $i$ (at the intersection of the dashed white lines), for five successive frames, are shown in the last five columns. The weighting functions adapt to underlying image content, and are shifted to track the motion of the stop sign. As the noise level $\sigma$ increases, their spatial extent grows, averaging out more of the noise while respecting object boundaries. The behavior is very similar to the corresponding filters of UDVD as shown in Fig~\ref{fig:jacobian_4}.}
\label{fig:jacobian_fastfast}
\end{figure*}

\begin{figure*}[ht]
    \def\f1ht{\linewidth}%
     
     \centering 
     \begin{tabular}{ >{\centering\arraybackslash}m{0.12\linewidth}
     >{\centering\arraybackslash}m{0.12\linewidth}
     >{\centering\arraybackslash}m{0.12\linewidth}
     >{\centering\arraybackslash}m{0.12\linewidth}
     >{\centering\arraybackslash}m{0.12\linewidth}
     >{\centering\arraybackslash}m{0.12\linewidth}
     >{\centering\arraybackslash}m{0.12\linewidth}
     >{\centering\arraybackslash}m{0.03\linewidth}
     }
     \centering
     
     \includegraphics[width=\f1ht]{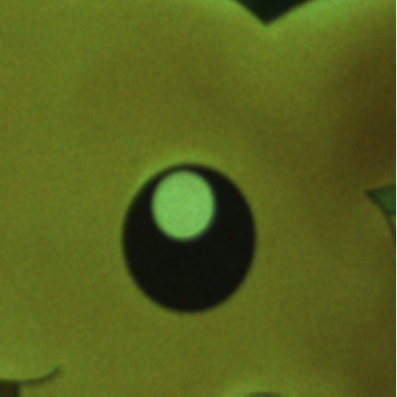} & 
     \includegraphics[width=\f1ht]{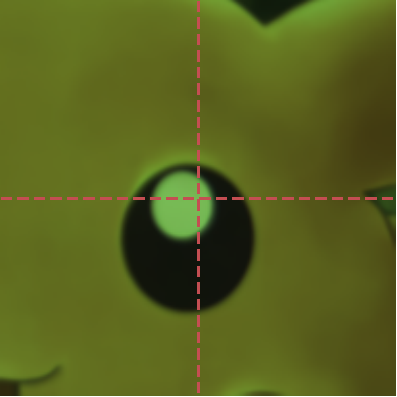} &
     \includegraphics[width=\f1ht]{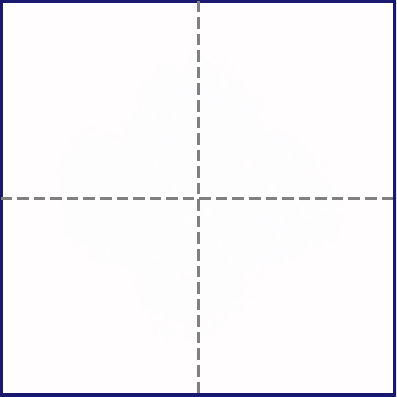} &
     \includegraphics[width=\f1ht]{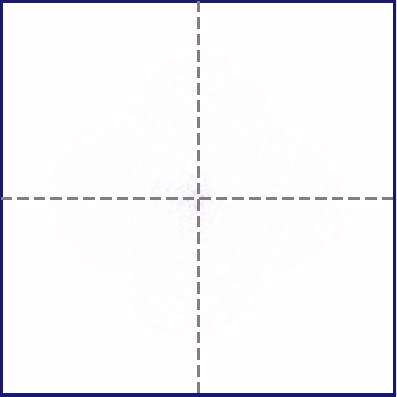} &
     \includegraphics[width=\f1ht]{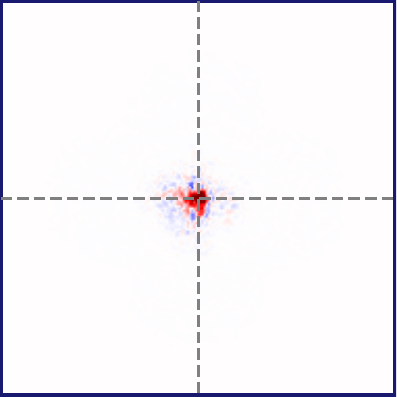} &
     \includegraphics[width=\f1ht]{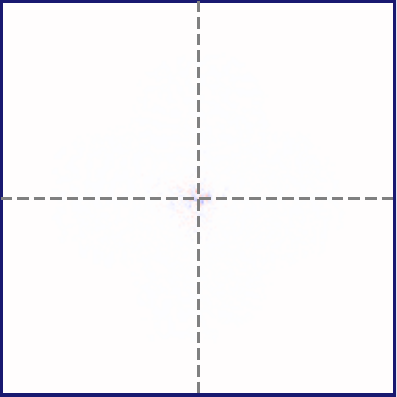} &
     \includegraphics[width=\f1ht]{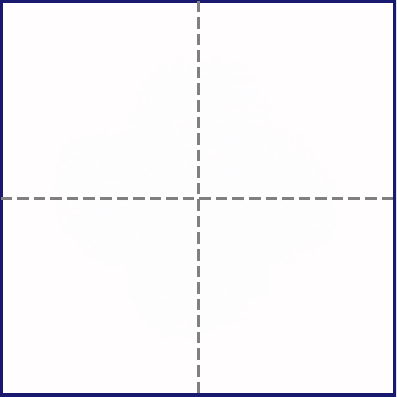} &
     \includegraphics[width=0.022\textwidth]{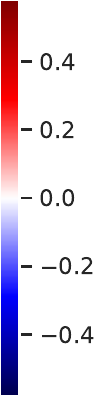} \\
     
     \includegraphics[width=\f1ht]{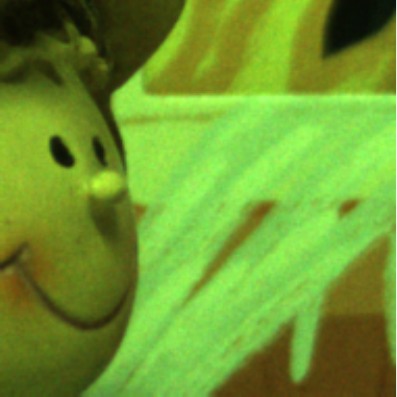} & 
     \includegraphics[width=\f1ht]{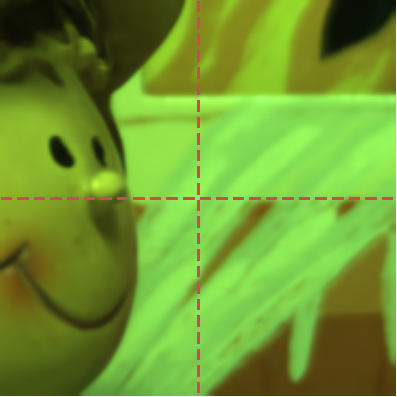} &
     \includegraphics[width=\f1ht]{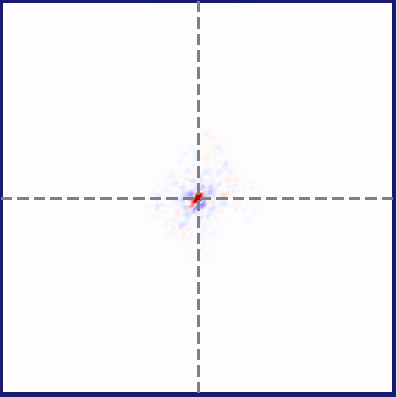} &
     \includegraphics[width=\f1ht]{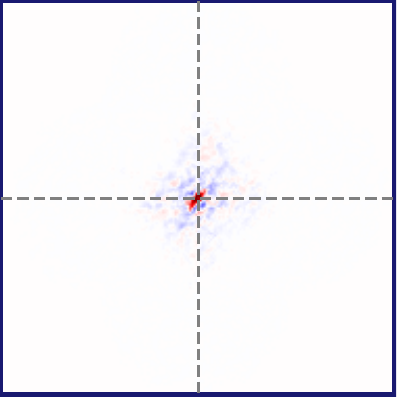} &
     \includegraphics[width=\f1ht]{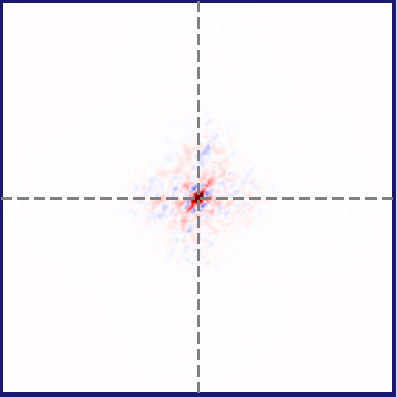} &
     \includegraphics[width=\f1ht]{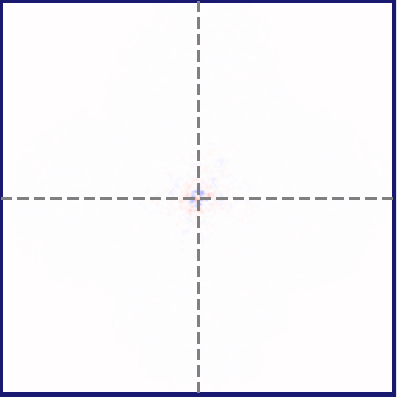} &
     \includegraphics[width=\f1ht]{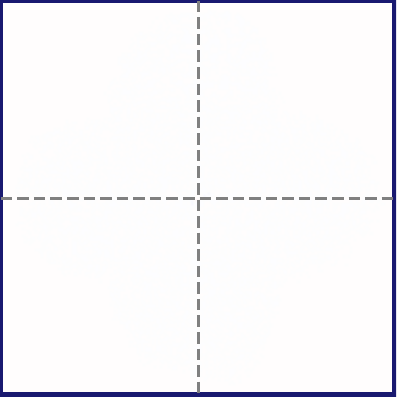} &
     \includegraphics[width=0.022\textwidth]{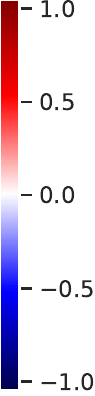} \\
     
     \includegraphics[width=\f1ht]{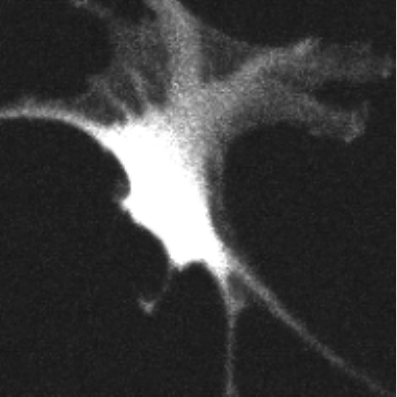} & 
     \includegraphics[width=\f1ht]{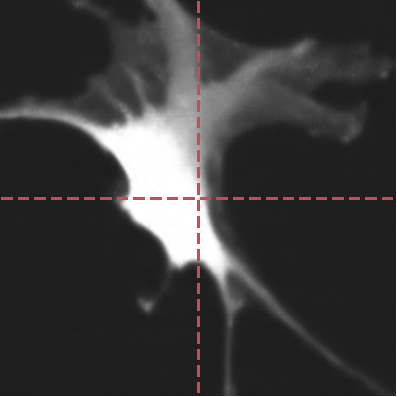} &
     \includegraphics[width=\f1ht]{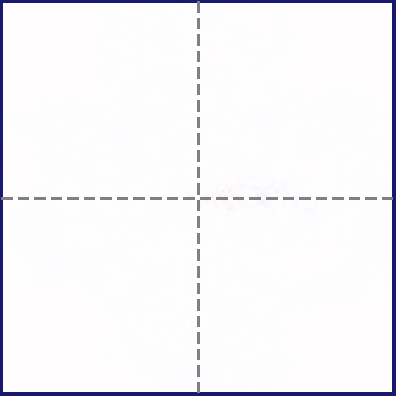} &
     \includegraphics[width=\f1ht]{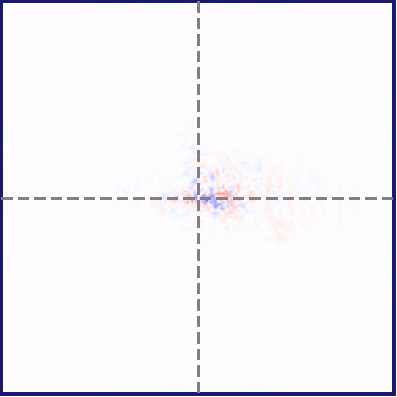} &
     \includegraphics[width=\f1ht]{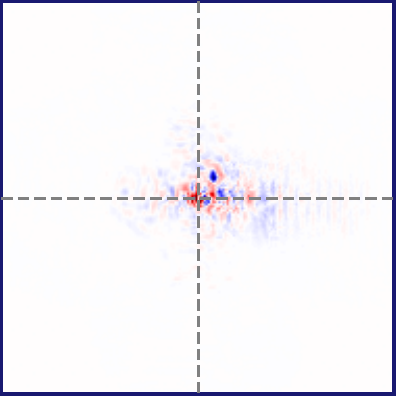} &
     \includegraphics[width=\f1ht]{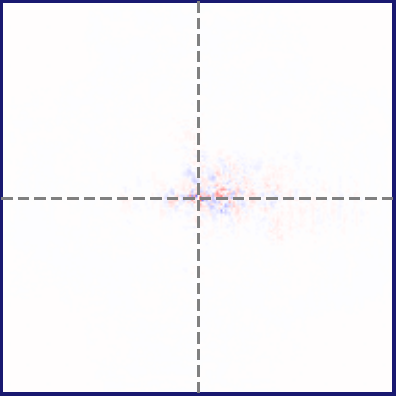} &
     \includegraphics[width=\f1ht]{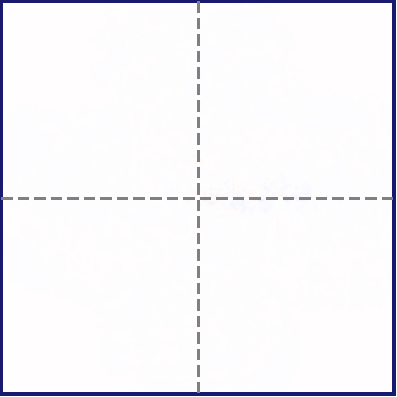} &
     \includegraphics[width=0.022\textwidth]{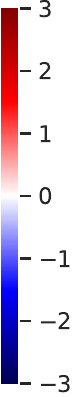} \\
     
     \includegraphics[width=\f1ht]{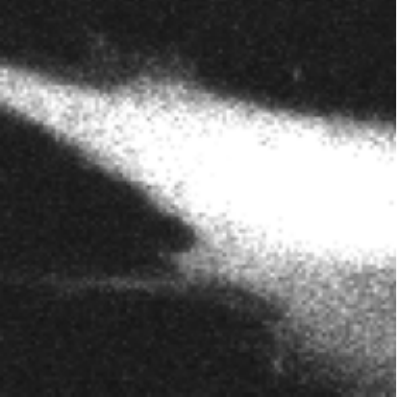} & 
     \includegraphics[width=\f1ht]{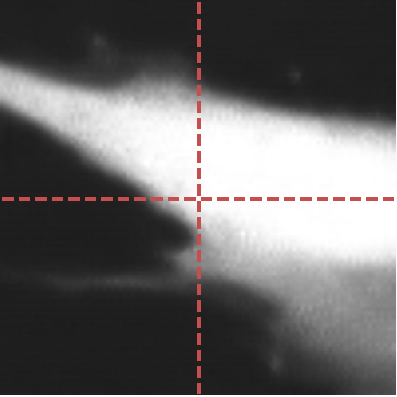} &
     \includegraphics[width=\f1ht]{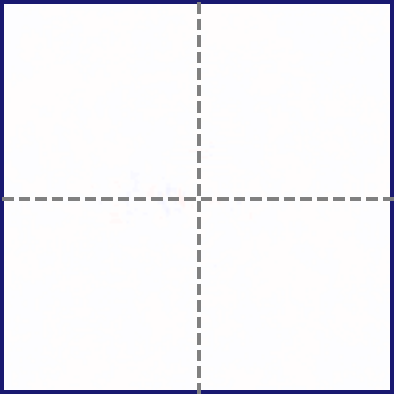} &
     \includegraphics[width=\f1ht]{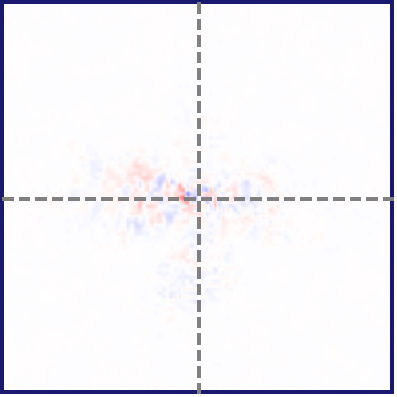} &
     \includegraphics[width=\f1ht]{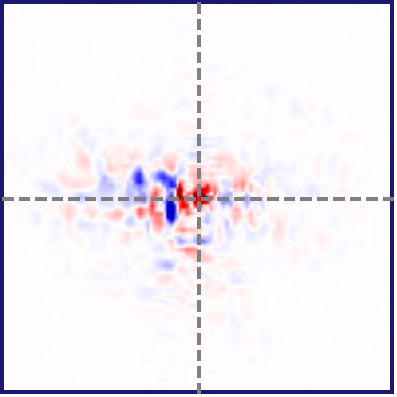} &
     \includegraphics[width=\f1ht]{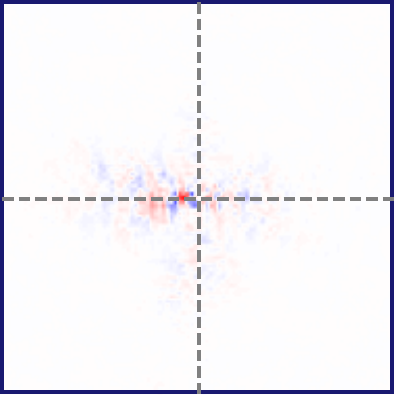} &
     \includegraphics[width=\f1ht]{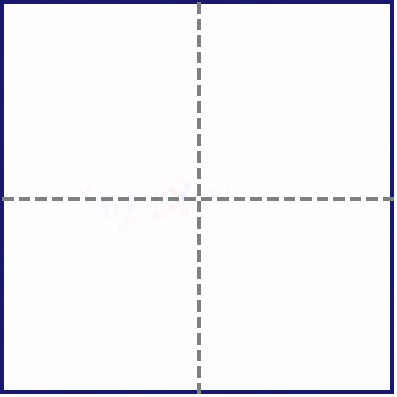} &
     \includegraphics[width=0.022\textwidth]{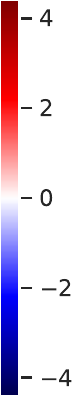} \\
     
     \includegraphics[width=\f1ht]{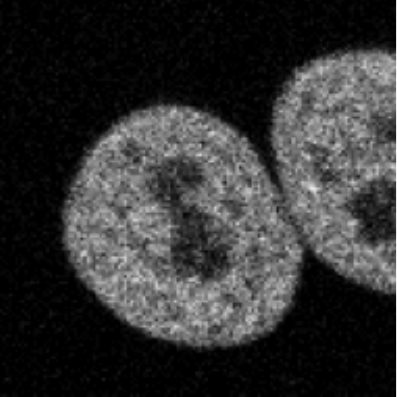} & 
     \includegraphics[width=\f1ht]{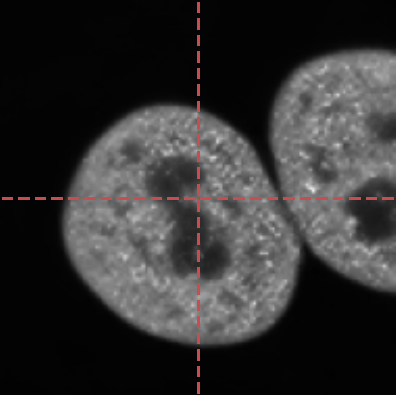} &
     \includegraphics[width=\f1ht]{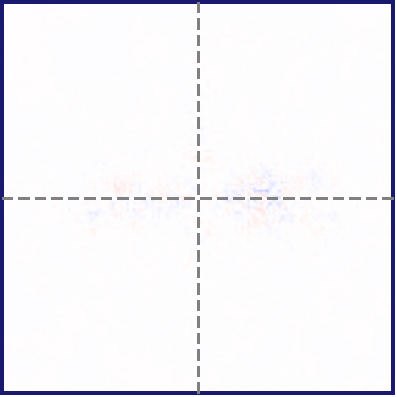} &
     \includegraphics[width=\f1ht]{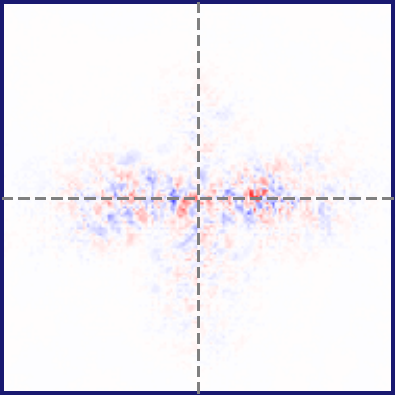} &
     \includegraphics[width=\f1ht]{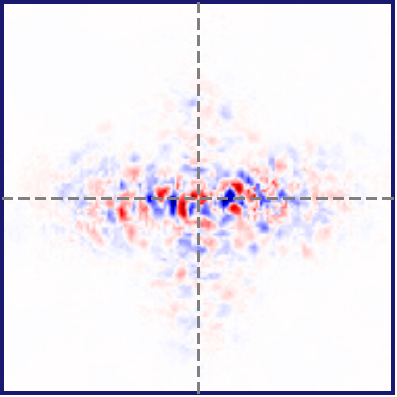} &
     \includegraphics[width=\f1ht]{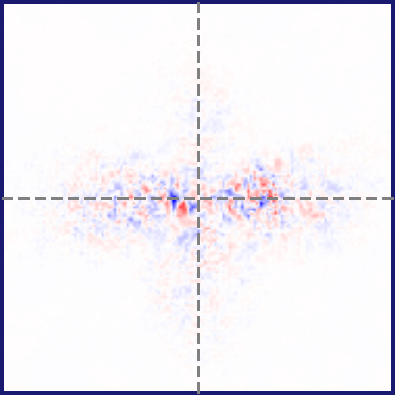} &
     \includegraphics[width=\f1ht]{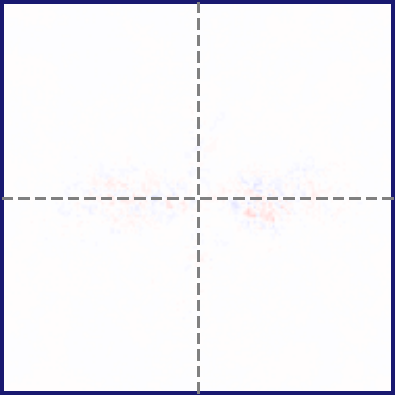} &
     \includegraphics[width=0.022\textwidth]{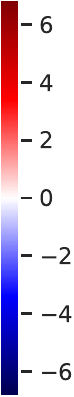} \\
     
     \includegraphics[width=\f1ht]{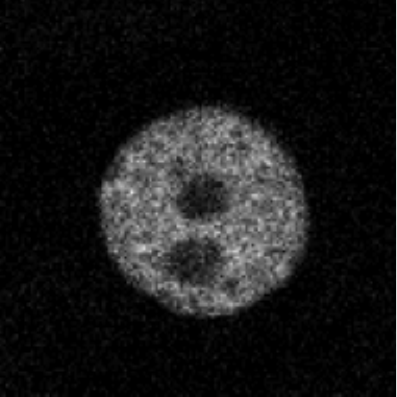} & 
     \includegraphics[width=\f1ht]{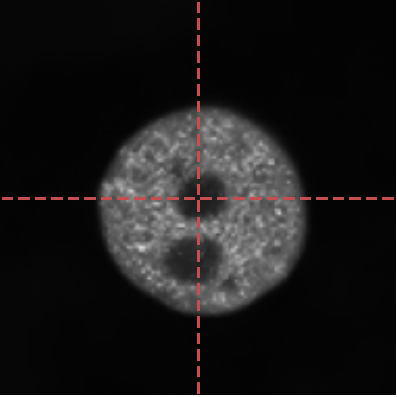} &
     \includegraphics[width=\f1ht]{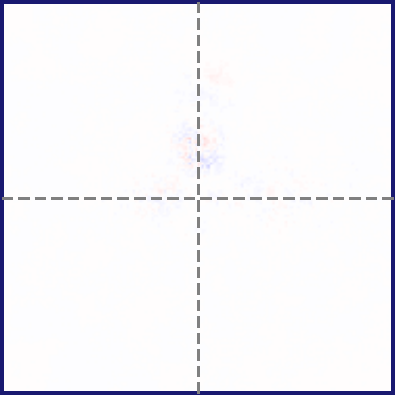} &
     \includegraphics[width=\f1ht]{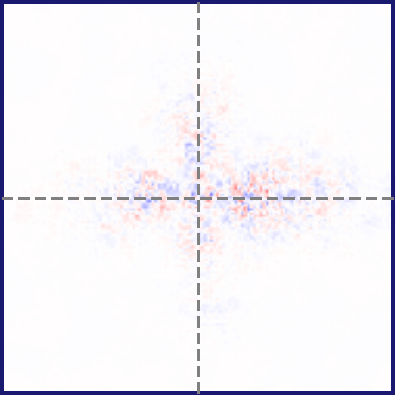} &
     \includegraphics[width=\f1ht]{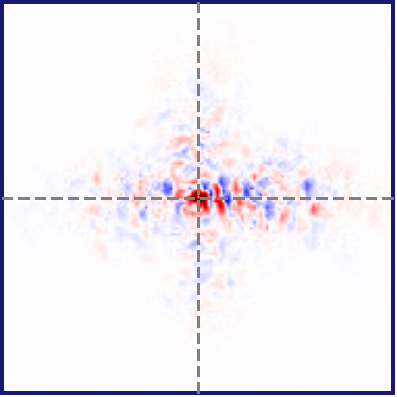} &
     \includegraphics[width=\f1ht]{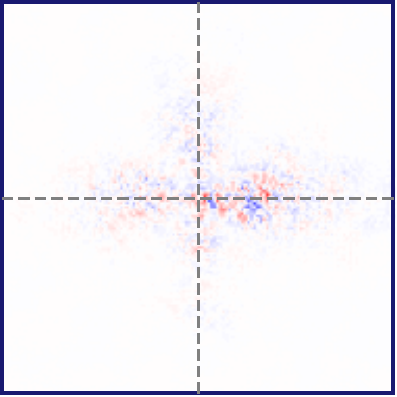} &
     \includegraphics[width=\f1ht]{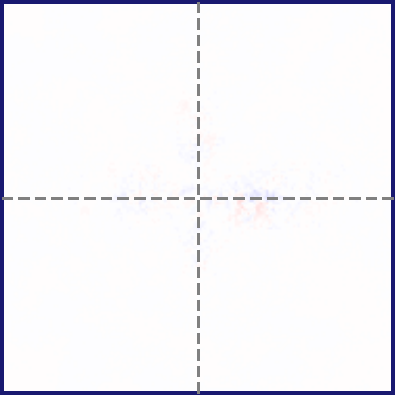} &
     \includegraphics[width=0.022\textwidth]{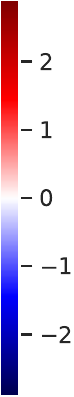} \\
     
     \includegraphics[width=\f1ht]{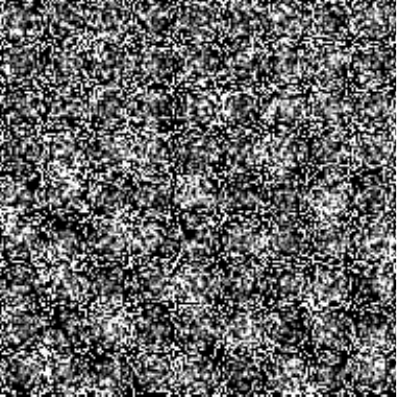} & 
     \includegraphics[width=\f1ht]{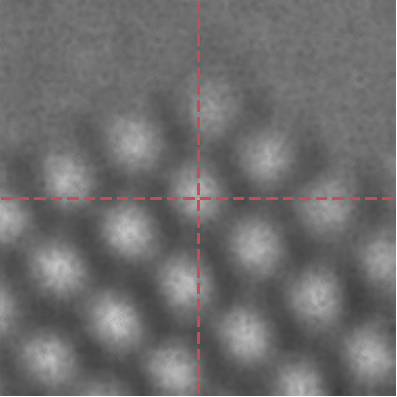} &
     \includegraphics[width=\f1ht]{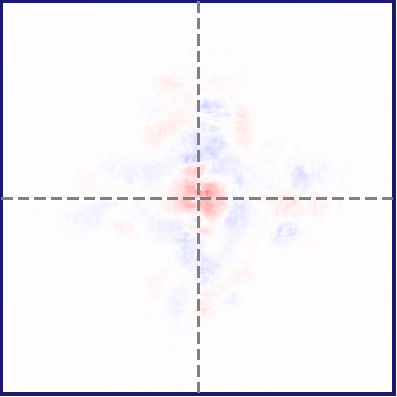} &
     \includegraphics[width=\f1ht]{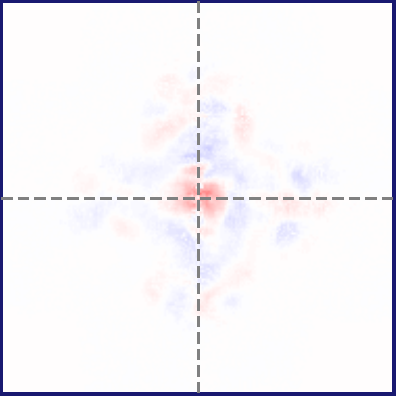} &
     \includegraphics[width=\f1ht]{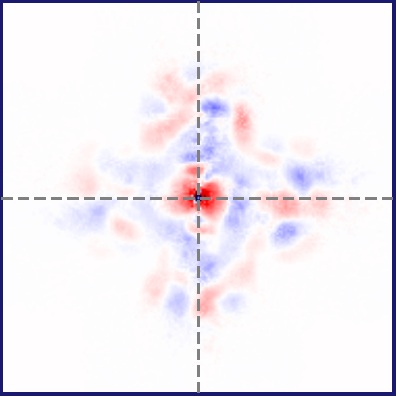} &
     \includegraphics[width=\f1ht]{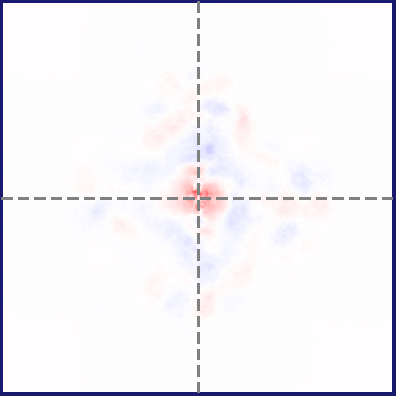} &
     \includegraphics[width=\f1ht]{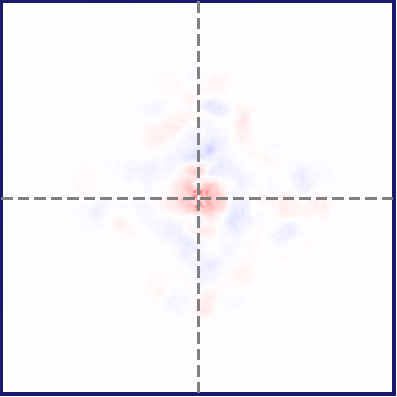} &
     \includegraphics[width=0.03\textwidth]{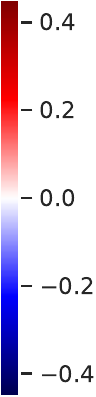} \\
     
     \includegraphics[width=\f1ht]{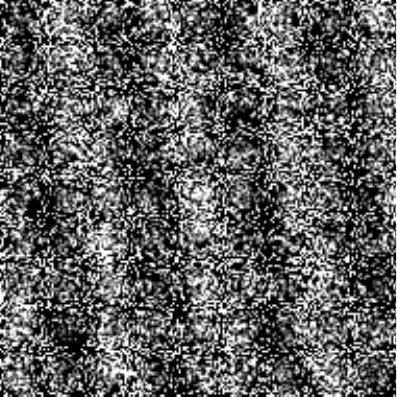} & 
     \includegraphics[width=\f1ht]{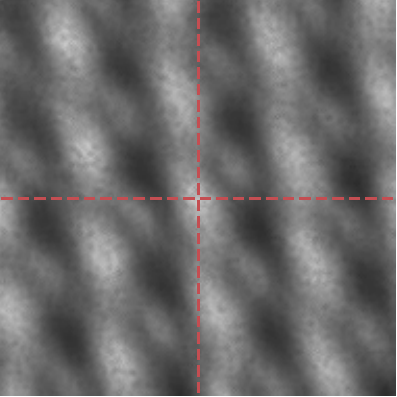} &
     \includegraphics[width=\f1ht]{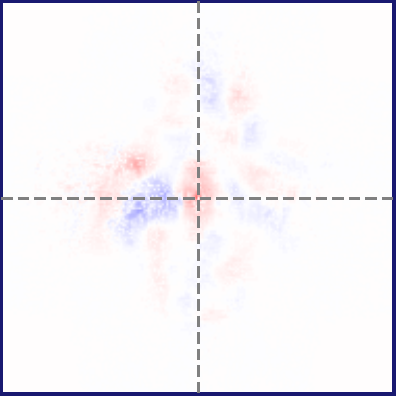} &
     \includegraphics[width=\f1ht]{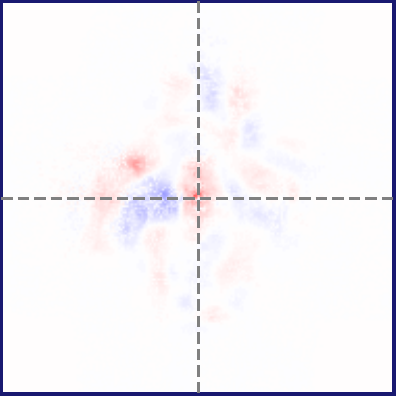} &
     \includegraphics[width=\f1ht]{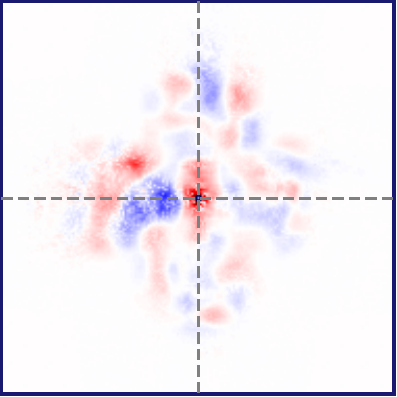} &
     \includegraphics[width=\f1ht]{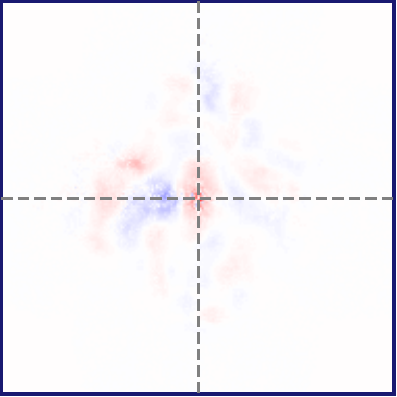} &
     \includegraphics[width=\f1ht]{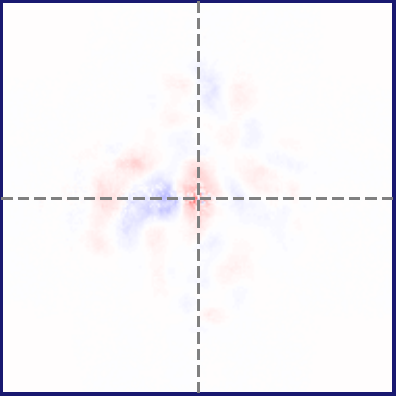} &
     \includegraphics[width=0.03\textwidth]{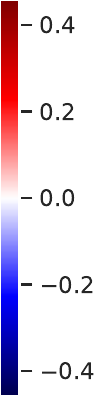} \\
     
     $y_t$ &  $d_t$ & $a(t-2, i)$ &  $a(t-1, i)$ & $a(t, i)$ & $a(t+1, i)$ & $a(t+2, i)$ & \\
     \end{tabular}
     
     \vspace{0.2cm}
     
\caption{ \textbf{Equivalent filters of UDVD when applied to real-world data}. Visualization of the linear weighting functions ($a(k,i)$, Section~6 of paper) of UDVD trained to denoise raw video, fluorescence and electron microscopy data. The left two columns show the noisy frame $y_t$ and the corresponding denoised frame, $d_t$.  Weighting functions $a(k, i)$ corresponding to the pixel $i$ (at the intersection of the dashed white lines), for five successive frames, are shown in the last five columns. In raw video data and fluorescence-microscopy data, the contributions from neighbouring frames are smaller. For electron-microscopy data they are larger (see also Fig~\ref{fig:filter_sum}). }
\label{fig:jacobian_micro}
\end{figure*}

\begin{figure*}[ht]
    \def\f1ht{\linewidth}%
     
     \centering 
     \begin{tabular}{ >{\centering\arraybackslash}m{0.02\linewidth}
     >{\centering\arraybackslash}m{0.23\linewidth}
     >{\centering\arraybackslash}m{0.23\linewidth}
     >{\centering\arraybackslash}m{0.23\linewidth}
     >{\centering\arraybackslash}m{0.23\linewidth}
     }
     \centering
     
     $\sigma$ &
     \footnotesize{(a) Noisy Frame} &
     \footnotesize{(b) DeepFlow on Clean Frame} &
     \footnotesize{(c) FastDVDnet} &
     \footnotesize{(d) Ours} \\
     
     30 & \includegraphics[width=\f1ht]{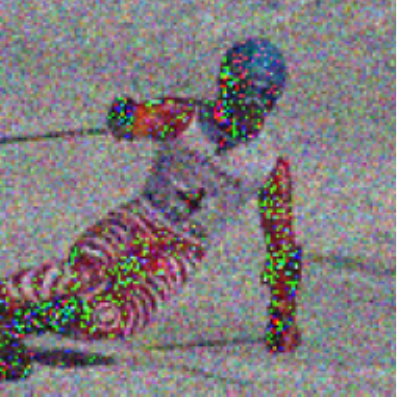} &
     \includegraphics[width=\f1ht]{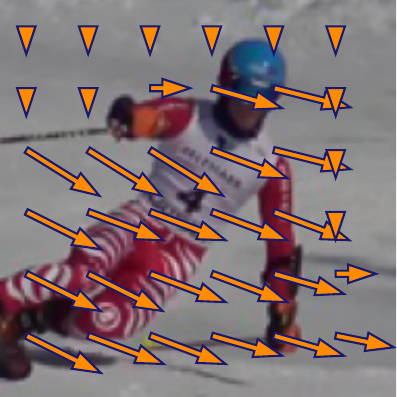} &
     \includegraphics[width=\f1ht]{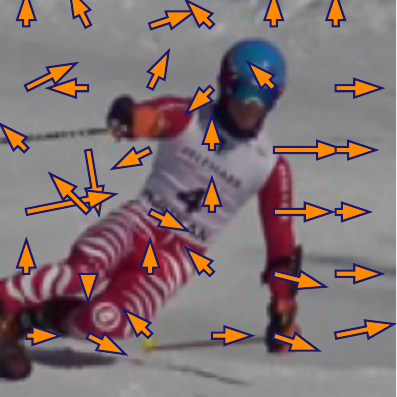} &
     \includegraphics[width=\f1ht]{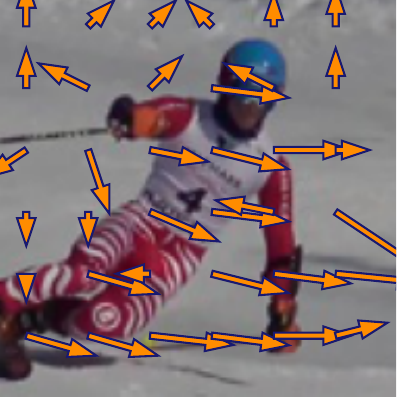} \\
     
     45 & \includegraphics[width=\f1ht]{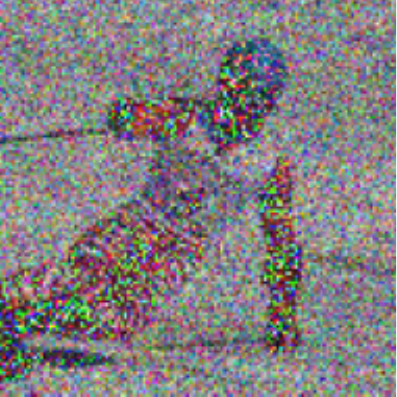} &
     \includegraphics[width=\f1ht]{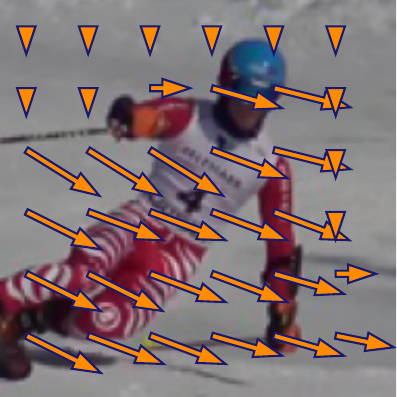} &
     \includegraphics[width=\f1ht]{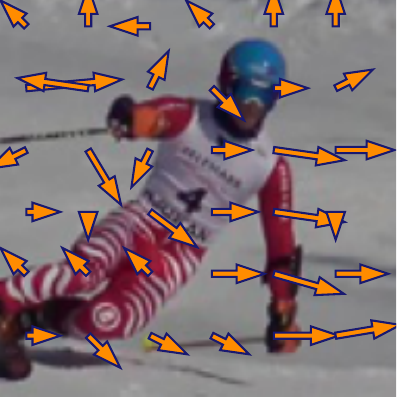} &
     \includegraphics[width=\f1ht]{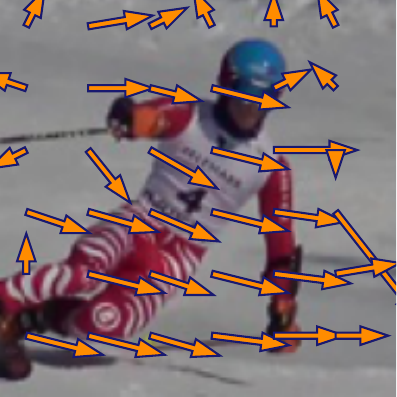} \\
     
     60 & \includegraphics[width=\f1ht]{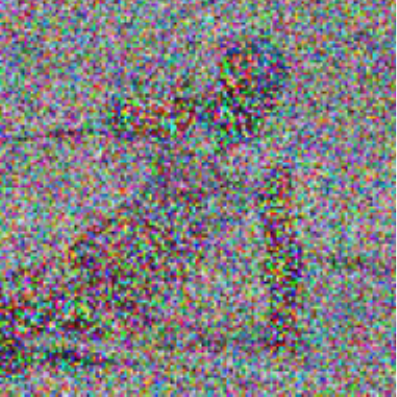} &
     \includegraphics[width=\f1ht]{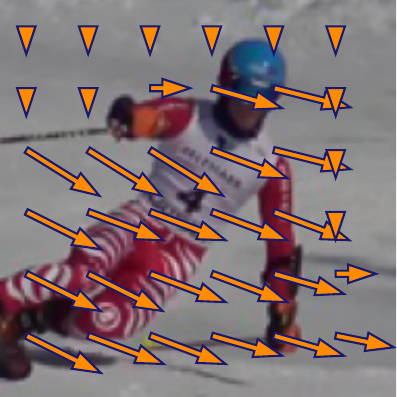} &
     \includegraphics[width=\f1ht]{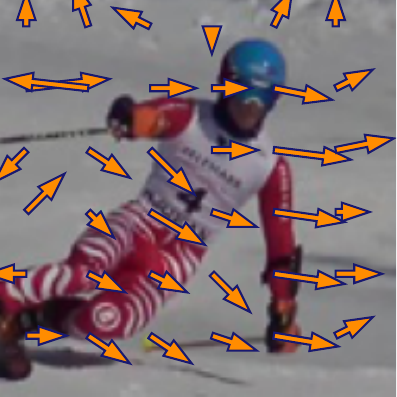} &
     \includegraphics[width=\f1ht]{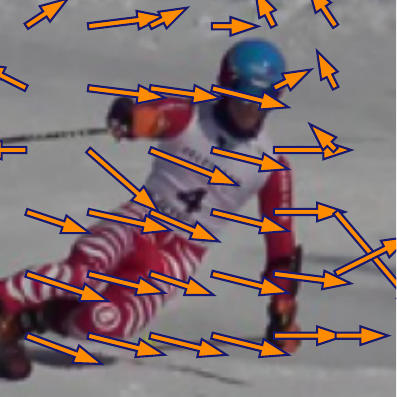} \\
     
     75 & \includegraphics[width=\f1ht]{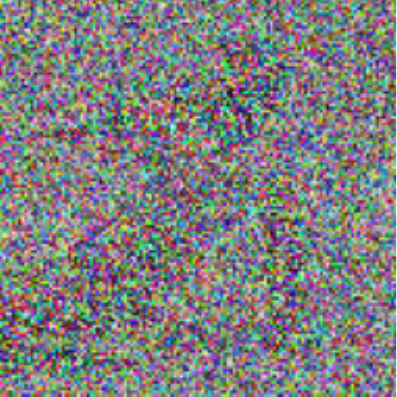} &
     \includegraphics[width=\f1ht]{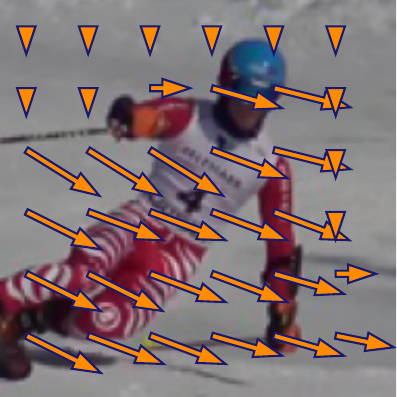} &
     \includegraphics[width=\f1ht]{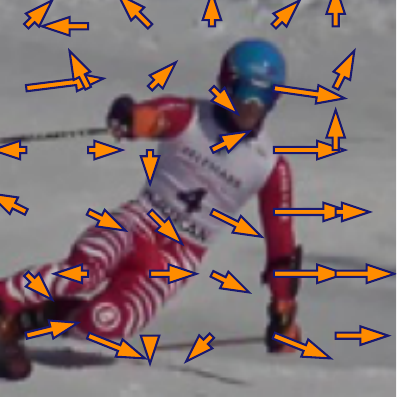} &
     \includegraphics[width=\f1ht]{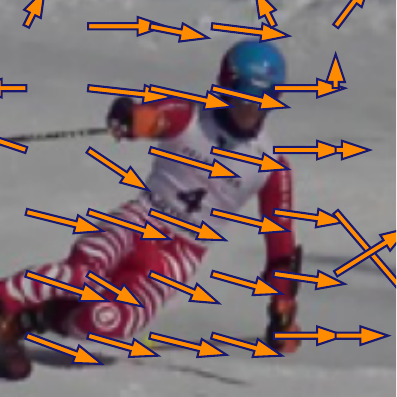} \\
     
     \end{tabular}
     
     \vspace{0.2cm}
     
\caption{\textbf{CNNs trained for denoising automatically learn to perform motion estimation}. (a) Noisy frame from \texttt{giant-slalom} video in the DAVIS dataset. (b) Optical flow direction at multiple locations of the image obtained using a state-of-the-art algorithm applied \emph{to the clean video}. Optical flow direction estimated from the shift of the adaptive filter obtained from the gradients of (c) FastDVDnet and (d) UDVD, both of which are trained with no optical flow information. FastDVDnet is trained with supervision. Optical flow estimates are well-matched to those in (b), but are not as accurated at oriented features, and in homogeneous regions where local motion is not well defined (e.g. in the background). Each row corresponds to a different noise levels. At higher noise levels, the networks perform averages over more frames, improving the motion estimation results. }
\label{fig:motion_1}
\end{figure*}

\begin{figure*}[ht]
    \def\f1ht{\linewidth}%
     
     \centering 
     \begin{tabular}{ >{\centering\arraybackslash}m{0.02\linewidth}
     >{\centering\arraybackslash}m{0.23\linewidth}
     >{\centering\arraybackslash}m{0.23\linewidth}
     >{\centering\arraybackslash}m{0.23\linewidth}
     >{\centering\arraybackslash}m{0.23\linewidth}
     }
     \centering
     
     $\sigma$ &
     \footnotesize{(a) Noisy Frame} &
     \footnotesize{(b) DeepFlow on Clean Frame} &
     \footnotesize{(c) FastDVDnet} &
     \footnotesize{(d) Ours} \\
     
     30 & \includegraphics[width=\f1ht]{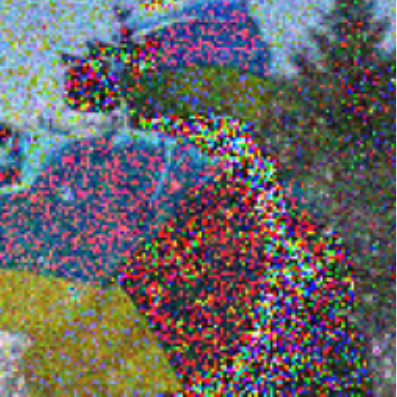} &
     \includegraphics[width=\f1ht]{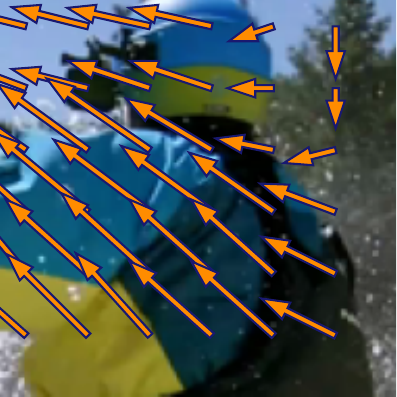} &
     \includegraphics[width=\f1ht]{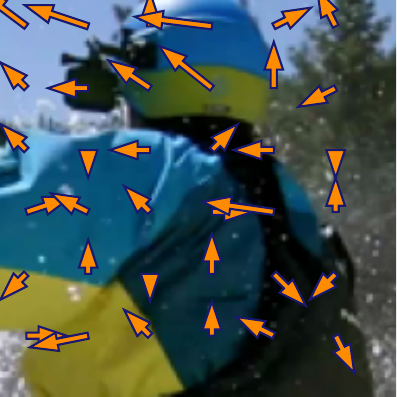} &
     \includegraphics[width=\f1ht]{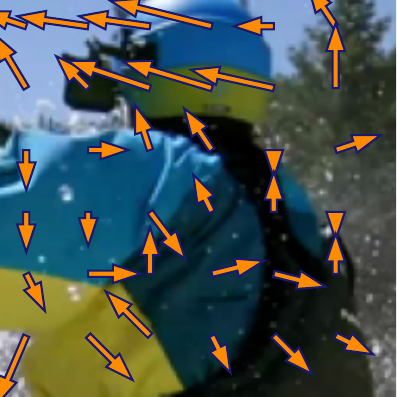} \\
     
     45 & \includegraphics[width=\f1ht]{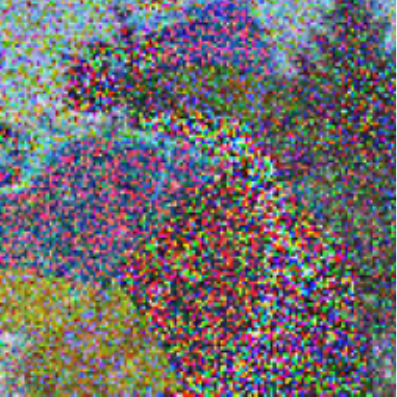} &
     \includegraphics[width=\f1ht]{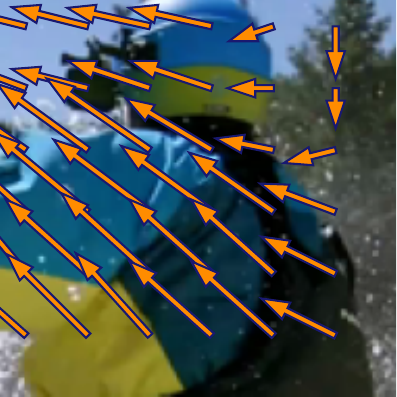} &
     \includegraphics[width=\f1ht]{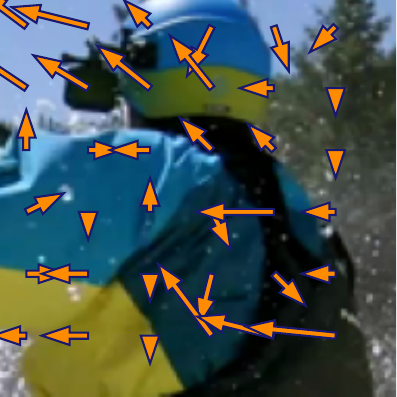} &
     \includegraphics[width=\f1ht]{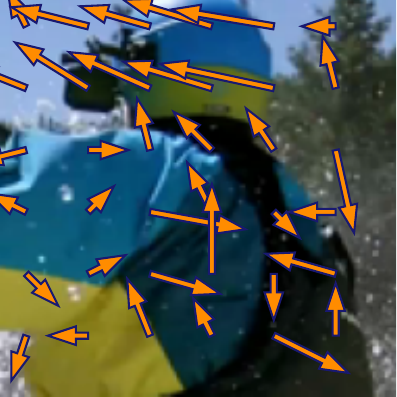} \\
     
     60 & \includegraphics[width=\f1ht]{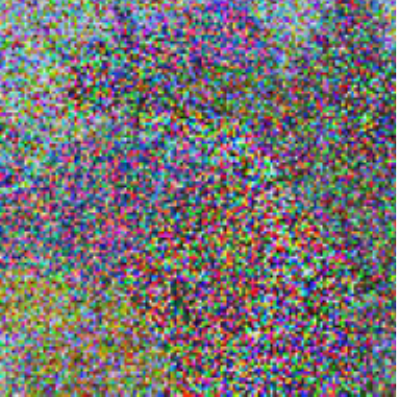} &
     \includegraphics[width=\f1ht]{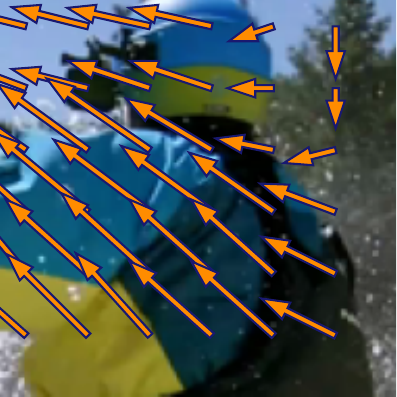} &
     \includegraphics[width=\f1ht]{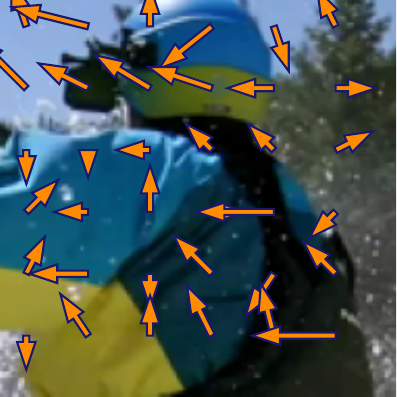} &
     \includegraphics[width=\f1ht]{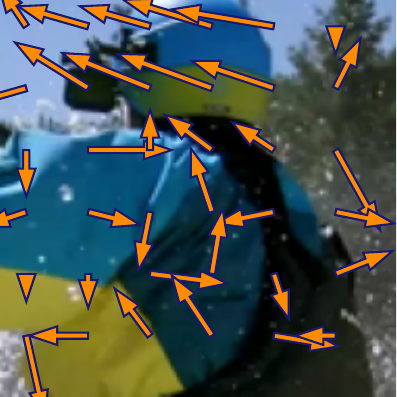} \\
     
     75 & \includegraphics[width=\f1ht]{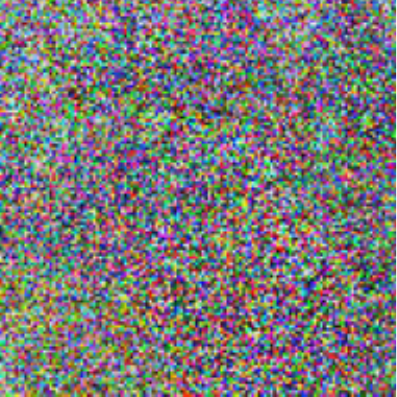} &
     \includegraphics[width=\f1ht]{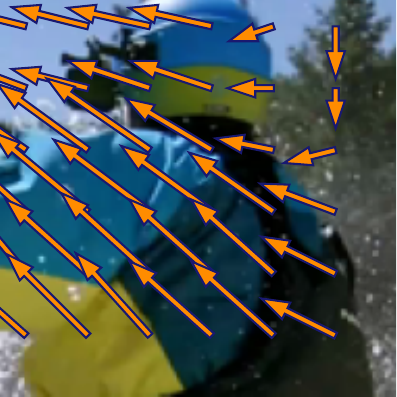} &
     \includegraphics[width=\f1ht]{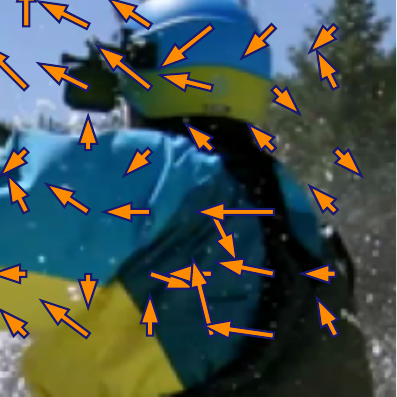} &
     \includegraphics[width=\f1ht]{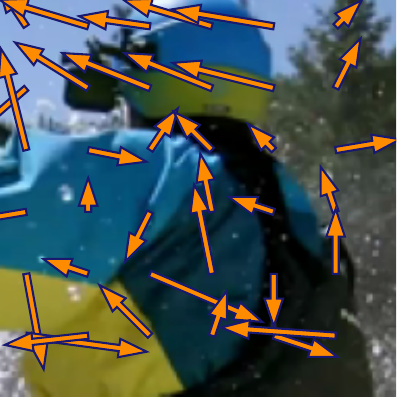} \\
     
     \end{tabular}
     
     \vspace{0.2cm}
     
\caption{\textbf{CNNs trained for denoising automatically learn to perform motion estimation; \texttt{rafting} video from Set8}. Motion estimated from the gradients of UDVD and FastDVDnet. See description of Figure~\ref{fig:motion_1}.}
\label{fig:motion_2}
\end{figure*}

\begin{figure*}[ht]
    \def\f1ht{\linewidth}%
     
     \centering 
     \begin{tabular}{ >{\centering\arraybackslash}m{0.02\linewidth}
     >{\centering\arraybackslash}m{0.23\linewidth}
     >{\centering\arraybackslash}m{0.23\linewidth}
     >{\centering\arraybackslash}m{0.23\linewidth}
     >{\centering\arraybackslash}m{0.23\linewidth}
     }
     \centering
     
     $\sigma$ & 
     \footnotesize{(a) Noisy Frame} &
     \footnotesize{(b) DeepFlow on Clean Frame} &
     \footnotesize{(c) FastDVDnet} &
     \footnotesize{(d) Ours} \\
     
     30 & \includegraphics[width=\f1ht]{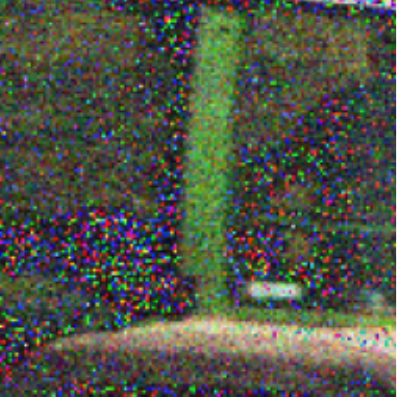} &
     \includegraphics[width=\f1ht]{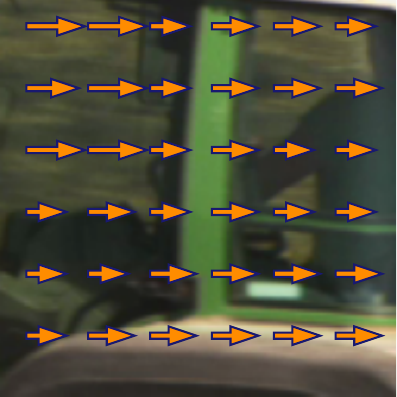} &
     \includegraphics[width=\f1ht]{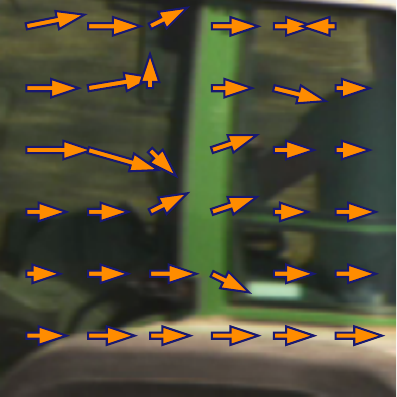} &
     \includegraphics[width=\f1ht]{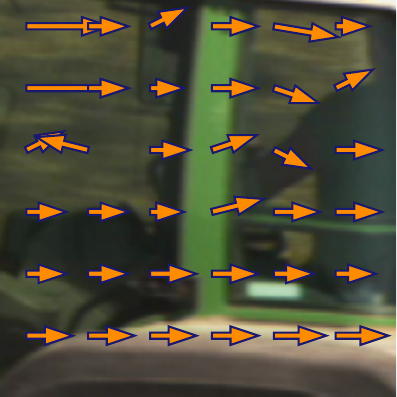} \\
     
     45 & \includegraphics[width=\f1ht]{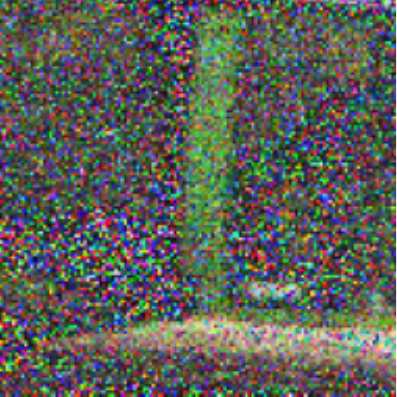} &
     \includegraphics[width=\f1ht]{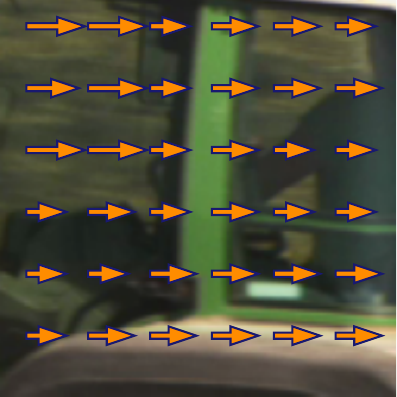} &
     \includegraphics[width=\f1ht]{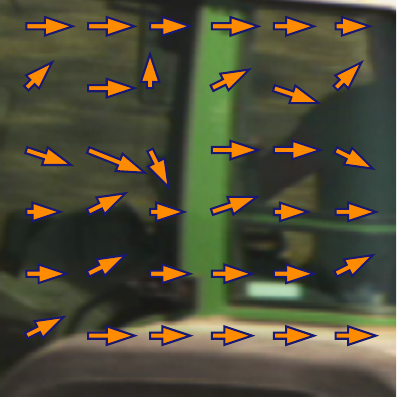} &
     \includegraphics[width=\f1ht]{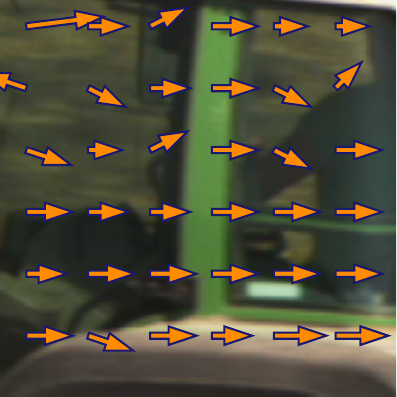} \\
     
     60 & \includegraphics[width=\f1ht]{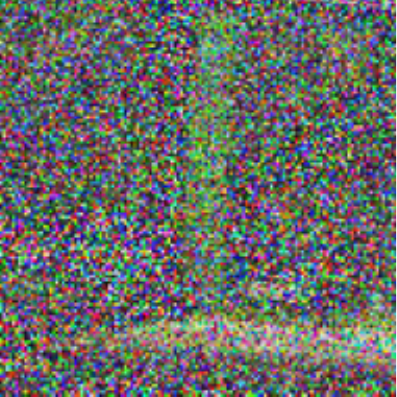} &
     \includegraphics[width=\f1ht]{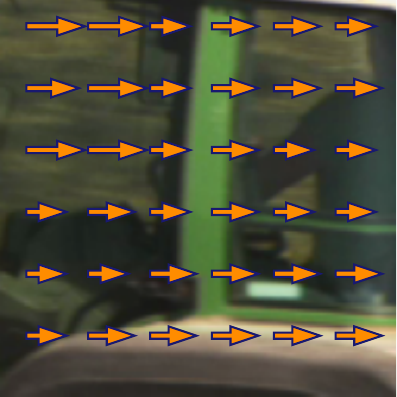} &
     \includegraphics[width=\f1ht]{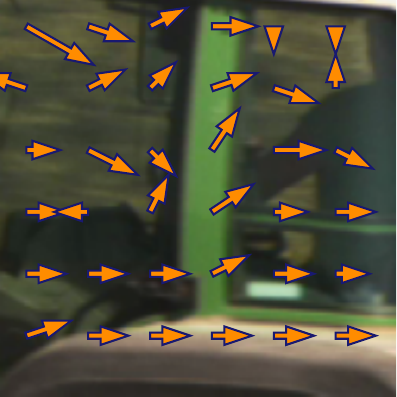} &
     \includegraphics[width=\f1ht]{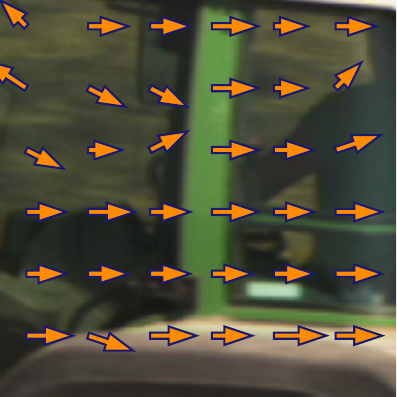} \\
     
     75 & \includegraphics[width=\f1ht]{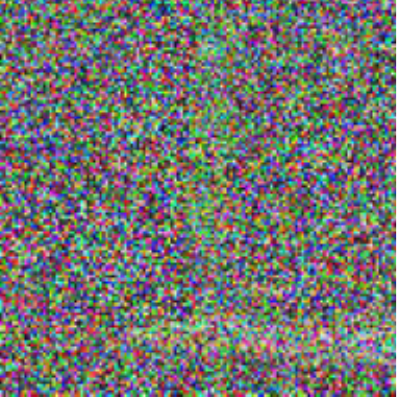} &
     \includegraphics[width=\f1ht]{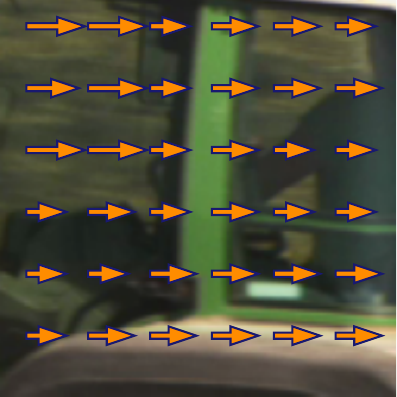} &
     \includegraphics[width=\f1ht]{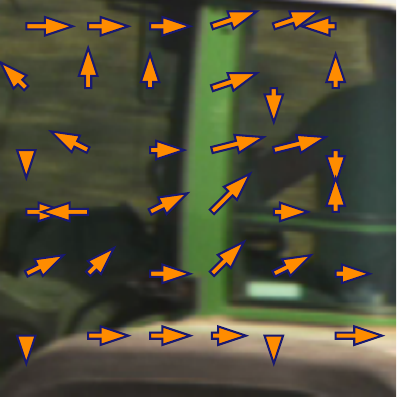} &
     \includegraphics[width=\f1ht]{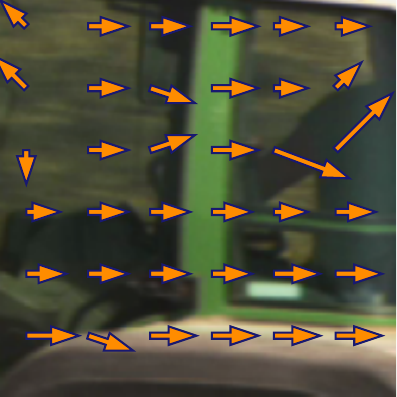} \\
     
     \end{tabular}
     
     \vspace{0.2cm}
     
\caption{\textbf{CNNs trained for denoising automatically learn to perform motion estimation; \texttt{tractor} video from Set8}.
Motion estimated from the gradients of UDVD and FastDVDnet. See description of Figure~\ref{fig:motion_1}.}
\label{fig:motion_3}
\end{figure*}

\begin{figure*}[ht]
    \def\f1ht{\linewidth}%
     
     \centering 
     \begin{tabular}{ >{\centering\arraybackslash}m{0.02\linewidth}
     >{\centering\arraybackslash}m{0.23\linewidth}
     >{\centering\arraybackslash}m{0.23\linewidth}
     >{\centering\arraybackslash}m{0.23\linewidth}
     >{\centering\arraybackslash}m{0.23\linewidth}
     }
     \centering
     
     $\sigma$ & 
     \footnotesize{(a) Noisy Frame} &
     \footnotesize{(b) DeepFlow on Clean Frame} &
     \footnotesize{(c) FastDVDnet} &
     \footnotesize{(d) Ours} \\
     
     30 & \includegraphics[width=\f1ht]{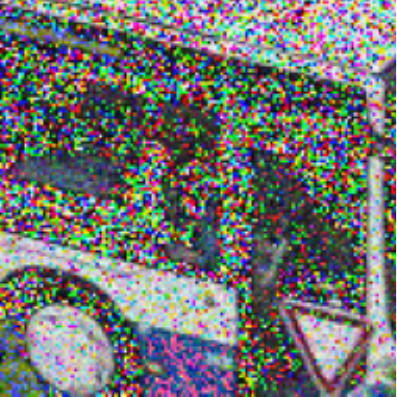} &
     \includegraphics[width=\f1ht]{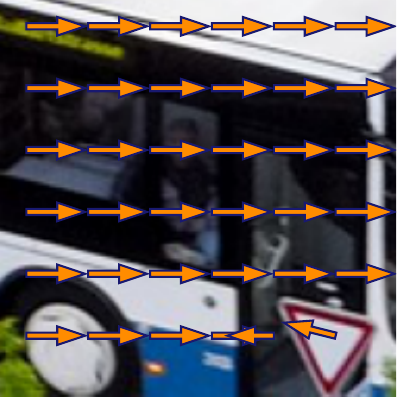} &
     \includegraphics[width=\f1ht]{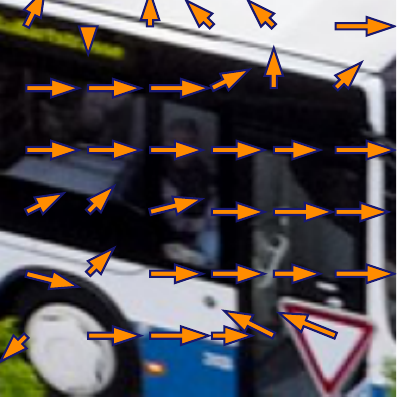} &
     \includegraphics[width=\f1ht]{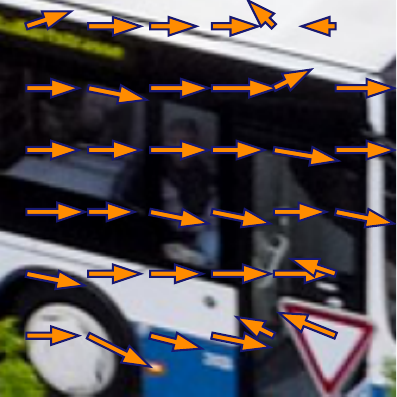} \\
     
     45 & \includegraphics[width=\f1ht]{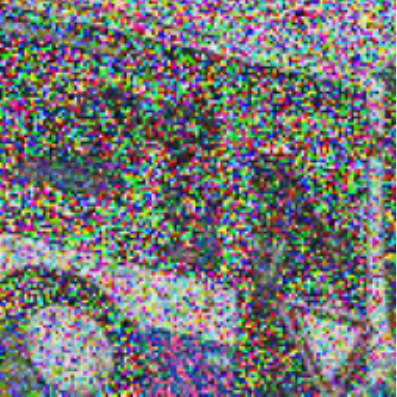} &
     \includegraphics[width=\f1ht]{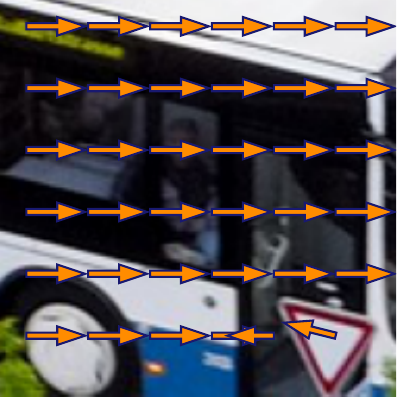} &
     \includegraphics[width=\f1ht]{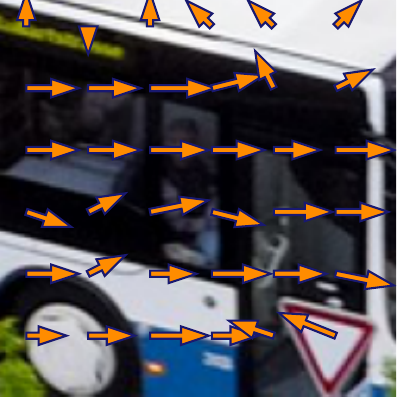} &
     \includegraphics[width=\f1ht]{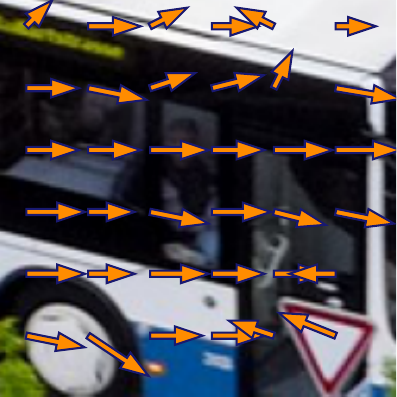} \\
     
     60 & \includegraphics[width=\f1ht]{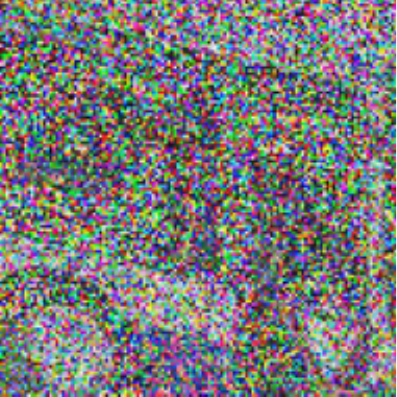} &
     \includegraphics[width=\f1ht]{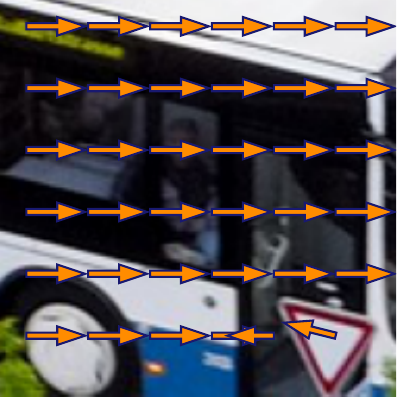} &
     \includegraphics[width=\f1ht]{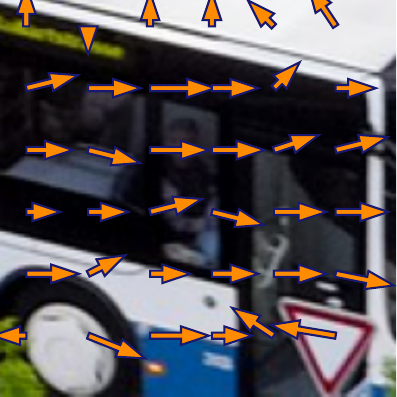} &
     \includegraphics[width=\f1ht]{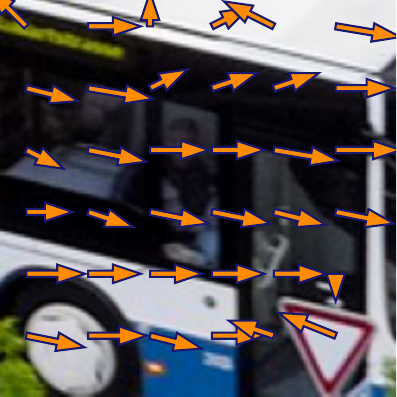} \\
     
     75 & \includegraphics[width=\f1ht]{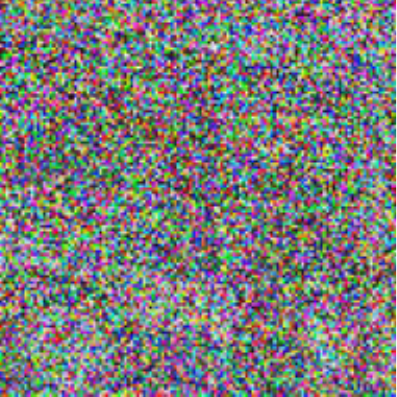} &
     \includegraphics[width=\f1ht]{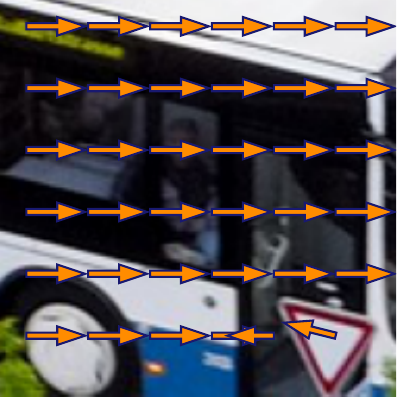} &
     \includegraphics[width=\f1ht]{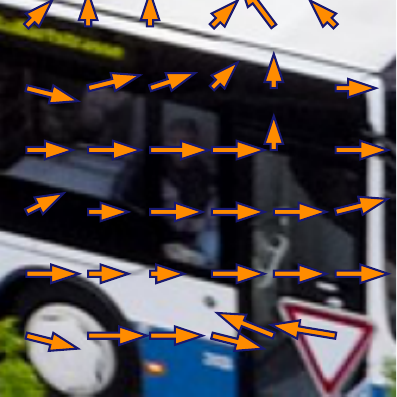} &
     \includegraphics[width=\f1ht]{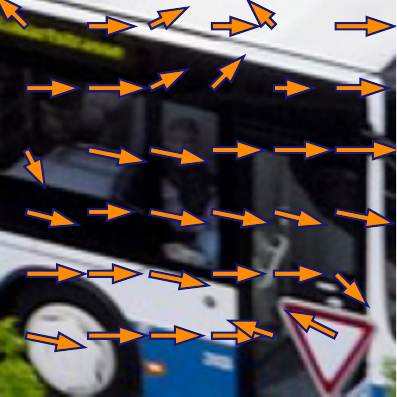} \\
     
     \end{tabular}
     
     \vspace{0.2cm}
     
\caption{\textbf{CNNs trained for denoising automatically learn to perform motion estimation; \texttt{bus} video from the DAVIS dataset}. 
Motion estimated from the gradients of UDVD and FastDVDnet. See description of Figure~\ref{fig:motion_1}.}
\label{fig:motion_4}
\end{figure*}

Figures~\ref{fig:jacobian_1}, \ref{fig:jacobian_2}, \ref{fig:jacobian_3} and \ref{fig:jacobian_4} show that the equivalent filters in adjoining frames are automatically shifted spatially to account for the movement of objects in the videos. We extracted motion information using the shift as explained in Section~6. Figures~\ref{fig:motion_1}, \ref{fig:motion_2}, \ref{fig:motion_3} and \ref{fig:motion_4} show additional examples for UDVD and FastDVDnet. The estimated optical flow is mostly consistent with the estimated obtained by DeepFlow \cite{deepflow} applied on the clean videos. The motion estimates obtained from the equivalent filters tends to be less accurate for pixels near strongly correlated features or highly homogeneous regions where the local motion is ambiguous. 
\label{sec:motion}

\end{document}